\UseRawInputEncoding
\documentclass[fleqn,usenatbib,twocolumn,usenatbib]{mnras}
\usepackage{amsmath}
\usepackage{subfigure}
\usepackage{amsfonts}
\usepackage{epstopdf}
\usepackage{longtable}
\usepackage{rotating}
\usepackage{lineno}
\usepackage{attachfile}
\usepackage{url}
\usepackage{amssymb}
\usepackage{subfigure}
\usepackage{graphicx}
\usepackage{graphicx,natbib}
\usepackage{times}
\usepackage{multirow}
\usepackage{natbib}
\usepackage{amssymb,amsmath}
\usepackage{tablefootnote}
\usepackage{lineno}
\usepackage{epstopdf}
\usepackage{appendix}
\usepackage{color}
\usepackage{multicol}        % Multi-column entries in tables
\usepackage{bm}		% Bold maths symbols, including upright Greek
\usepackage{pdflscape}	% Landscape pages
\usepackage{supertabular}	% Landscape pages
\usepackage{booktabs}

\def\mnras{MNRAS}
\def\aap{A\&A}
\def\araa{ARA\&A}
\def\aj{AJ}
\def\apjl{ApJ}
\def\apj{ApJ}
\def\apjs{ApJS}
\def\pasp{PASP}

\def\nat{Nature}

\def\aaps{A\&AS}

\title{Simultaneous Multi-band Optical Follow-up Observations of a Gamma-Ray Flare in BL Lacertae}

\author[X. Chang et al.]{
X. Chang \href{https://orcid.org/0009-0008-1701-2792}{,$^{1}$}
D. R. Xiong \href{https://orcid.org/0000-0002-6809-9575}{,$^{2}$}\thanks{E-mail: xiongdingrong@ynao.ac.cn}
Chenxu Liu \href{https://orcid.org/0000-0001-5561-2010}{,$^{1}$}
J. R. Xu \href{https://orcid.org/0000-0003-2477-3430}{,$^{7}$}
G. Bhatta \href{https://orcid.org/0000-0002-0705-6619}{,$^{6}$}
T. F. Yi \href{https://orcid.org/0000-0001-8920-0073}{,$^{3,4}$}
J. Zhang \href{https://orcid.org/0000-0003-2017-9159}{,$^{2}$}\and
Y. Pan \href{https://orcid.org/0009-0002-7625-2653}{,$^{1}$}
X. Z. Zou \href{https://orcid.org/0009-0006-5847-9271}{,$^{1}$}
X. L. Chen \href{https://orcid.org/0009-0000-4068-1320}{,$^{1}$}
Y. P. Yang \href{https://orcid.org/0000-0001-6374-8313}{,$^{1}$}
J. H. Zhang \href{https://orcid.org/0000-0002-2510-6931}{,$^{1}$}
X. K. Liu \href{https://orcid.org/0000-0003-0394-1298}{,$^{1}$}
Y. Fang \href{https://orcid.org/0009-0006-1010-1325}{,$^{1}$}\and
G. W. Du \href{https://orcid.org/0000-0002-8109-7152}{,$^{1}$}
T. Wang \href{https://orcid.org/0009-0005-8762-0871}{,$^{1}$}
X. F. Zhu \href{https://orcid.org/0009-0003-6936-7548}{,$^{1}$}
Y. L. Gong \href{https://orcid.org/0000-0003-3326-2173}{,$^{5}$}
Z. X. Wang \href{https://orcid.org/0000-0003-1984-3852}{,$^{5}$}
X. W. Liu \href{https://orcid.org/0000-0003-1295-2909}{,$^{1}$}\thanks{E-mail: x.liu@ynu.edu.cn}
\\
$^{1}$ South-Western Institute for Astronomy Research, Yunnan Key Laboratory of Survey Science, Yunnan University, Kunming, Yunnan 650504, People's Republic of China\\
$^{2}$ Yunnan Observatories, Chinese Academy of Sciences, 396 Yangfangwang, Guandu District, Kunming, 650216, People's Republic of China\\
$^{3}$ Key Laboratory of Colleges and Universities in Yunnan Province for High-energy Astrophysics, Department of Physics, Yunnan Normal University, \\Kunming 650500, People's Republic of China\\
$^{4}$ Guangxi Key Laboratory for the Relativistic Astrophysics, Nanning 530004, People's Republic of China\\
$^{5}$ Department of Astronomy, School of Physics and Astronomy, Yunnan University, Kunming 650091, People's Republic of China\\
$^{6}$ Janusz Gil Institute of Astronomy, University of Zielona G{\'o}ra, ul. Szafrana 2, 65-516 Zielona G{\'o}ra, Poland\\
$^{7}$ CAS Key Laboratory of Space Astronomy and Technology, National Astronomical Observatories, Chinese Academy of Sciences, \\Beijing, 100101, People's Republic of China\\
}

\begin{document}
\maketitle

\begin{abstract}
On $2024$ October $5$, BL Lacertae ($2200+420$) experienced one of its brightest gamma-ray flares. We conducted simultaneous follow-up observations in the $u$, $v$, $g$, $r$, $i$, and $z$ bands from $2024$ October $17$ to November $21$ using the Mephisto telescope and its two $50$ cm twin auxiliary photometric telescopes of Yunnan University. Intraday variability (IDV) was detected in the $g$, $r$, $i$, and $z$ bands. The IDV duty cycle increased with observing frequency across these bands. The shortest variability time-scale, derived from auto-correlation analysis, constrains the upper limit of the black hole mass to be $M_{\bullet} \lesssim 10^{8.29} M_{\odot}$ assuming a Kerr black hole, and $M_{\bullet} \lesssim 10^{8.77} M_{\odot}$ assuming a Schwarzschild black hole. The emission region responsible for the observed variability has a size of $R \le 3.51 \times 10^{14}$ cm and is located at a distance of $R_H \le 2.83 \times 10^{15}$ cm from the central supermassive black hole. This distance is approximately three orders of magnitude smaller than the typical radius of the broad-line region, indicating that the emission region lies well within it. A general bluer-when-brighter (BWB) trend was detected on intraday time-scales, suggesting that shock-accelerated relativistic electrons enhance the high-energy particle population, leading to spectral hardening. A potential quasi-periodic oscillation (QPO) with a period of $\sim 100.77$ minutes was detected with $>99.99$ per cent confidence, consistent with predictions from the magnetic reconnection model. These observed optical intraday variabilities and colour variations of BL Lacertae can be well explained by the turbulent jet model.
\end{abstract}

\begin{keywords}
galaxies: active --- galaxies: photometry --- BL Lacertae objects: general --- BL Lacertae objects: individual (2200+420)
\end{keywords}

\section{Introduction}\label{sec:intro}
Blazars are a type of active galactic nucleus (AGN) distinguished by relativistic jets oriented nearly along the observer's line of sight \citep{Urry95}. Their emission, dominated by the nuclear region, exhibits strong optical variability across a wide range of time-scales (from minutes to years), accompanied by sporadic and violent flares, and shows high and variable degrees of polarization \citep{Ulri97}. Their radiation spans the entire electromagnetic spectrum, with wavelengths ranging from the microwave/radio bands all the way up to the very-high-energy (VHE) gamma-ray regime \citep{Urry95}. Blazars are classified into BL Lacertae objects (BL Lacs) and flat-spectrum radio quasars (FSRQs) based on the presence or absence of prominent emission lines and/or thermal accretion disc features in their spectra \citep{Foss98,Urry95,Weav20}. BL Lacs are characterized by the absence of significant emission lines or the presence of only weak emission lines in their optical spectra \citep{Stic91}, whereas FSRQs exhibit prominent broad emission lines \citep{Fan03}. BL Lacertae (redshift $z = 0.0686 \pm 0.0004$) \citep{Verm95,Mill78,MacL68,Oke74} is classified as a Low-Synchrotron-Peaked BL Lac (LSP BL Lac) object \citep{Nils18}. This source is renowned for its intense and rapid optical variability on extremely short time-scales (ranging from several minutes to a few hours) as well as intraday time-scales \citep{Mill89, Fan98, Agar15}. Additionally, it exhibits significant variations in polarization degree \citep{Mars08,Gaur14}. Many researchers have carried out extensive studies based on long-term optical monitoring of BL Lacertae, aiming to investigate its flux variability, spectral evolution, and potential periodic signals in its light curves
\citep[e.g.,][]{Raci70,Spez98,Fan01,Hu06}.

BL Lacertae is notable for its prominent short-term variability, with a particular emphasis on its intraday variability (IDV) \citep[e.g.,][]{Mill89, Meng17, Fang22}. BL Lacertae has been monitored in the optical band for nearly a century, revealing extreme optical variability characteristics \citep{Sitk85}. \citet{Meng17} detected significant IDV in different optical bands during $13$ nights of observations between $2012$ and $2016$. Using optical data from the Transiting Exoplanet Survey Satellite (TESS), \citet{Weav20} reported ultra-short-timescale variability of approximately 30 minutes, alongside an X-ray variability period of $14.5$ hours. Furthermore, \citet{Kali23} found that the amplitude of intraday variability can reach up to $\sim 30\%$, with the variability displaying a clear dependence on observational wavelengths. \citet{Fang22} recorded IDV across all three optical bands during four nights of observations and confirmed that its variability characteristics are independent of the optical state of the blazar. The minimum variability time-scale is a critical parameter for estimating the size of the emission region and the mass of the central black hole \citep[e.g.,][]{Mill89, Liu15}. \citet{Xie02} reported a minimum variability time-scale of $45.9$ minutes, while \citet{Meng17} measured a time-scale of $42.5$ minutes. Based on this, \citet{Meng17} estimated the black hole mass to be $1.91 \times 10^7 M_{\odot}$, whereas \citet{Xie02} obtained a higher value of $4.9 \times 10^7 M_{\odot}$. Notably, these estimates are highly sensitive to the assumed Doppler factor ($\delta$). The systematic underestimation of black hole masses in the aforementioned studies can be attributed to the use of conservative Doppler factors, such as $\delta \approx 1.23$ in \citet{Xie02}. These results can be explained by particle acceleration mechanisms induced by shocks within the jet. We employed the turbulent jet model \citep{Bhat13} to interpret the observed IDV/micro-variation phenomena. Within this model, a plane shock wave interacts with a turbulent cell, accelerating and energizing the electrons within it. These electrons subsequently cool down through synchrotron radiation. During this process, a flux pulse is generated, which manifests as a flare on the light curve \citep{Webb21}. The observed IDV arises from the aggregation of individual pulses that originate from turbulent cells with diverse properties \citep{Agar23}. The investigation of optical flares plays a critically vital role in gaining an in-depth comprehension of the physical mechanisms that trigger or sustain these phenomena \citep{Kali23}.

Flux variability in blazars is often accompanied by significant changes in colour (or spectral) characteristics. Specifically, the colour variations can typically exhibit two distinct modes:``bluer when brighter" (BWB) or ``redder when brighter" (RWB) \citep{Bonn12,Meng17,Li21,Fang22,Chan23}. Additionally, some blazars may display achromatic behavior, where their colour or spectral features remain relatively stable despite changes in brightness \citep{Vill02,Rait03}. BL Lacertae typically exhibits the BWB trend in its colour index evolution \citep{Meng17,Li21,Fang22}. This BWB behavior is primarily attributed to the dominance of non-thermal emission from the relativistic jet \citep{Rait13,Agar15,Xion17}. \citet{Vill02} further emphasized that the observed BWB trend is a direct manifestation of the jet's intrinsic radiative properties rather than contamination from the host galaxy. The spectral variability of BL Lacertae typically exhibits a significantly enhanced BWB component on intraday time-scales, while presenting mildly chromatic behavior on time-scales of days to months \citep{Vill04}. The mildly chromatic behavior is believed to result from Doppler factor fluctuations caused by the dynamical state changes of the jet. Observations indicate that this source may also exhibit a BWB evolution trend on long-term time-scales (months to years) \citep{Meng17,Gaur19}.

Some sources exhibit intraday periodic oscillations. \citet{Kinz88} analysed the $7$ mm wavelength light curve of OJ 287 and identified a periodic signal with a duration of $35$ minutes. Meanwhile, \citet{Valt85} detected a potential oscillatory signal with a period of $15.7$ minutes in the radio band for the same source. In the case of PKS  0735+178, \citet{Yuan21} reported, based on the optical light curve data from January $6$, $2016$, the presence of an oscillation with a period of $66.9 \pm 4.1$ minutes. For BL Lacertae, \citet{Jors22} conducted an analysis incorporating optical flux, optical linear polarization, and $\gamma$-ray flux measurements, and subsequently reported the discovery of a quasi-periodic oscillation (QPO) with an approximate period of $13$ hours. More recently, \citet{Jors22} again reported the detection of a transient periodic signal of $0.55$ days in the $R$-band light curve generated by the Whole Earth Blazar Telescope. The study of QPOs provides valuable insights into the accretion disc, relativistic jets, and the dynamic processes occurring within the central engine of blazars \citep{Li23}. To investigate the presence of QPOs in BL Lacertae and the underlying physical mechanisms, we conducted a detailed search and analysis of its light curves for QPO signatures.

The study of blazars provides crucial scientific insights into the intricate physical processes within jets, including particle acceleration mechanisms, the origins of flaring activity, and the structure and dynamical evolution of the emission region \citep{Mill89,Gupt09,Agar15,Liu15,Xion17}. Despite the progress made over the years, a fully self-consistent theoretical model capable of comprehensively explaining all observed features has yet to be developed. Therefore, a systematic study of the IDV of BL Lacertae can significantly advance the improvement of existing models, particularly when the colour variations are simultaneously investigated. In this work, we strive to elucidate the possible physical processes underlying the observed variations in BL Lacertae, including its multi-band flux and colour variability, as well as the potential presence of QPOs.

The structure of this paper is organized as follows: Section \ref{sec_data} provides a detailed description of the observations and data reduction. Section \ref{sec_results} presents the research results. Sections \ref{sec_discuss} and \ref{sec_conclusion} offer the discussion and conclusion, respectively.

\section{Observations and Data Reduction}
\label{sec_data}
The Fermi Large Area Telescope (LAT) detected gamma-ray flaring activity from BL Lacertae. On October $5$, $2024$, the source was observed to be in a state of enhanced gamma-ray emission. The daily averaged gamma-ray flux (for E $>100$ MeV) was measured to be $ (10.4 \pm 0.5) \times 10^{-8} $ photons cm$^{-2}$ s$^{-1}$ (ATel \#16849 \footnote{\url{https://www.astronomerstelegram.org/?read=16849}}) \citep{Van24}. VERITAS (Very Energetic Radiation Imaging Telescope Array System) reported the detection of very high-energy (VHE; E $> 100$ GeV) gamma-ray flaring activity from BL Lacertae. The data indicated that the flaring activity was prominent between October $5$ and $6$, with energies exceeding $200$ GeV. The preliminary flare flux was measured as $F = (4.1 \pm 0.7) \times 10^{-7}$ m$^{-2}$ s$^{-1}$. By October $7$, the flux had dropped significantly (ATel \#16854 \footnote{\url{ https://www.astronomerstelegram.org/?read=16854}}) \citep{Coll24}. Swift-XRT also recorded the highest X-ray flux during October $5$, measured as $ (7.9 \pm 0.8) \times 10^{-11} $ erg cm$^{-2}$ s$^{-1}$. After October $6$, the X-ray flux decreased significantly to $ (4.4 \pm 0.4) \times 10^{-11}$ erg cm$^{-2}$ s$^{-1}$ (ATel \#16855 \footnote{\url{https://www.astronomerstelegram.org/?read=16855}}) \citep{Noza24}. The $1.3$-meter Devasthal Fast Optical Telescope (DFOT), located at the Aryabhatta Research Institute of Observational Sciences (ARIES), conducted $BVRI$ multi-band photometric monitoring of BL Lacertae. The results indicate an elevated activity state in the optical bands of BL Lacertae, consistent with the enhanced activity observed in other wavelength bands (ATel \#16856 \footnote{\url{https://www.astronomerstelegram.org/?read=16856}}) \citep{Kish24}. We subsequently conducted follow-up observations. Due to limitations in telescope availability and weather conditions, our observations commenced on October $17$.

During the period from $2024$ October $17$ to November $21$, we conducted a $23$-night monitoring on BL Lacertae. The monitoring was carried out using the $1.6$-meter telescope (Mephisto) located at the Lijiang Observatory of Yunnan University in China, as well as the auxiliary photometric telescopes of Mephisto, which are two $50$-centimeter twin telescopes. Mephisto is a wide-field multi-channel telescope that utilizes a Ritchey-Chr{\'e}tien (RC) system with correctors and film-coated cubic prisms. The telescope features a $1.6$ m primary mirror. It is equipped with three-channel single-chip CCD cameras, covering 1/4 of the field of view (FOV). By utilizing a dichroic prism, we conducted simultaneous multi-colour observations, enabling simultaneous measurements in the $uv$, $gr$, and $iz$ bands. The twin telescopes, model Alluna RC $20$, are equipped with a flat-field corrector and feature an optical aperture of $505$ mm along with a focal ratio of $f/8.1$. Each telescope is equipped with an FLI ML $50100$ CCD camera, featuring a resolution of $8176 \times 6132$ pixels with a pixel size of $6$ $\mu$m. The observations were conducted simultaneously using the twin telescopes, with filters in the $g$, $r$, $i$, and $z$ bands. The filter systems on these facilities are identical to those on Mephisto, thus allowing for coordinated observations. To achieve a high signal-to-noise ratio (SNR) and ensure intensive sampling, an exposure time of $10$ to $100$ seconds was adopted for the $1.6$ m telescope, while an exposure time of $100$ to $120$ seconds was used for the $50$ cm telescopes. During this period, we collected a total of $539$ data points ($20$ days) across six optical bands using the $1.6$ m telescope, specifically: $41$ in $u$-band, $41$ in $v$-band, $166$ in $g$-band, $160$ in $r$-band, $67$ in $i$-band, and $64$ in $z$-band. We gathered $1479$ data points ($16$ days) across four optical bands of the $50$ cm telescopes, specifically: $371$ in $g$-band, $371$ in $r$-band, $369$ in $i$-band, $368$ in $z$-band, respectively.

The raw frames were processed using a custom pipeline developed for Mephisto, which included bias and dark subtraction, flat-field correction, and cosmic ray removal. Source detection was performed with \texttt{SExtractor} \footnote{\url{https://www.astromatic.net/software/sextractor/}} \citep{Bert96}. Astrometric calibration utilized reference stars from the Gaia DR3 catalog \citep[DR3;][]{Gaia23}, achieving uncertainties better than $0.1$ pixels. Aperture photometry was then carried out using \texttt{SExtractor} \citep{Bert96}. For photometric calibration, we employed recalibrated Gaia BP/RP (XP) spectra via the Synthetic Photometry Method \citep{De23,Mont23,Huan24}. Synthetic AB magnitudes were computed by convolving the XP spectra with the Mephisto telescope's filter transmission curves. We then determined, for each frame, the difference $\Delta m$ between instrumental magnitudes and these synthetic magnitudes for high-quality, non-variable stars. To account for any gain measurement inaccuracies, we calculated a weighted average of $\Delta m$ for each CCD output and applied this correction to all targets on that output. The weights incorporated uncertainties from both synthetic and instrumental magnitude measurements. The root-mean-square (RMS) error $\sigma$ of the photometry is calculated using the following formula \citep{Xion17}:
\begin{equation}
\sigma=\sqrt\frac{\sum(m_i-\overline{m})^2}{N-1},
\label{eq:LebsequeIp1}
\end{equation}
where $m_i = (m_{S1} - m_{S2})_i$ denotes the differential magnitude between comparison star ($S1$) and check star ($S2$) for a specific observation, while $\overline m =\overline {m_{S1} - m_{S2}}$ represents the average of $m_i$ for a given night's dataset. The star $S1$ and star $S2$ are taken from the finding chart of \cite{Smit91}\footnote{\url{https://www.lsw.uni-heidelberg.de/projects/extragalactic/charts/2200+420.html}}, corresponding to star ${\text{B}}$ and star ${\text{C}}$ in the finding chart, respectively. The amplitude (Amp) of intranight variability is provided by \citet{Heid96}. The formula is as follows:
\begin{equation}
 \mathrm{Amp}=100\times\sqrt{(A_{\text{max}}-A_{\text{min}})^2-2\sigma^2} \; \mathrm{percent},
\label{eq:Lebseque2}
\end{equation}
where $A_{\text{max}}$ and $A_{\text{min}}$ represent the maximum and minimum values, respectively, in the calibrated light curve of the blazar, and $\sigma$ denotes the RMS error. The results of the observations are given in Table \ref{table1}. The light curves are given in Figures \ref{1} $\sim$ \ref{5}, respectively.

\begin{table*}
   \centering
   \caption{The log of observations. Column 1 is the universal time (UT) of the observations, column 2 is the corresponding Julian day (JD), column 3 is the magnitudes, column 4 is the RMS errors, and column 5 is the observed band.}
   %\label{tab:tab1}
   \renewcommand\arraystretch{1.2}
  \setlength{\tabcolsep}{6.2mm}
   \begin{tabular}{cccccccccccc} % four columns, alignment for each
      \hline
Date (UT) & JD & Mag & $\sigma$ & Band \\\hline
2024 Oct 17	&	2460601.138	&	13.874	&	0.003	&	g	\\
2024 Oct 17	&	2460601.139	&	13.882	&	0.003	&	g	\\
...	&	...	&	...	&	...	&	...	\\\hline
\multicolumn{5}{l}{(This table is available in its entirety in machine-readable form.)}\\
\label{table1}
   \end{tabular}
\end{table*}

\section{Results}
\label{sec_results}
\subsection{Variability Analysis}
\label{idv_analysis}
To investigate the IDV/micro-variation properties of BL Lacertae, we conducted a search for IDV utilizing the F-test and one-way analysis of variance (ANOVA) \citep{Dieg10,Hu14,Hong18,Yuan21}.
The F-test \citep{Dieg10,Josh11,Hu14,Agar15} is an effective tool for quantifying variability across different time-scales. This method compares the differential variance of the source with that of the check star. The F value is defined as:
\begin{equation}
F_1=\frac{Var(BL-StarS1)}{Var(StarS1-StarS2)},
\label{eq:LebsequeIp1}
\end{equation}
\begin{equation}
F_2=\frac{Var(BL-StarS2)}{Var(StarS1-StarS2)},
\label{eq:LebsequeIp2}
\end{equation}
where $Var(BL-StarS1)$ and $Var(BL-StarS2)$ respectively denote the measured variances of the differential instrumental magnitudes between the blazar BL Lacertae and the comparison $S1$ and check star $S2$, while $Var(StarS1-StarS2)$ represents the measured variance of the differential instrumental magnitudes between the comparison star $S1$ and the check star $S2$. The calculated mean F value is compared with the critical value, $F^{\alpha}_{\nu_{bl},\nu_\ast}$, where $\nu_{bl}$ and $\nu_\ast$ denote the degrees of freedom for the blazar and the comparison star, respectively ($\nu = N - 1$). The significance level $\alpha$ is set to $0.01$, corresponding to a $99\%$ confidence level ($2.6 \sigma$) \citep{Xion17}. If the mean F value exceeds the critical value, it can be concluded with $99\%$ confidence that the blazar exhibits variability \citep{Dieg10,Xion17}.

The one-way analysis of variance (ANOVA) test is a robust and powerful statistical method for detecting IDV/micro-variation in blazars, as it identifies micro-variations based on the inherent variability of the blazar rather than on measurement errors \citep{Dieg10}. In the ANOVA test, the data sample is divided into several groups, and the variances between these subsamples are analysed \citep{Li24}. The ANOVA test statistic is calculated as follows \citep{Dieg10}:
\begin{equation}
F_{\mathrm{(ANOVA)}}=\frac{{\textstyle \sum_{j=1}^{k}} n_j(\bar{{y_i}}-\bar{y})^2/(k-1)}{{\textstyle \sum_{j=1}^{k}}  {\textstyle \sum_{i=1}^{n_j}} (y_{ij}-\bar{y_j})^2/(N-k)},
\label{eq:LebsequeIp3}
\end{equation}
where $k$ represents the number of groups, $N$ denotes the total data set, and $n_j$ is the $j$th group. $y_{ij}$, $\bar{y_j}$, and $\bar{y}$ denote the $i$th observation in the $j$th group, the average of the $j$th group, and the overall average of the data set, respectively. Taking into account the time intervals and durations of the observations, we bin the data in $k$ groups of three or five observations each \citep{Dieg10,Xion17,Hong18}. Should the measurements in the last group fall short of three or five, they will be combined with the preceding group. The critical value $F$$^{\alpha}_{\nu_1,\nu_2}$ is determined based on the $F$-statistic, where $\alpha = 0.01, \nu_1 = k - 1$, and $\nu_2 = N - k$. If the F value calculated from the equation exceeds the critical value $F$$^{\alpha}_{\nu_1,\nu_2}$, the blazar is considered variable with a confidence level exceeding $99\%$ \citep{Hu14}.

\begin{figure*}
\centering
\includegraphics[scale=.28]{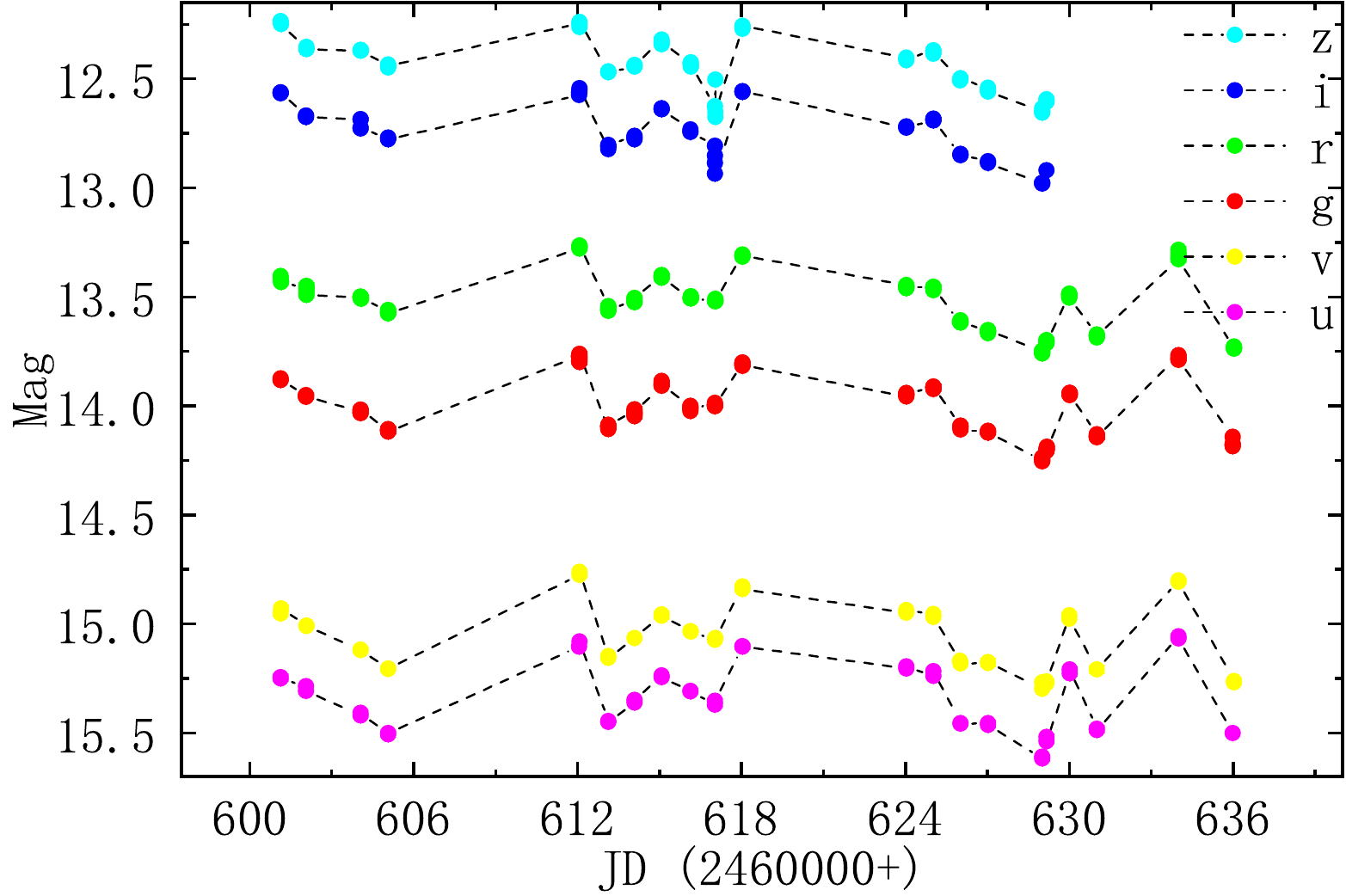}
\includegraphics[scale=.28]{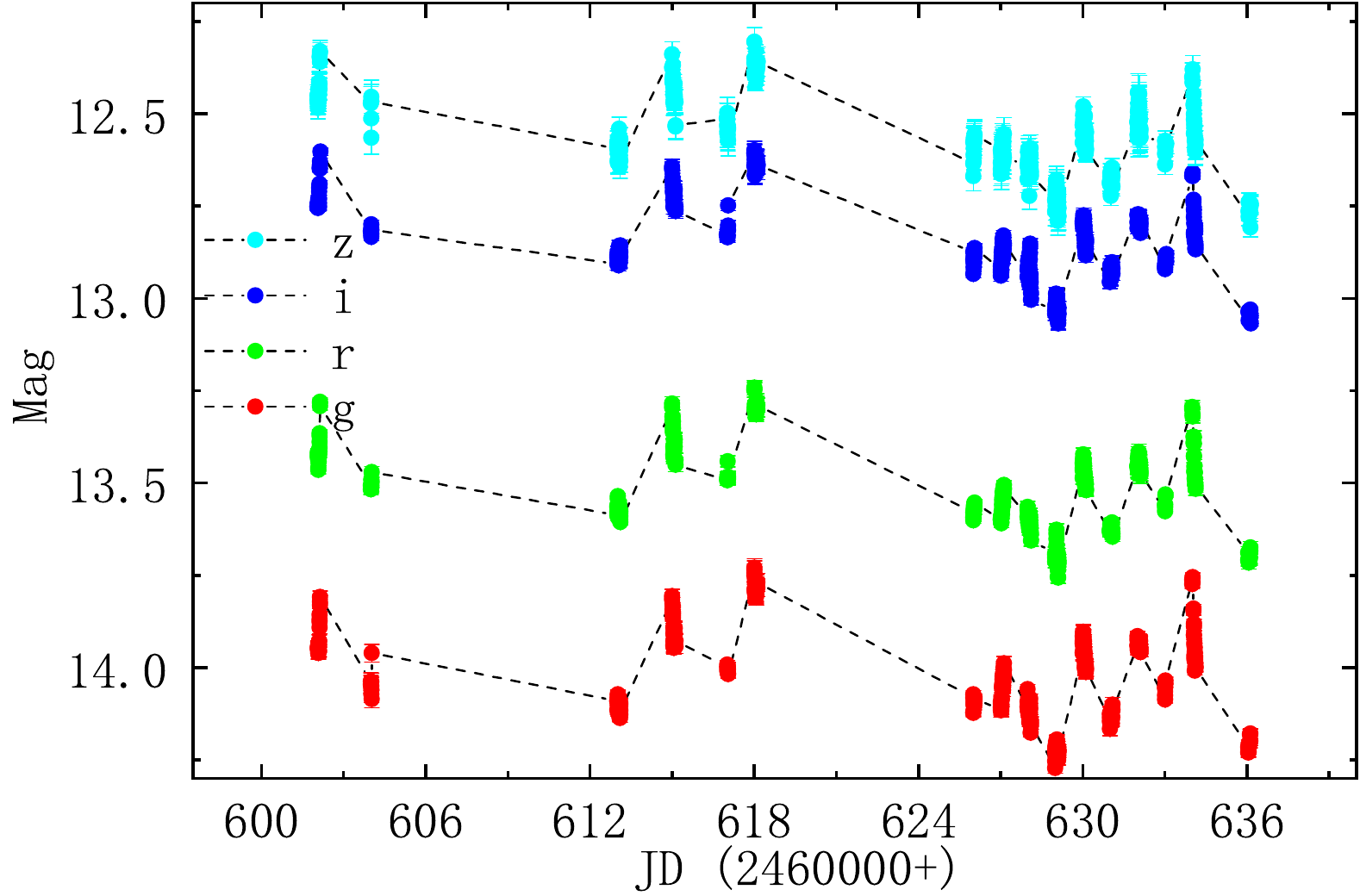}
\caption{$1.6$ m (left) and $50$ cm (right) Light curves of BL Lacertae observed from 2024 October to November. \label{1}}
\end{figure*}

\begin{figure*}
\centering
\includegraphics[scale=.64]{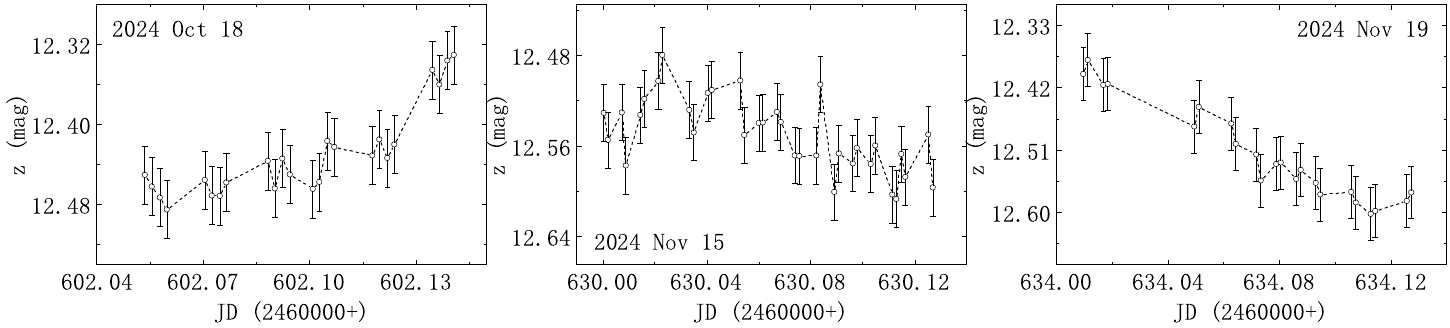}
\caption{$50$ cm IDV Light curves of BL Lacertae in the $z$-band.
\label{5}}
\end{figure*}

The blazar is classified as variable (V) if its light curve satisfies both test criteria on a given night. If not (i.e., one or all criteria are unmet), it is classified as non-variable (N). Table \ref{table2} $\sim$ \ref{table5} present the results of the variability analysis. Both testing methods indicate that IDVs were detected in different bands ($8$ days in the $g$-band, $7$ days in the $r$-band, $6$ days in the $i$-band, and $3$ days in the $z$-band). The corresponding light curves are presented in Figures \ref{1} $\sim$ \ref{5}. The duty cycle (DC) of BL Lacertae is calculated based on the definition provided by \citet{Rome99}:
\begin{equation}
DC = 100  \ \ \frac{{\textstyle \sum_{i=1}^{n}} N_i(1/\Delta T_i)}{{\textstyle \sum_{i=1}^{n}} (1/\Delta T_i)}  \ \  \mathrm{percent},
\label{eq:LebsequeIp4}
\end{equation}
where $\Delta T_i = \Delta T_{i,obs}(1+z)^{-1}$ represents the corrected duration of the monitoring session for the source on the $i$th night, with the correction based on its cosmological redshift $z$. Since the monitoring durations are not always equal on different nights, the calculation of the DC is weighted by the actual monitoring duration $\Delta T_i$ on the $i$th night. $N_i$ is set to $1$ if IDV is detected by the two tests (as shown in Table \ref{table2} $\sim$ \ref{table5}), otherwise $N_i = 0$ \citep{Goya13}. During this period, the DCs of BL Lacertae in different bands are as follows: $46.83\%$ in the $g$-band, $41.33\%$ in the $r$-band, $36.13\%$ in the $i$-band, and $19.27\%$ in the $z$-band.

\begin{table*}
   \centering
   \caption{Results of the $g$-band IDV analysis. Column 1: The date of the observation; Column 2: The number of data points; Column 3: The F value of F-test; Column 4: The critical F value of F-test with 99\% confidence level; Column 5: The F value of ANOVA; Column 6: The critical F value of ANOVA with 99\% confidence level; Column 7: The variability status (V: variable, N: non-variable); Column 8: The daily average magnitudes and errors; Column 9: The daily amplitude.}
   %\label{tab:tab1}
   \renewcommand\arraystretch{1.2}
  \setlength{\tabcolsep}{4.2mm}
   \begin{tabular}{cccccccccccc} % four columns, alignment for each
      \hline
Date    &   Number  &   $F$ &   $F_c(99)$   &   $F_a$   &   $F_a(99)$   &   V/N &   Ave(mag) & Amp \\\hline
2024 Oct 18	&	22	&	8.12	&	2.86	&	77.30	&	4.28	&	V	&	13.91 (0.017)	& 15.02\\
2024 Oct 29	&	34	&	2.43	&	2.29	&	12.71	&	3.18	&	V	&	14.11 (0.011)	& 6.28\\
2024 Oct 31	&	31	&	5.56	&	2.39	&	30.41	&	3.67	&	V	&	13.89 (0.018)	& 13.74\\
2024 Nov 02	&	12	&	0.65	&	4.46	&	21.90	&	7.59	&	N	&	14.00 (0.012)	& 1.89\\
2024 Nov 03	&	28	&	4.93	&	2.51	&	15.43	&	3.60	&	V	&	13.78 (0.024)	& 6.95\\
2024 Nov 11	&	14	&	1.60	&	3.91	&	1.32	&	6.42	&	N	&	14.10 (0.010)	& 4.90\\
2024 Nov 12	&	36	&	17.95	&	2.23	&	17.18	&	3.09	&	V	&	14.06 (0.019)	& 12.45\\
2024 Nov 13	&	28	&	3.67	&	2.51	&	12.51	&	3.60	&	V	&	14.12 (0.013)	& 11.60\\
2024 Nov 14	&	32	&	1.44	&	2.35	&	14.28	&	3.31	&	N	&	14.22 (0.014)	& 7.42\\
2024 Nov 15	&	34	&	4.54	&	2.29	&	34.45	&	3.18	&	V	&	13.96 (0.019)	& 10.49\\
2024 Nov 16	&	24	&	0.91	&	2.72	&	1.91	&	4.03	&	N	&	14.13 (0.019)	& 5.89\\
2024 Nov 17	&	32	&	1.24	&	2.35	&	16.74	&	3.31	&	N	&	13.94 (0.013)	& 3.91\\
2024 Nov 19	&	22	&	97.97	&	2.86	&	41.58	&	4.28	&	V	&	13.91 (0.012)	& 25.32\\\hline
\label{table2}
   \end{tabular}
\end{table*}

\begin{table*}
   \centering
   \caption{Results of the $r$-band IDV analysis. Columns 1 to 9 are identical to those in Table 2.}
   %\label{tab:tab1}
   \renewcommand\arraystretch{1.2}
  \setlength{\tabcolsep}{4.2mm}
   \begin{tabular}{cccccccccccc} % four columns, alignment for each
      \hline
Date    &   Number  &   $F$ &   $F_c(99)$   &   $F_a$   &   $F_a(99)$   &   V/N &   Ave(mag) & Amp \\\hline
2024 Oct 18	&	22	&	2.92	&	2.86	&	103.01	&	4.28	&	V	&	13.40 (0.010)	& 18.43\\
2024 Oct 29	&	34	&	4.20	&	2.29	&	14.25	&	3.18	&	V	&	13.58 (0.008)	& 6.84\\
2024 Oct 31	&	30	&	3.97	&	2.42	&	44.62	&	3.46	&	V	&	13.38 (0.019)	& 16.28\\
2024 Nov 02	&	12	&	1.10	&	4.46	&	6.59	&	7.59	&	N	&	13.48 (0.015)	& 4.62\\
2024 Nov 03	&	28	&	1.78	&	2.51	&	33.81	&	3.60	&	N	&	13.29 (0.020)	& 6.68\\
2024 Nov 11	&	14	&	0.62	&	3.91	&	0.73	&	6.42	&	N	&	13.58 (0.009)	& 4.50\\
2024 Nov 12	&	36	&	2.85	&	2.23	&	15.99	&	3.09	&	V	&	13.56 (0.012)	& 10.18\\
2024 Nov 13	&	29	&	2.91	&	2.46	&	20.07	&	3.52	&	V	&	13.60 (0.015)	& 8.76\\
2024 Nov 14	&	32	&	2.28	&	2.35	&	11.72	&	3.31	&	N	&	13.70 (0.017)	& 12.67\\
2024 Nov 15	&	34	&	2.34	&	2.29	&	53.70	&	3.18	&	V	&	13.48 (0.017)	& 9.43\\
2024 Nov 16	&	24	&	0.72	&	2.72	&	3.42	&	4.03	&	N	&	13.62 (0.014)	& 3.21\\
2024 Nov 17	&	32	&	0.84	&	2.35	&	10.28	&	3.31	&	N	&	13.45 (0.021)	& 5.63\\
2024 Nov 19	&	22	&	13.71	&	2.86	&	48.97	&	4.28	&	V	&	13.43 (0.018)	& 21.97\\\hline
\label{table3}
   \end{tabular}
\end{table*}

\begin{table*}
   \centering
   \caption{Results of the $i$-band IDV analysis. Columns 1 to 9 are identical to those in Table 2.}
   %\label{tab:tab1}
   \renewcommand\arraystretch{1.2}
  \setlength{\tabcolsep}{4.2mm}
   \begin{tabular}{cccccccccccc} % four columns, alignment for each
      \hline
Date    &   Number  &   $F$ &   $F_c(99)$   &   $F_a$   &   $F_a(99)$   &   V/N &   Ave(mag) & Amp \\\hline
2024 Oct 18	&	25	&	5.77	&	2.66	&	40.90	&	3.89	&	V	&	12.72 (0.010)	& 15.18\\
2024 Oct 29	&	33	&	1.14	&	2.32	&	4.35	&	3.26	&	N	&	12.89 (0.015)	& 4.88\\
2024 Oct 31	&	29	&	3.18	&	2.46	&	19.06	&	3.52	&	V	&	12.72 (0.020)	& 11.72\\
2024 Nov 02	&	12	&	1.19	&	4.46	&	3.21	&	7.59	&	N	&	12.82 (0.014)	& 8.39\\
2024 Nov 03	&	28	&	0.74	&	2.51	&	3.40	&	3.60	&	N	&	12.63 (0.024)	& 5.95\\
2024 Nov 11	&	14	&	1.83	&	3.91	&	0.95	&	6.42	&	N	&	12.89 (0.011)	& 6.84\\
2024 Nov 12	&	32	&	3.34	&	2.35	&	15.33	&	3.31	&	V	&	12.88 (0.015)	& 10.59\\
2024 Nov 13	&	30	&	3.70	&	2.42	&	13.29	&	3.46	&	V	&	12.92 (0.014)	& 15.08\\
2024 Nov 14	&	32	&	0.86	&	2.35	&	6.35	&	3.31	&	N	&	13.02 (0.018)	& 7.66\\
2024 Nov 15	&	34	&	2.47	&	2.29	&	35.92	&	3.18	&	V	&	12.82 (0.019)	& 10.51\\
2024 Nov 16	&	24	&	1.12	&	2.72	&	0.60	&	4.03	&	N	&	12.93 (0.018)	& 4.75\\
2024 Nov 17	&	32	&	0.93	&	2.35	&	3.26	&	3.31	&	N	&	12.79 (0.015)	& 4.49\\
2024 Nov 19	&	22	&	61.40	&	2.86	&	18.01	&	4.28	&	V	&	12.78 (0.008)	& 20.69\\\hline
\label{table4}
   \end{tabular}
\end{table*}

\begin{table*}
   \centering
   \caption{Results of the $z$-band IDV analysis. Columns 1 to 9 are identical to those in Table 2.}
   %\label{tab:tab1}
   \renewcommand\arraystretch{1.2}
  \setlength{\tabcolsep}{4.2mm}
   \begin{tabular}{cccccccccccc} % four columns, alignment for each
      \hline
Date    &   Number  &   $F$ &   $F_c(99)$   &   $F_a$   &   $F_a(99)$   &   V/N &   Ave(mag) & Amp \\\hline
2024 Oct 18	&	24	&	3.12	&	2.72	&	6.50	&	4.03	&	V	&	12.43 (0.029)	& 14.86\\
2024 Oct 29	&	33	&	2.24	&	2.32	&	2.46	&	3.26	&	N	&	12.60 (0.032)	& 9.27\\
2024 Oct 31	&	29	&	1.66	&	2.46	&	14.21	&	3.52	&	N	&	12.44 (0.034)	& 19.02\\
2024 Nov 02	&	12	&	2.24	&	4.46	&	15.35	&	7.59	&	N	&	12.54 (0.042)	& 4.45\\
2024 Nov 03	&	28	&	1.35	&	2.51	&	5.58	&	3.60	&	N	&	12.37 (0.039)	& 7.45\\
2024 Nov 11	&	14	&	0.92	&	3.91	&	0.79	&	6.42	&	N	&	12.59 (0.040)	& 9.77\\
2024 Nov 12	&	32	&	1.40	&	2.35	&	0.66	&	3.31	&	N	&	12.61 (0.044)	& 8.89\\
2024 Nov 13	&	30	&	1.46	&	2.42	&	4.13	&	3.46	&	N	&	12.64 (0.036)	& 11.98\\
2024 Nov 14	&	32	&	1.33	&	2.35	&	2.14	&	3.31	&	N	&	12.73 (0.038)	& 9.79\\
2024 Nov 15	&	34	&	3.26	&	2.29	&	3.27	&	3.18	&	V	&	12.55 (0.025)	& 12.22\\
2024 Nov 16	&	24	&	2.35	&	2.72	&	1.26	&	4.03	&	N	&	12.68 (0.025)	& 6.75\\
2024 Nov 17	&	32	&	0.87	&	2.35	&	5.62	&	3.31	&	N	&	12.52 (0.051)	& 10.26\\
2024 Nov 19	&	22	&	3.52	&	2.86	&	35.35	&	4.28	&	V	&	12.51 (0.038)	& 21.52\\\hline
\label{table5}
   \end{tabular}
\end{table*}

\begin{figure*}
\centering
\includegraphics[scale=.64]{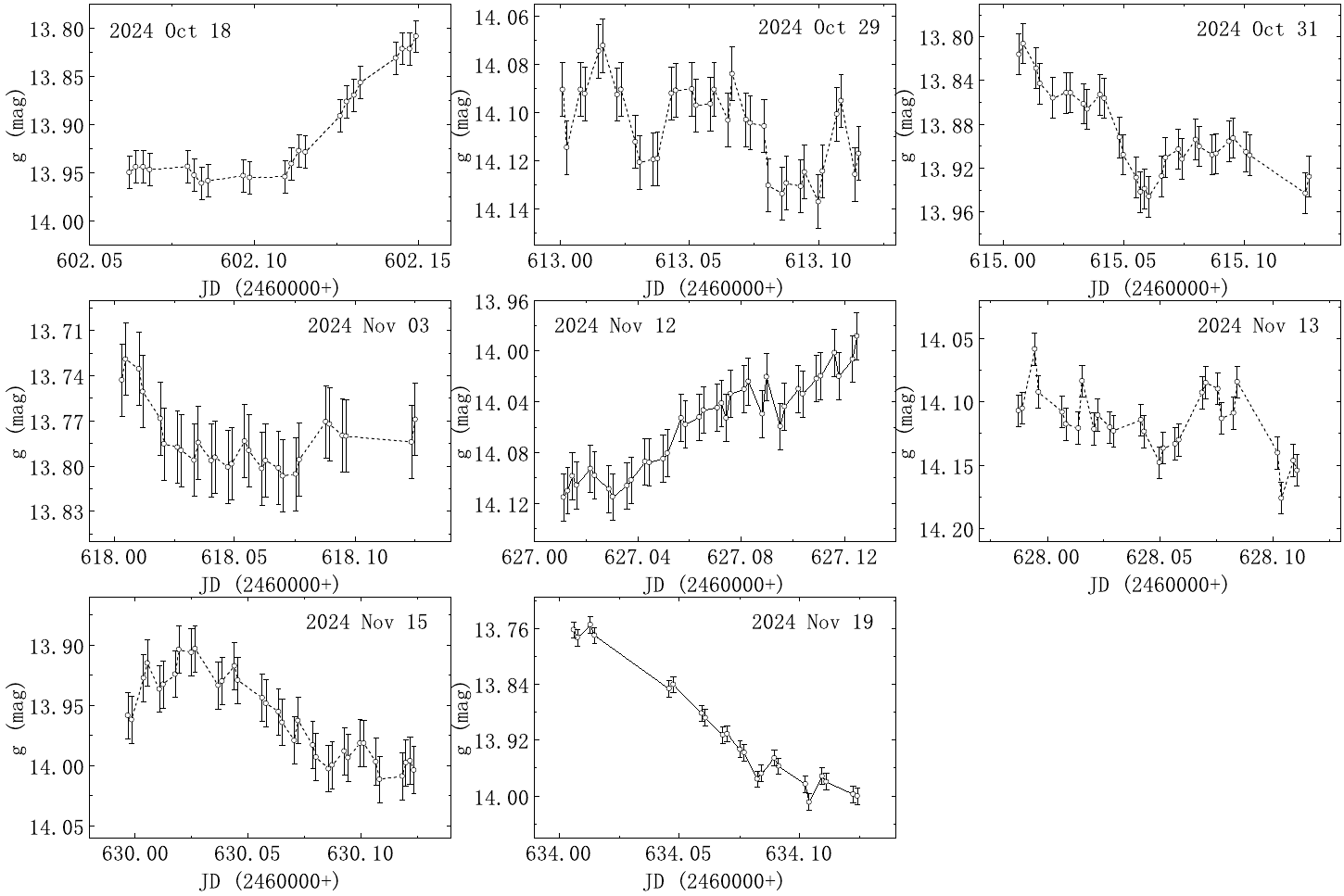}
\caption{$50$ cm IDV Light curves of BL Lacertae in the $g$-band.
\label{2}}
\end{figure*}

\begin{figure*}
\centering
\includegraphics[scale=.64]{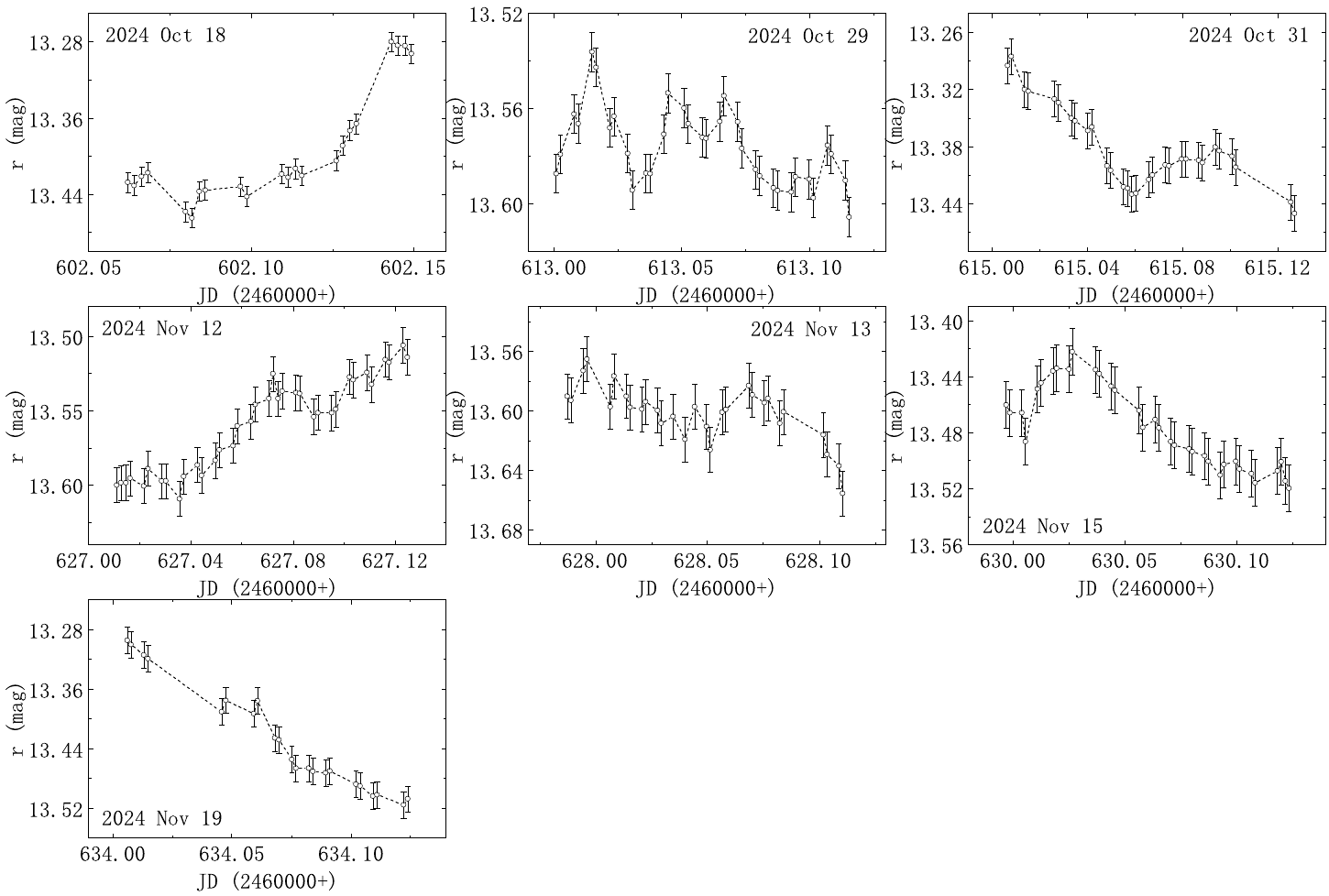}
\caption{$50$ cm IDV Light curves of BL Lacertae in the $r$-band.
\label{3}}
\end{figure*}

\begin{figure*}
\centering
\includegraphics[scale=.64]{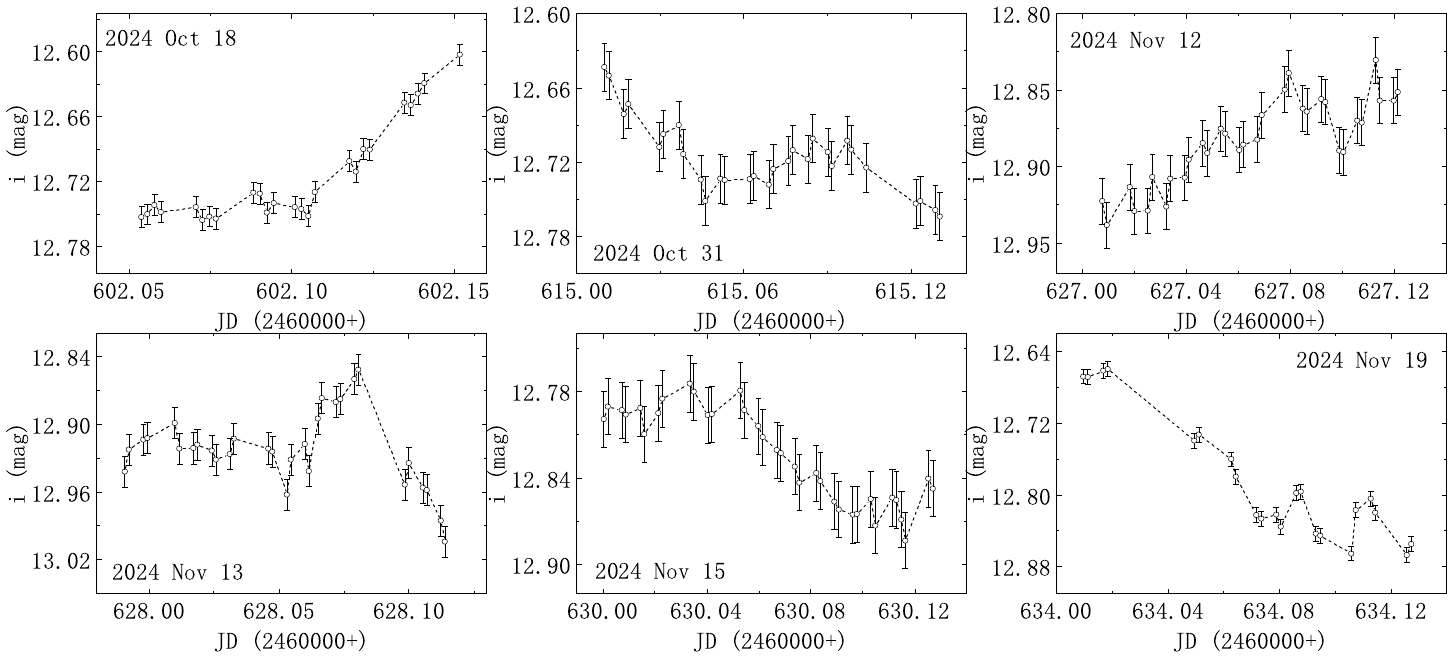}
\caption{$50$ cm IDV Light curves of BL Lacertae in the $i$-band.
\label{4}}
\end{figure*}

\subsection{Autocorrelation Analysis}
\label{sec_acf}
\begin{figure*}
\centering
\includegraphics[scale=.24]{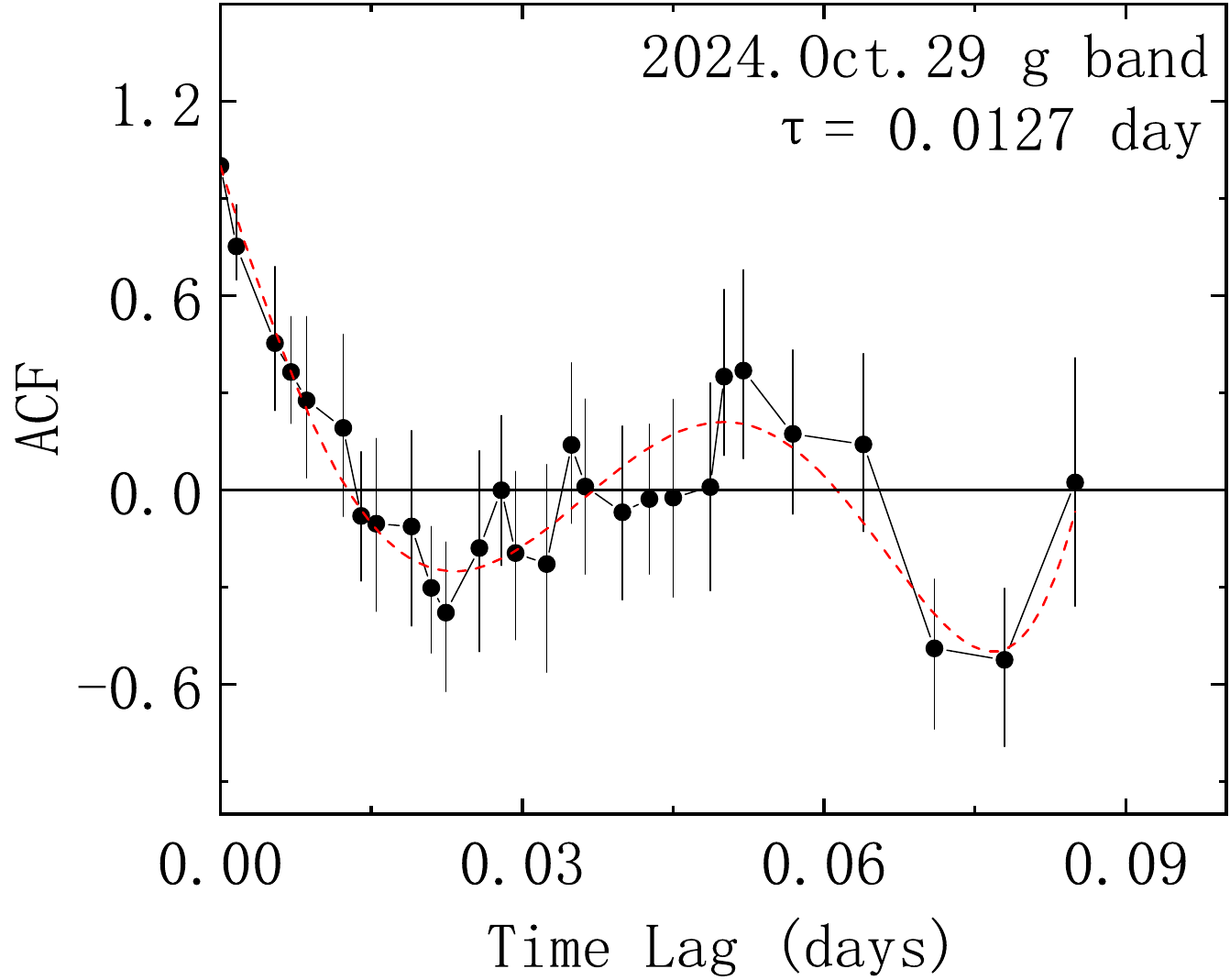}
\includegraphics[scale=.24]{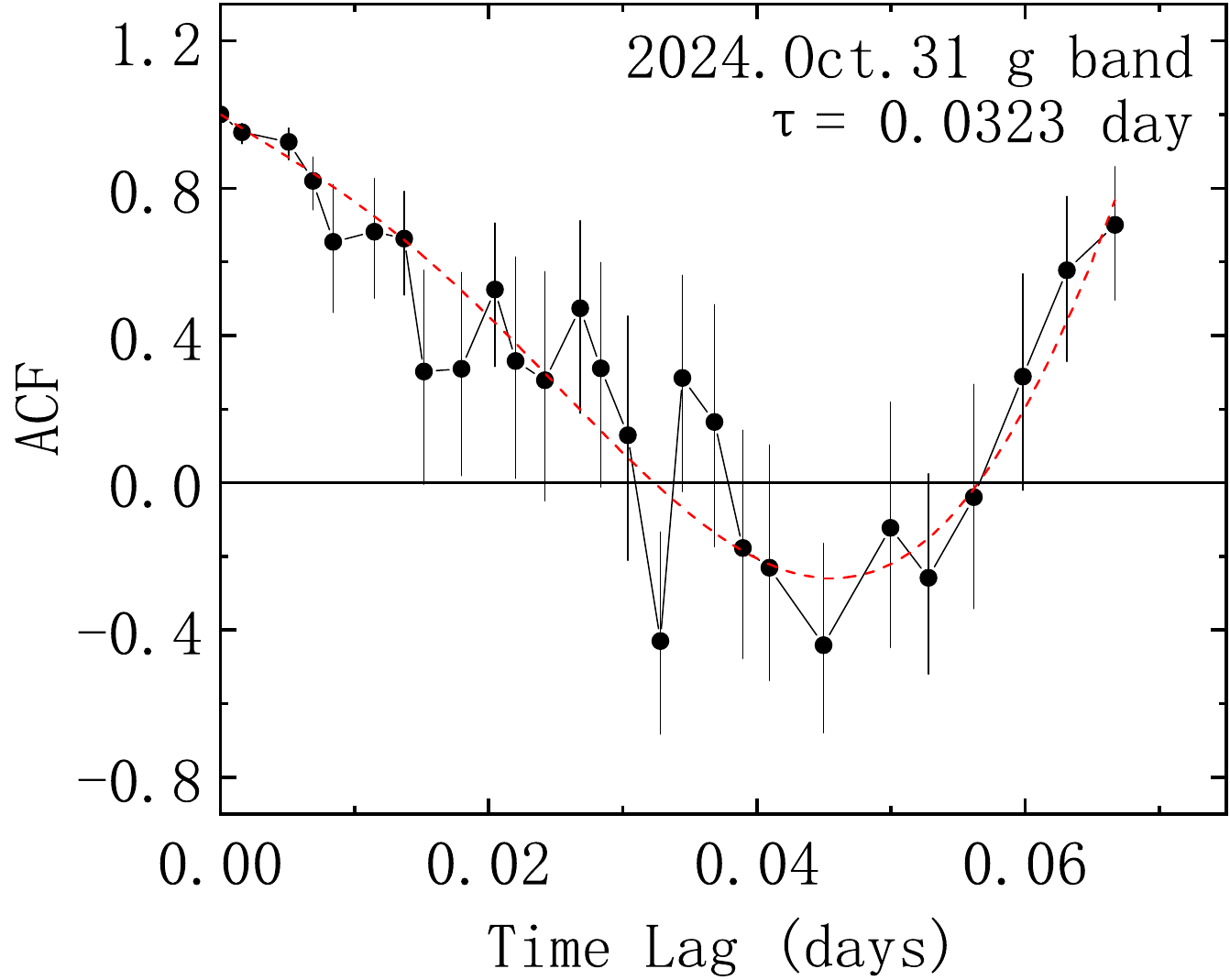}
\includegraphics[scale=.24]{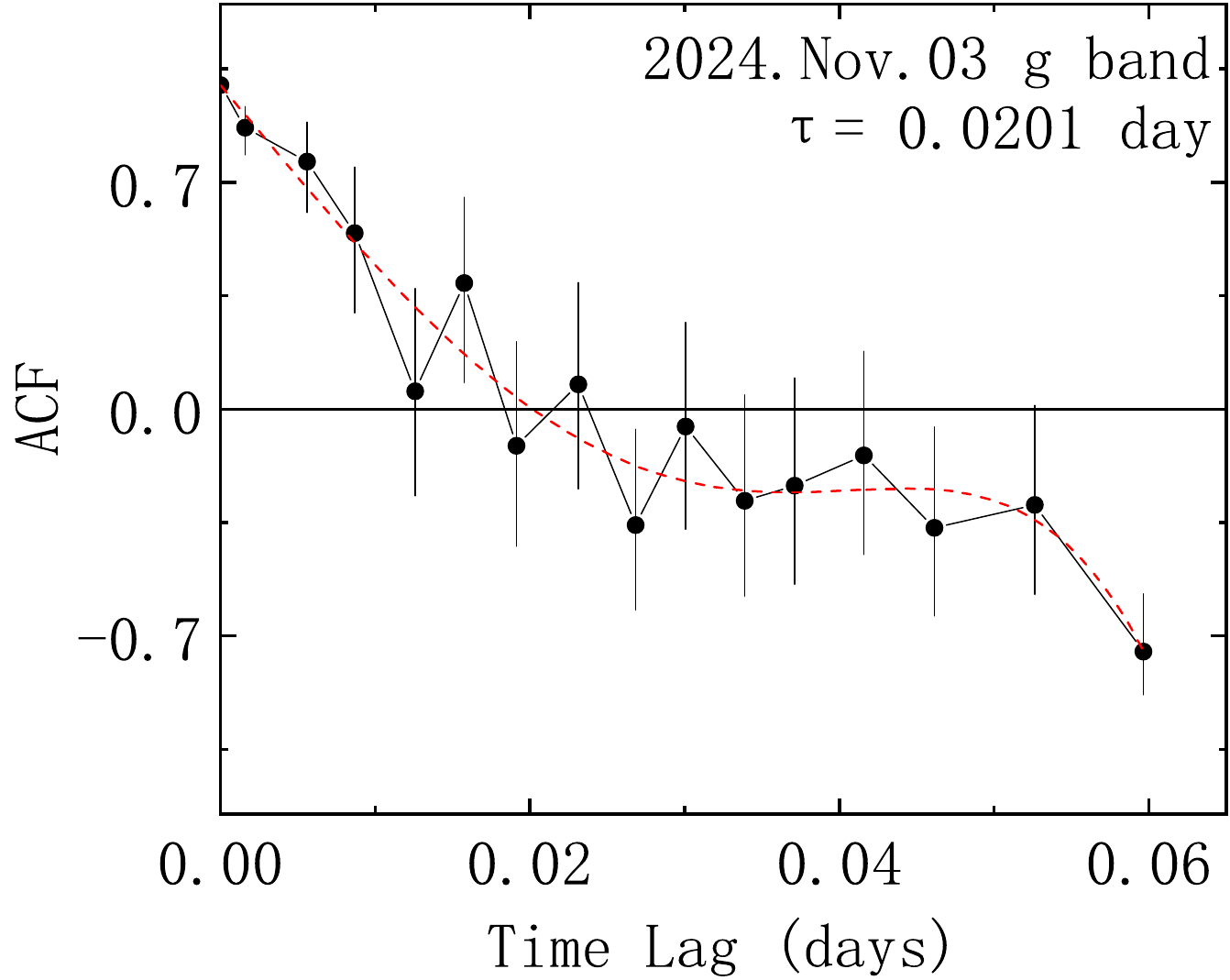}
\includegraphics[scale=.24]{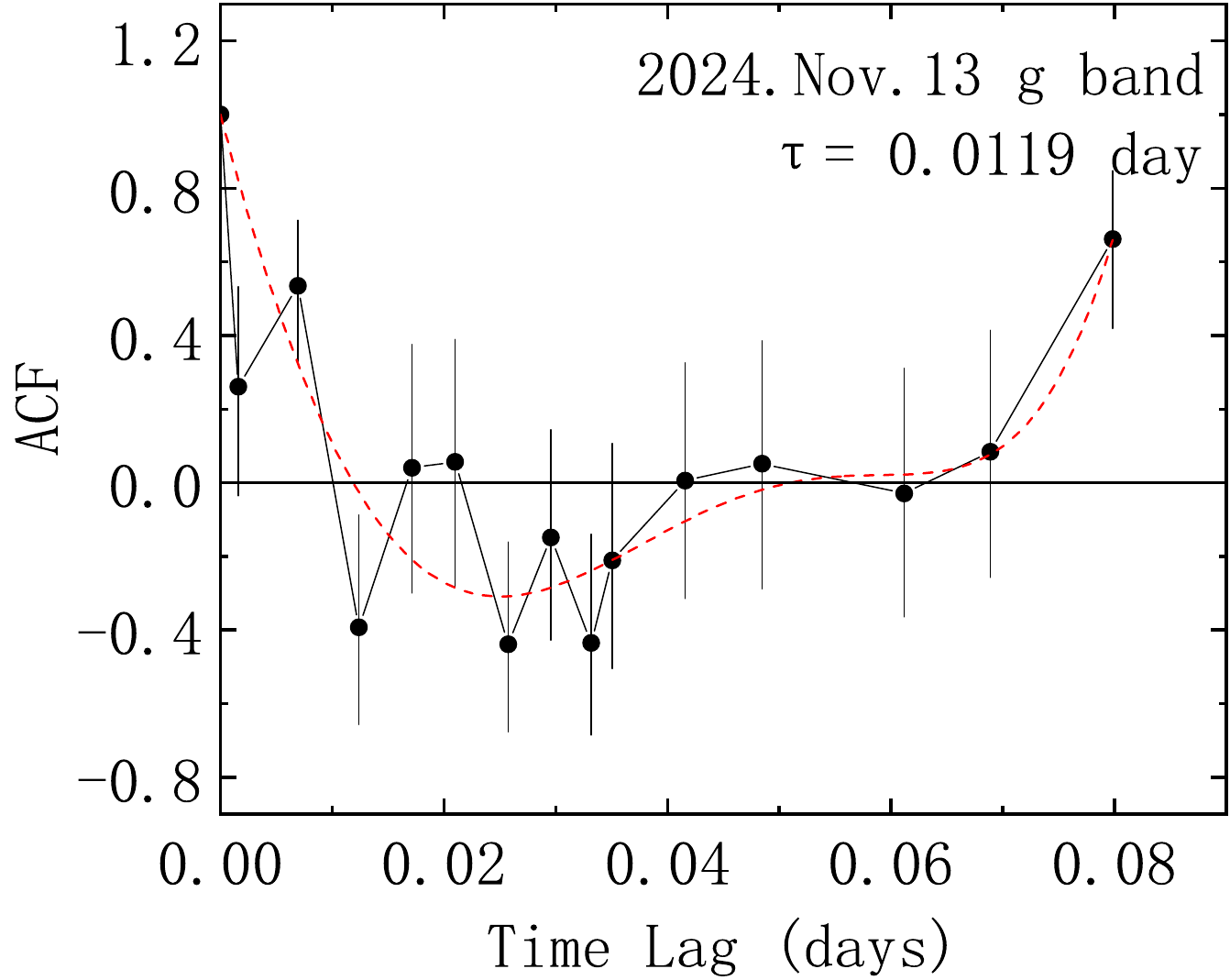}
\includegraphics[scale=.24]{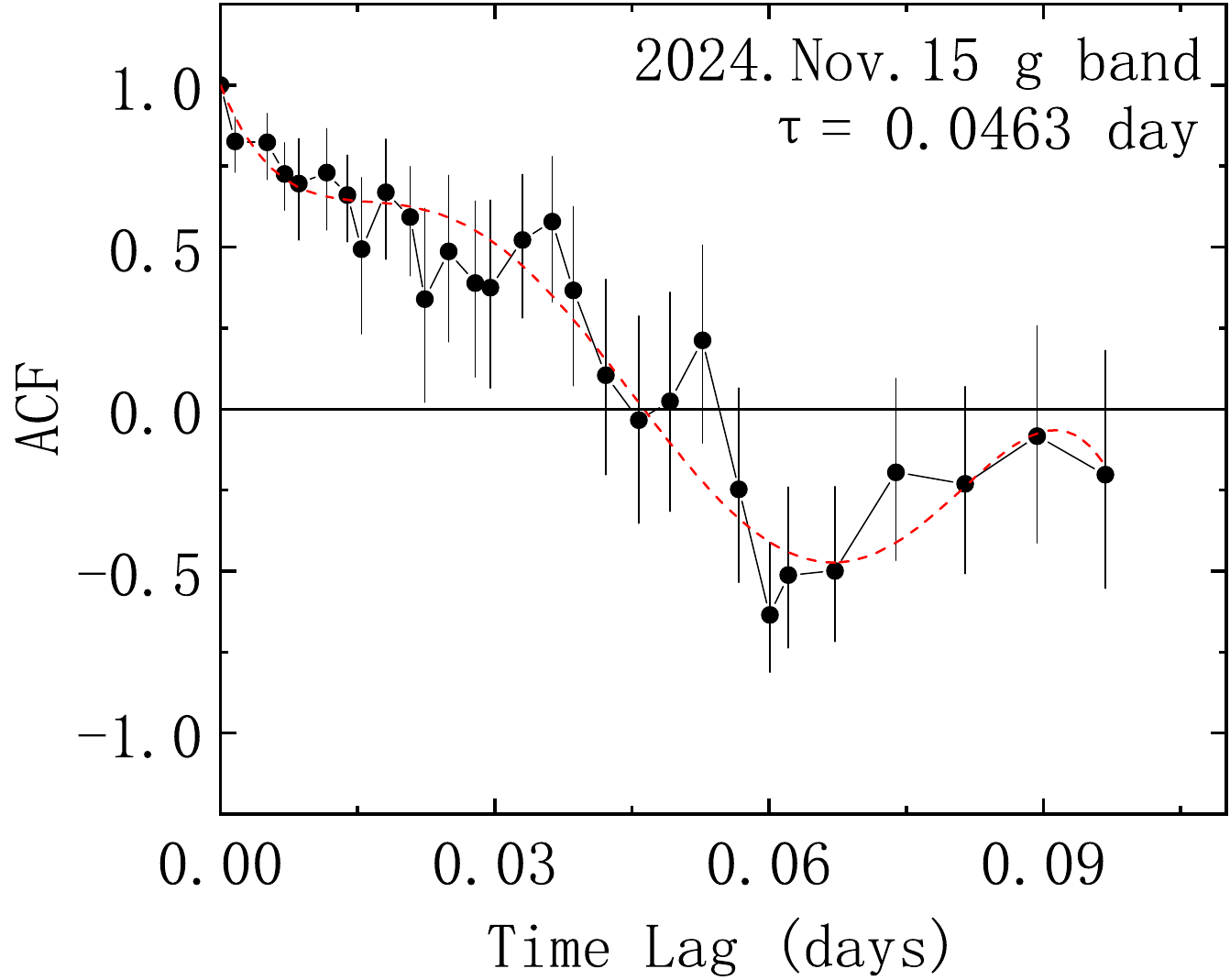}
\includegraphics[scale=.24]{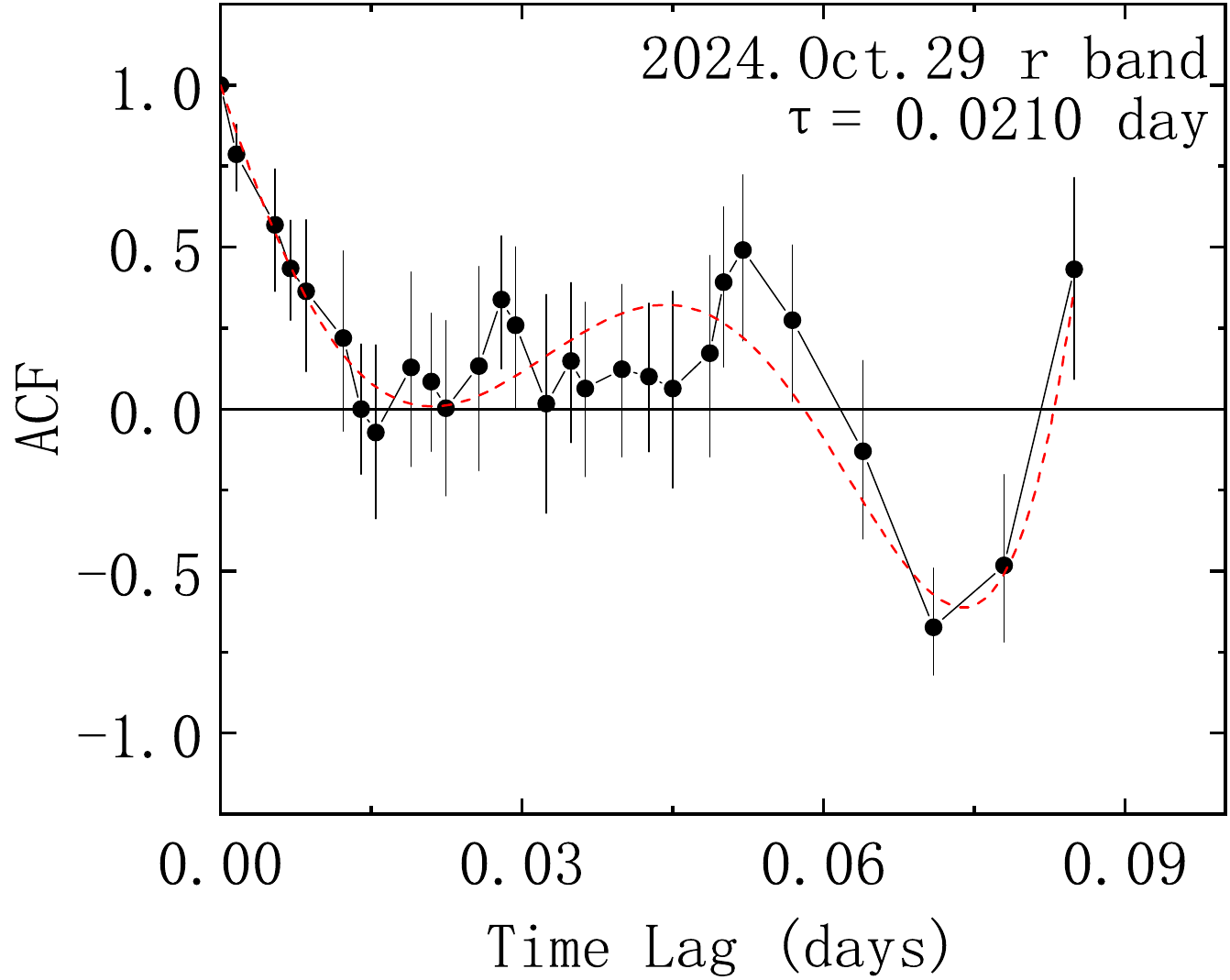}
\includegraphics[scale=.24]{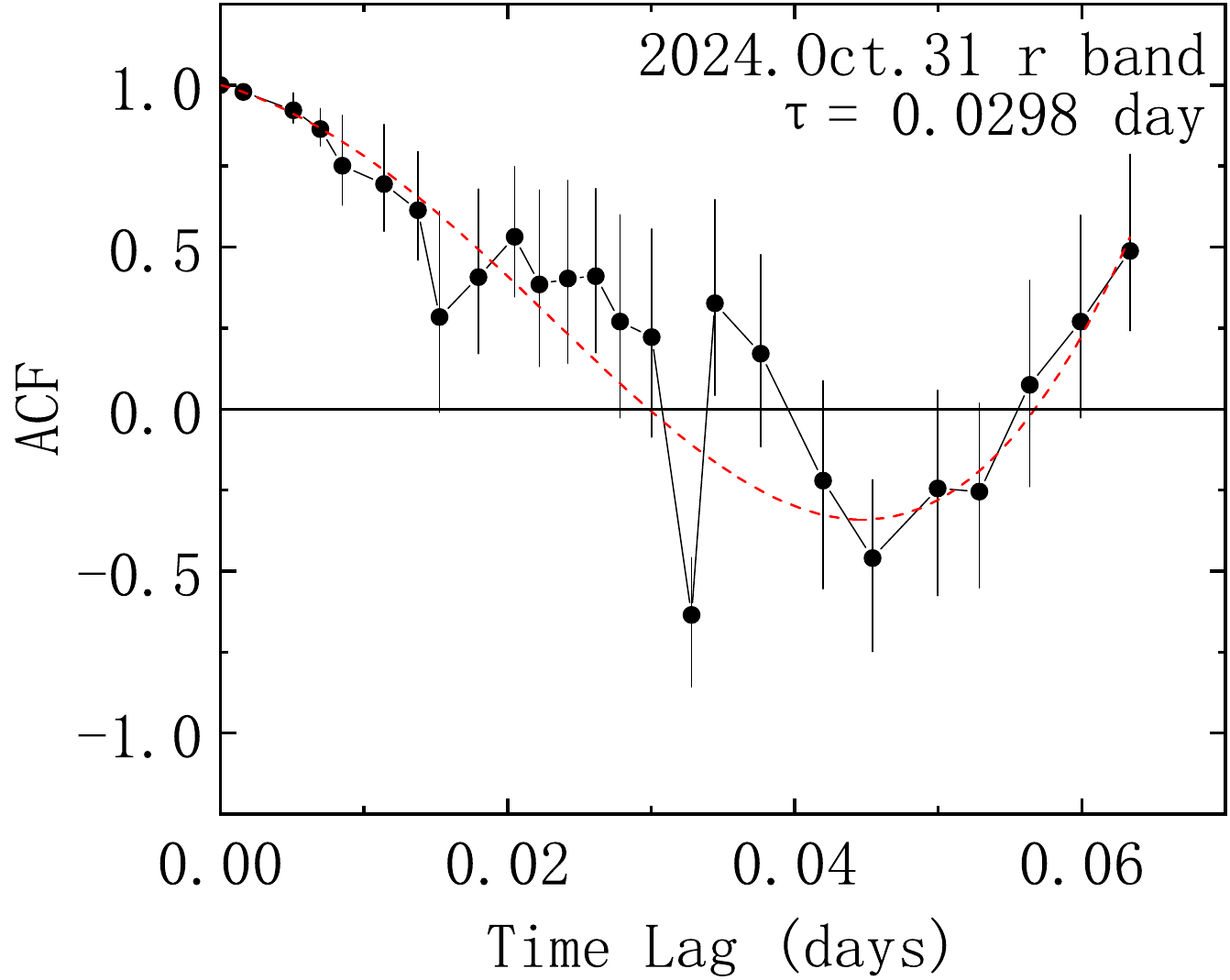}
\includegraphics[scale=.24]{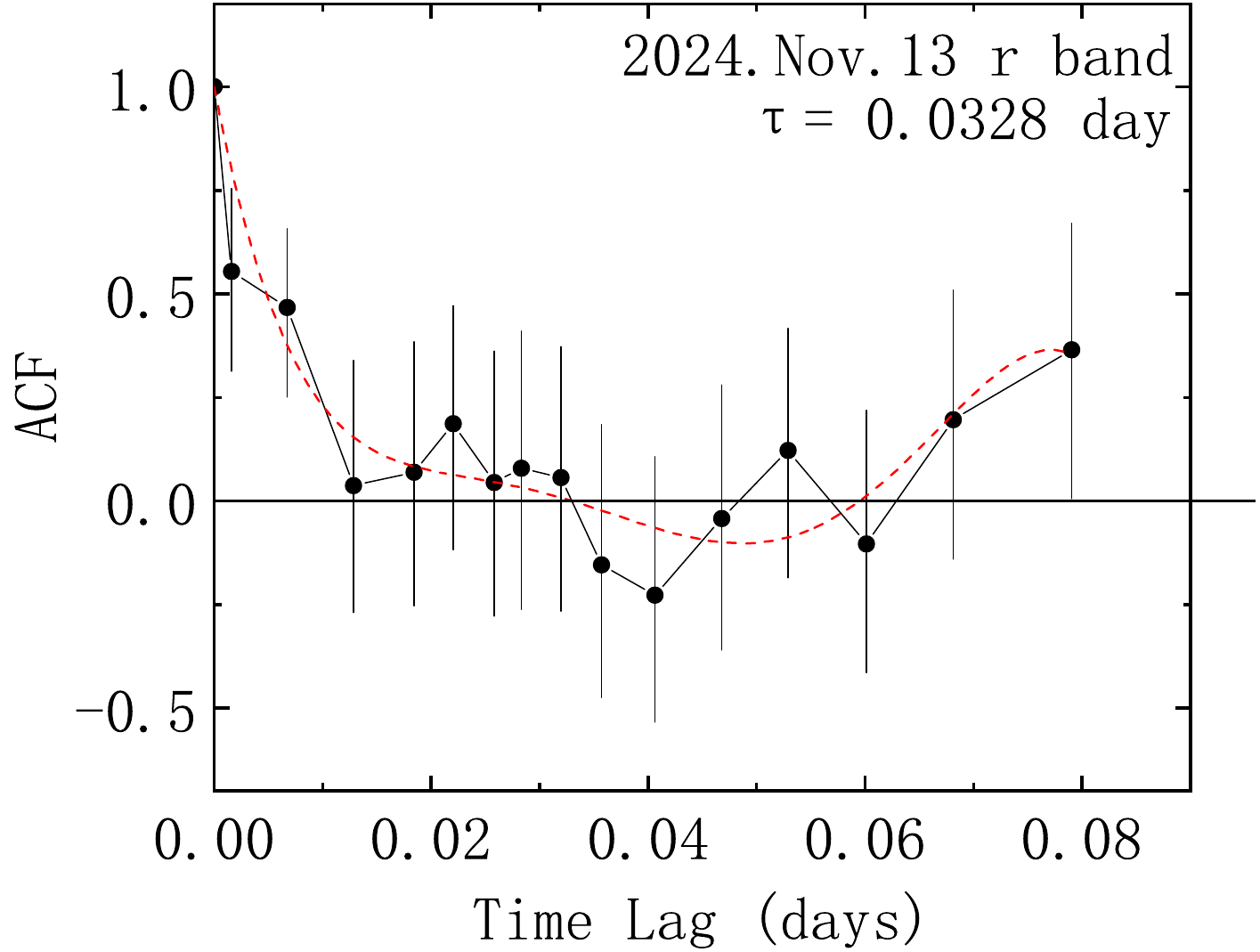}
\includegraphics[scale=.24]{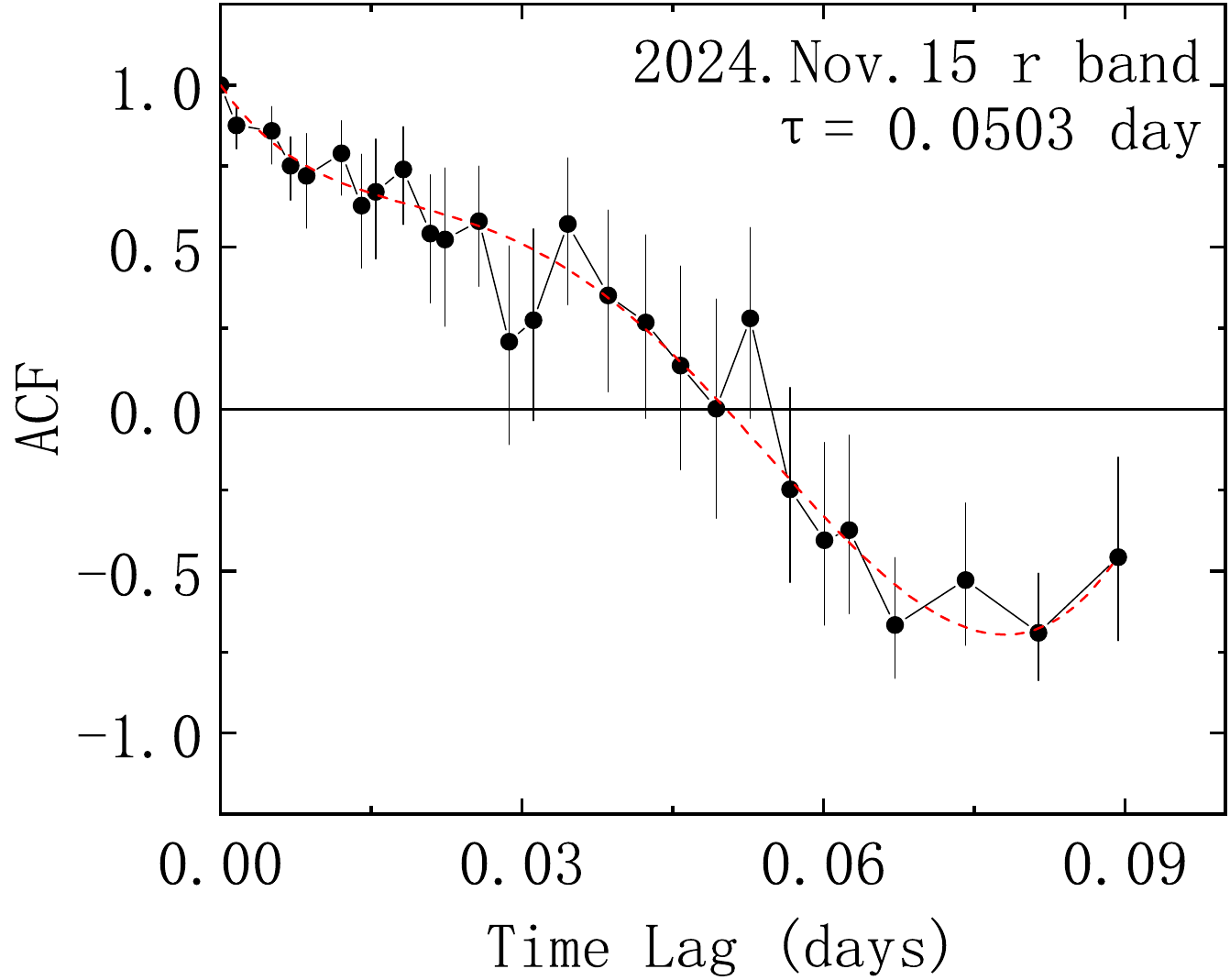}
\includegraphics[scale=.24]{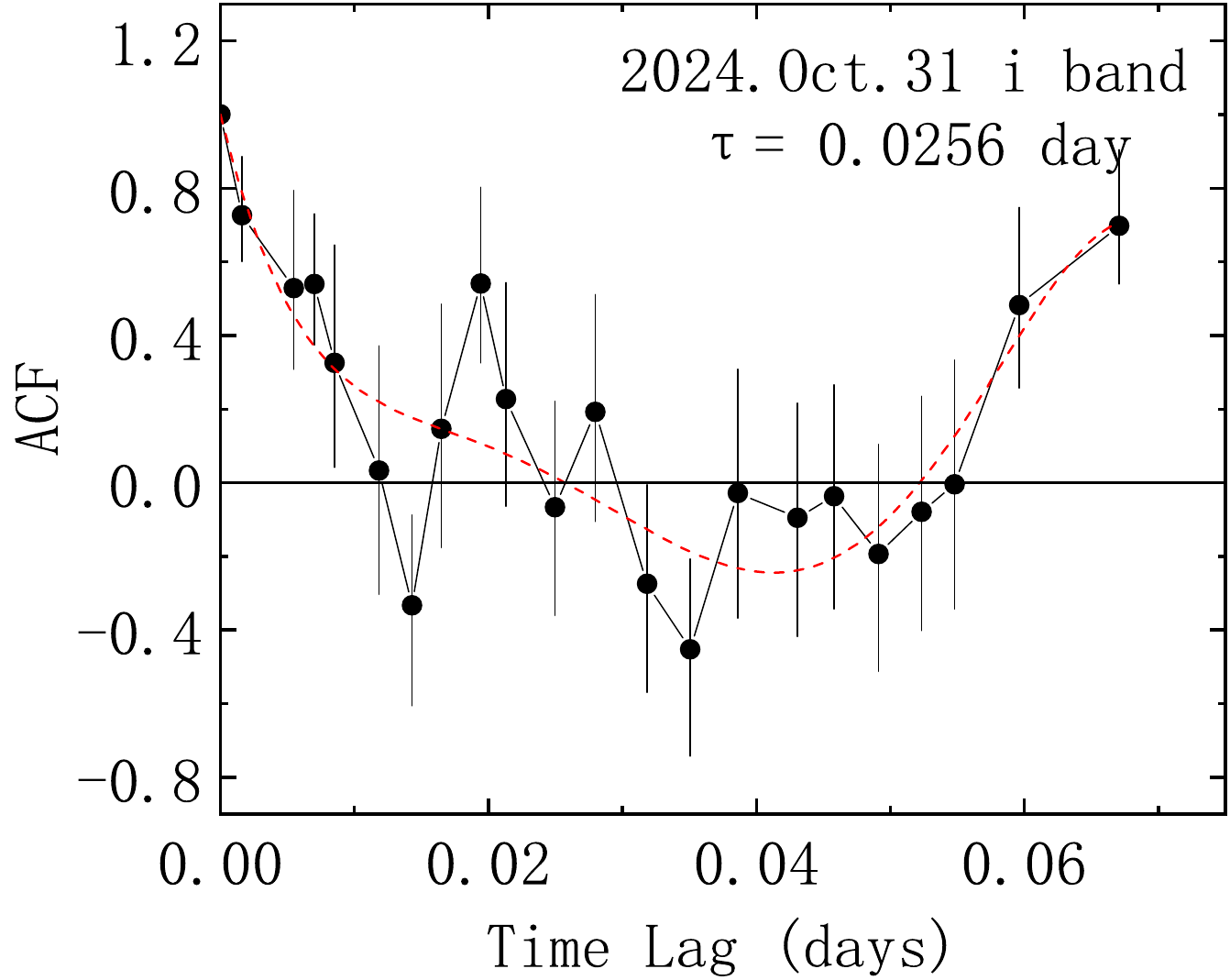}
\includegraphics[scale=.24]{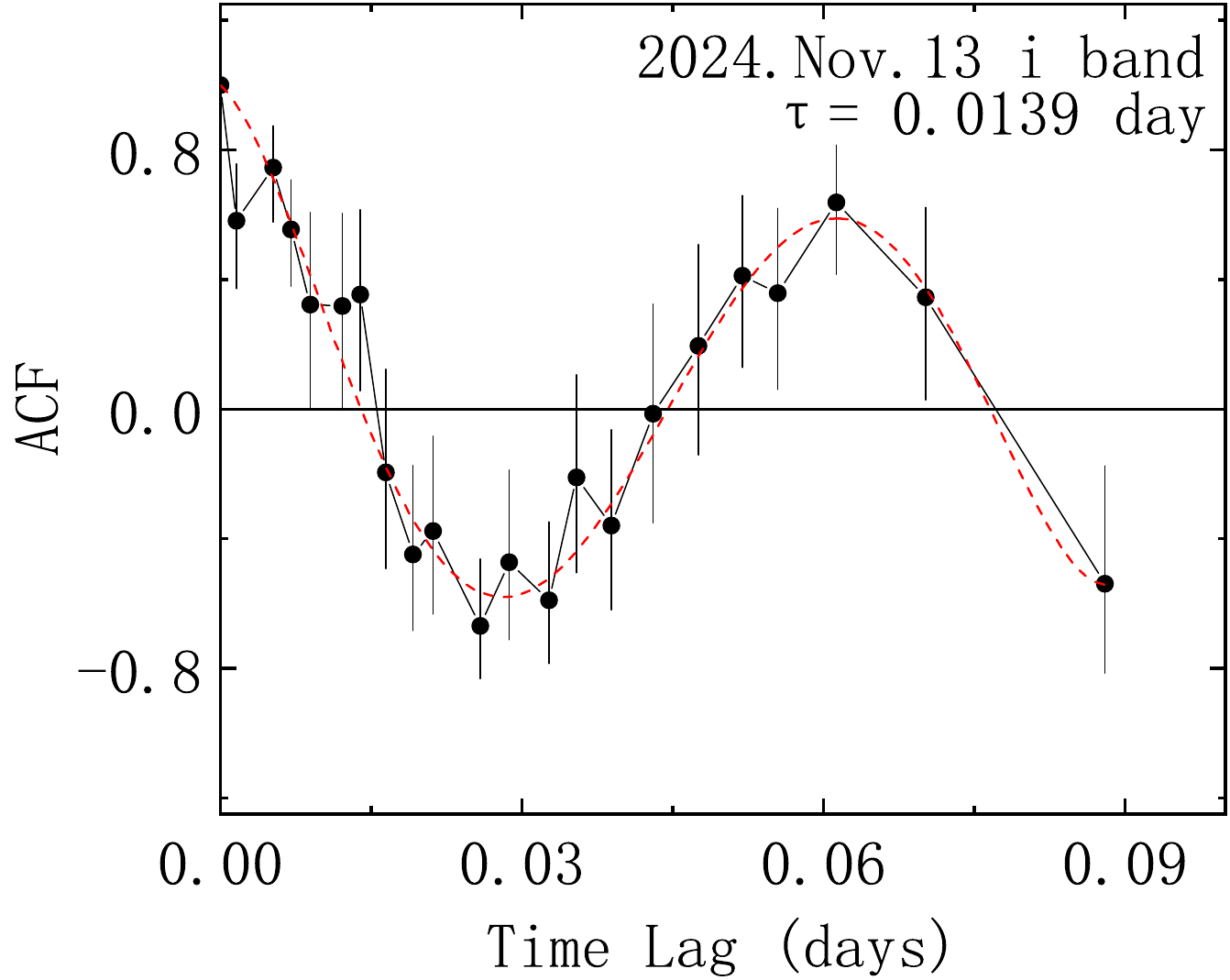}
\includegraphics[scale=.24]{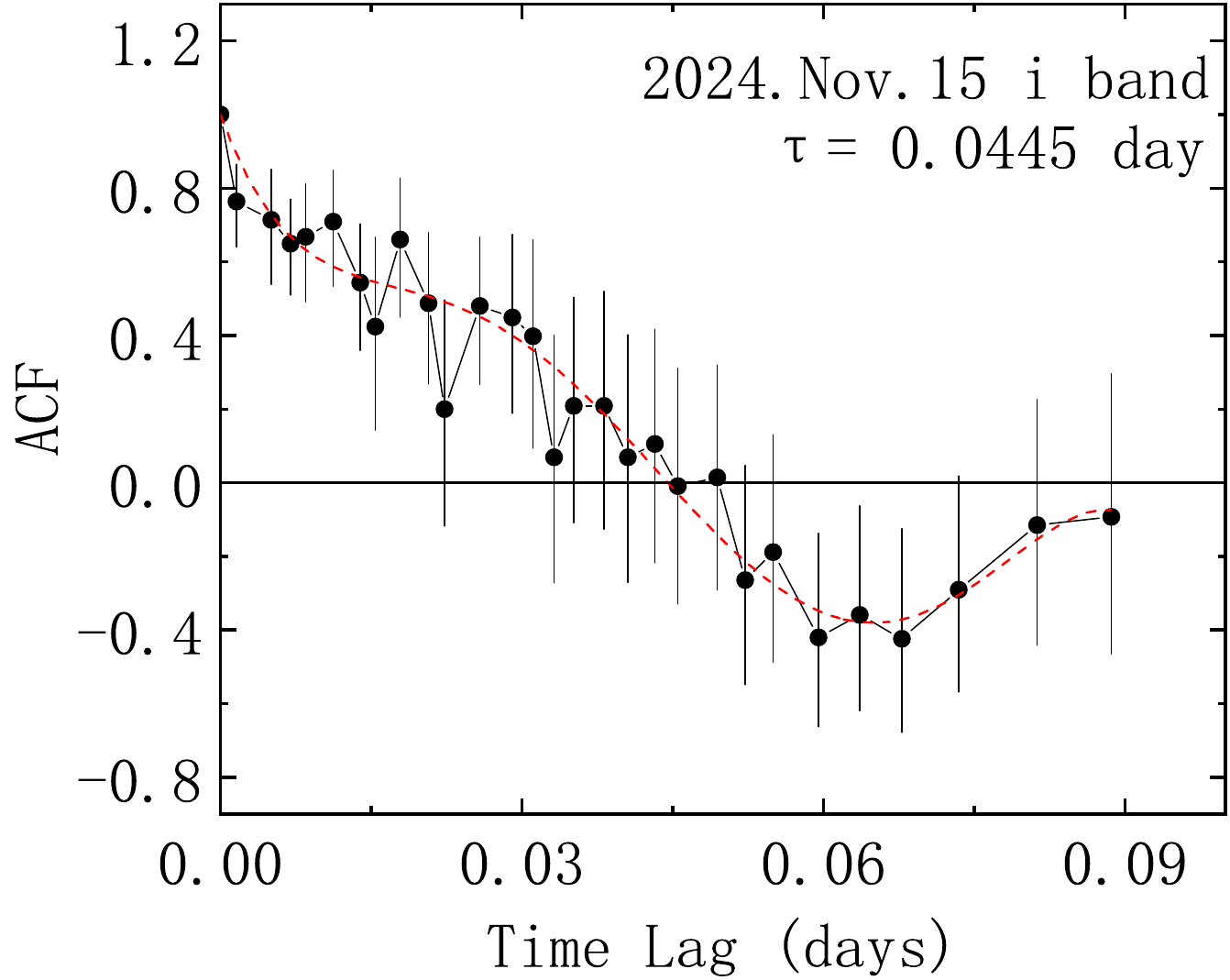}
\includegraphics[scale=.24]{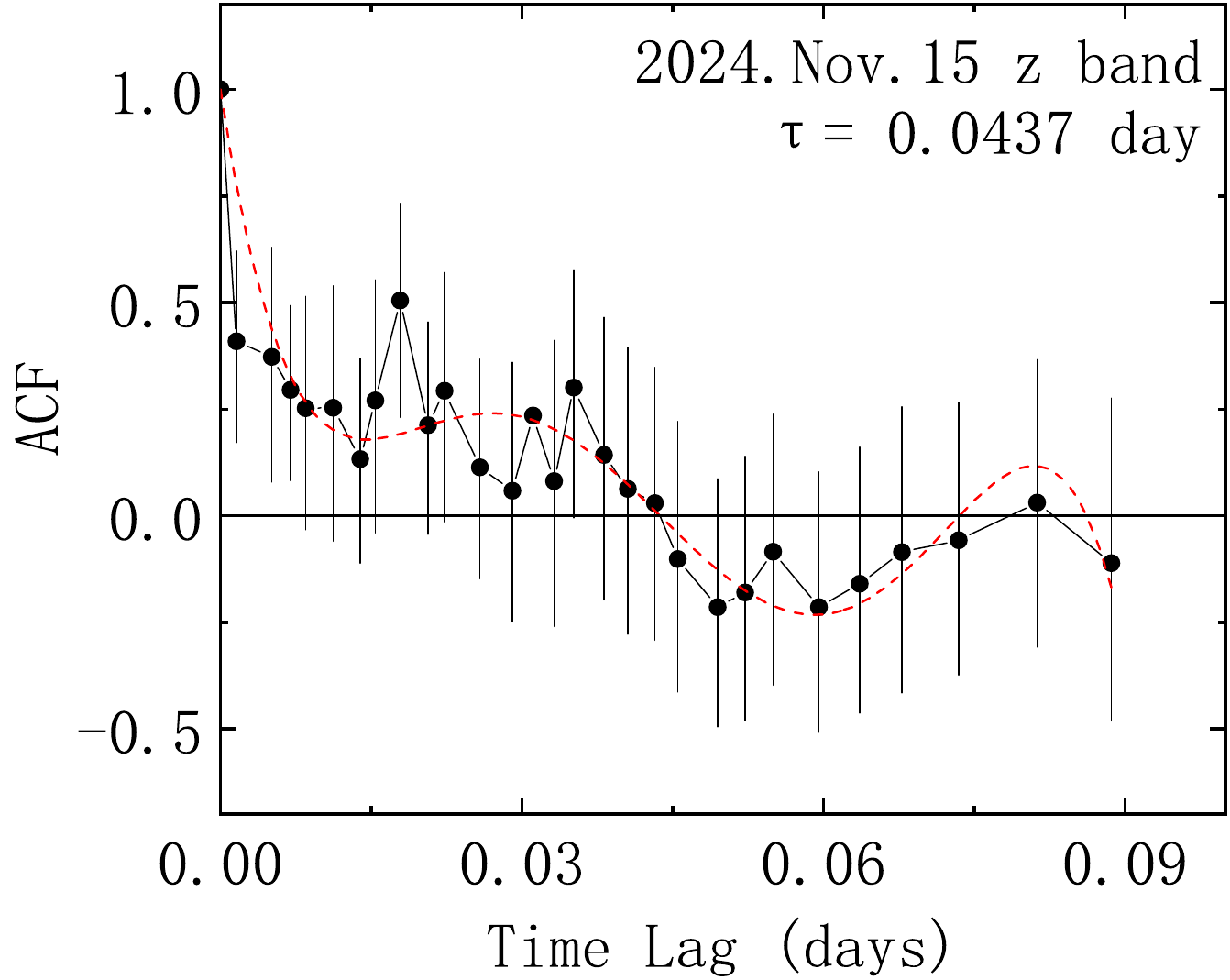}
\caption{Results of the ACF analysis for BL Lacertae. The red dashed line is a polynomial least-squares fit.
\label{6}}
\end{figure*}

The variability time-scale is a critical parameter in the study of blazars. Typically, the minimum variability time-scale is related to the size of the emission region and is most likely generated in the vicinity of the central black hole within blazars \citep{Xie02,Gupt09,Liu15}. Consequently, it is commonly employed to constrain the mass of black holes located at the centre of blazars \citep{Mill89,Liu15}. We conducted an autocorrelation function (ACF) analysis, as described by \citet{Alex97}, to identify the characteristic time-scale of variability. The ACF is given by:
\begin{equation}
\mathrm{ACF}(\tau)=\left\langle\left(m\left(t\right)-\left\langle m\right\rangle\right)\cdot\left(m\left(t+\tau\right)-\left\langle m\right\rangle\right)\right\rangle,
\label{eq:LebsequeIp5}
\end{equation}
where the brackets represent the time average. The ACF is used to measure the correlation of an optical light curve with its time-shifted self, as a function of the time lag $\tau$ \citep{Give99,Xion17}. If there is an underlying signal in the light curve with a typical variability time-scale, the width of the ACF peak near zero time lag is proportional to that time-scale \citep{Give99,Liu08}. The zero-crossing time, defined as the shortest time required for ACF to fall to zero \citep{Alex97}, can be used as the variability time-scale \citep{Give99,Liu08,Xion17}. The ACF was estimated using the code provided by \citet{Alex97}.
Only nights with detected IDV/micro-variability (with more than 30 data points per night) were analysed using the ACF. Following \citet{Give99}, we selected fifth-order polynomial least-squares fitting to determine the zero-crossing time, with the constraint that ACF ($\tau = 0$) = 1. When performing the polynomial fitting, we employed an error-weighted polynomial fitting method. The results of the ACF analysis, for which the variability time-scale has been detected, are shown in Figure \ref{6}.

The results indicate that the minimum variability time-scales ($t_{\text{var}}$) in the $g$, $r$, and $i$ optical bands are $17.14$ minutes (November $13$), $30.24$ minutes (October $29$), and $20.02$ minutes (November $13$), respectively. These results are highly consistent with the previous findings of \citet{Xie02} and \citet{Meng17}.

\subsection{Colour Analysis}
To investigate the optical spectral properties, we analysed the colour-magnitude diagrams (CMDs) on each intranight and all observation nights, respectively. The colour index was obtained from the observation data for $g$ and $r$ bands of the $1.6$ m telescope, and $g$, $r$, and $i$ bands of the $50$ cm telescopes. In this section, we focus on the correlations between the $g-r$/$g-i$ colour indices and the $g$-band magnitude, as well as between the $r-i$ colour index and the $r$-band magnitude. Based on the characteristics of the multi-channel telescope, we can obtain the simultaneous colour indices. The magnitudes were corrected for galactic extinction using the values obtained from the NASA/IPAC Extragalactic Database (NED \footnote{\url{http://ned.ipac.caltech.edu/}}) \citep{Schl98}. We converted the extinction-corrected magnitudes to fluxes using the zero-point values provided by \citet{Will18}. Based on the host galaxy fluxes in the $g$, $r$, and $i$ bands, which are $2.62$, $2.48$, and $2.91$ mJy, respectively \citep{Kali23}, we subtracted the host galaxy contribution from both the observed magnitudes and fluxes during colour and spectral studies to prevent contamination of the colour indices.

\begin{figure*}
\centering
\includegraphics[scale=.98]{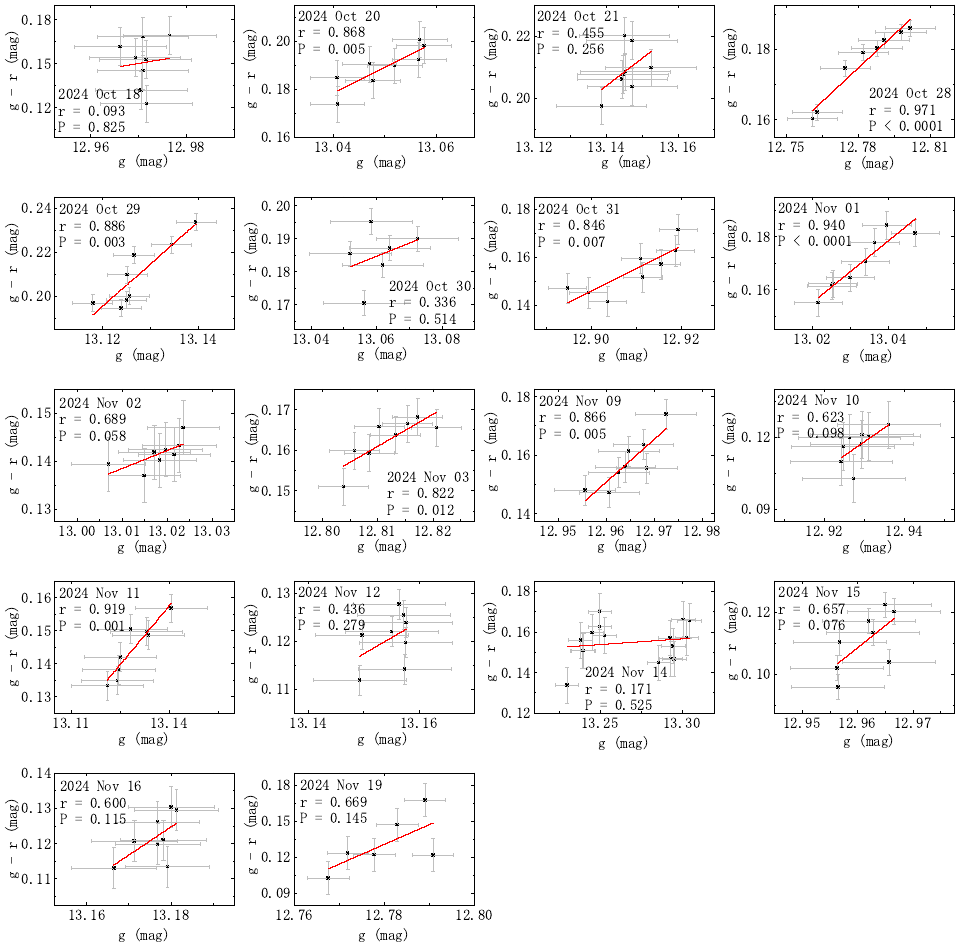}
\caption{$1.6$ m correlations between the $g - r$ index and $g$ magnitude for intraday variability. The red solid lines are the results of linear regression analysis. $r$ is the correlation coefficient, and $p$ is the chance probability.
\label{7}}
\end{figure*}

\begin{figure*}
\centering
\includegraphics[scale=.24]{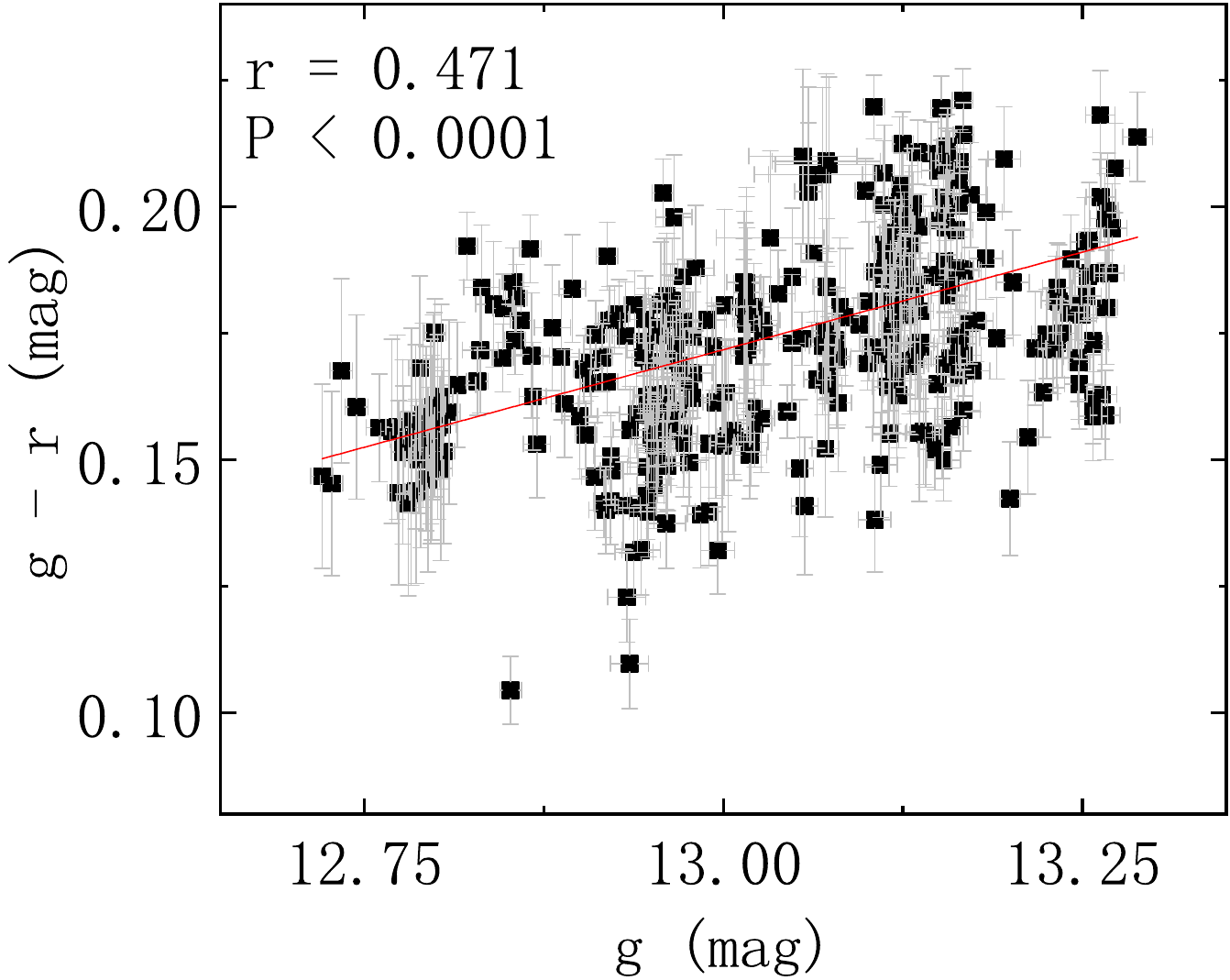}
\includegraphics[scale=.24]{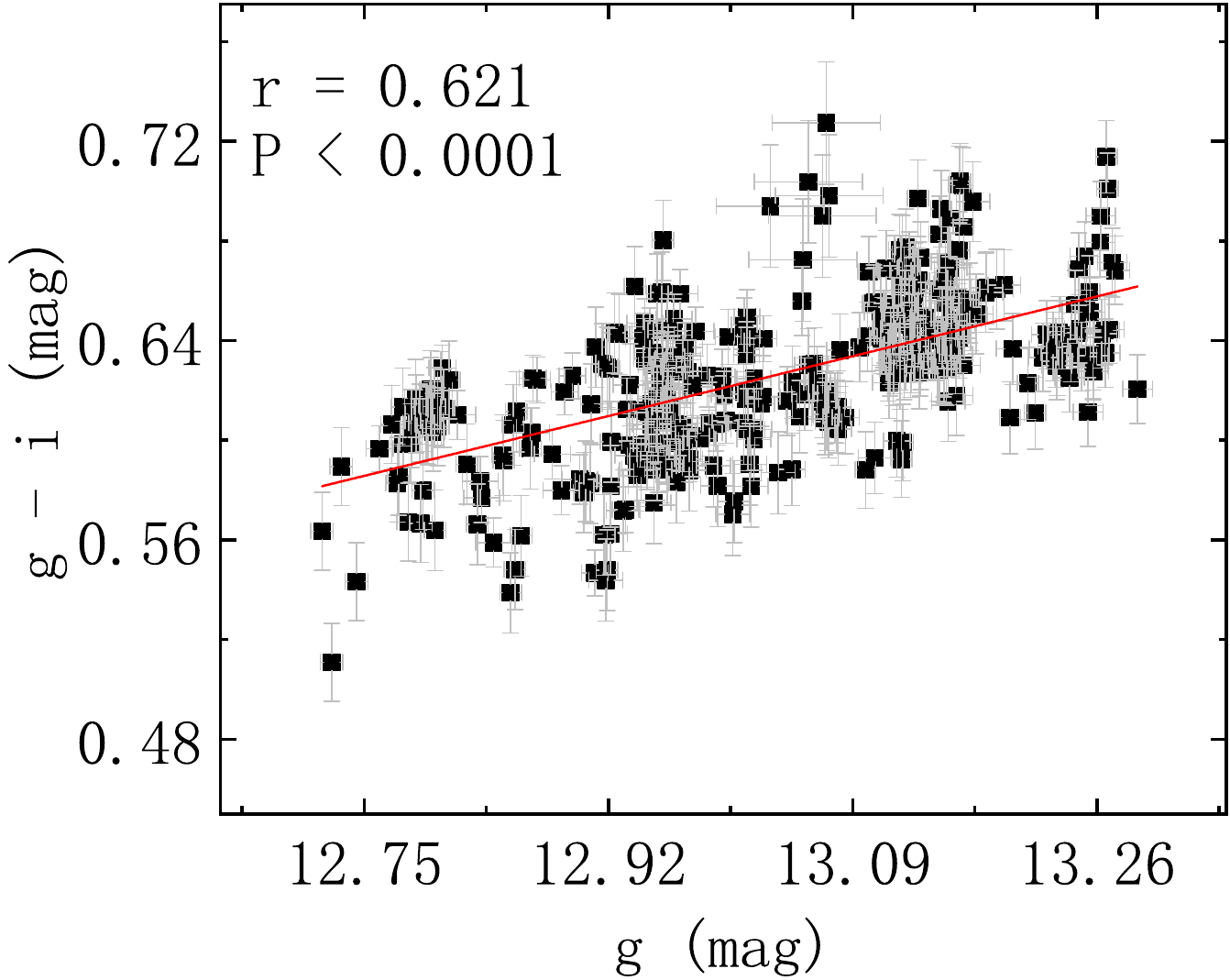}
\includegraphics[scale=.24]{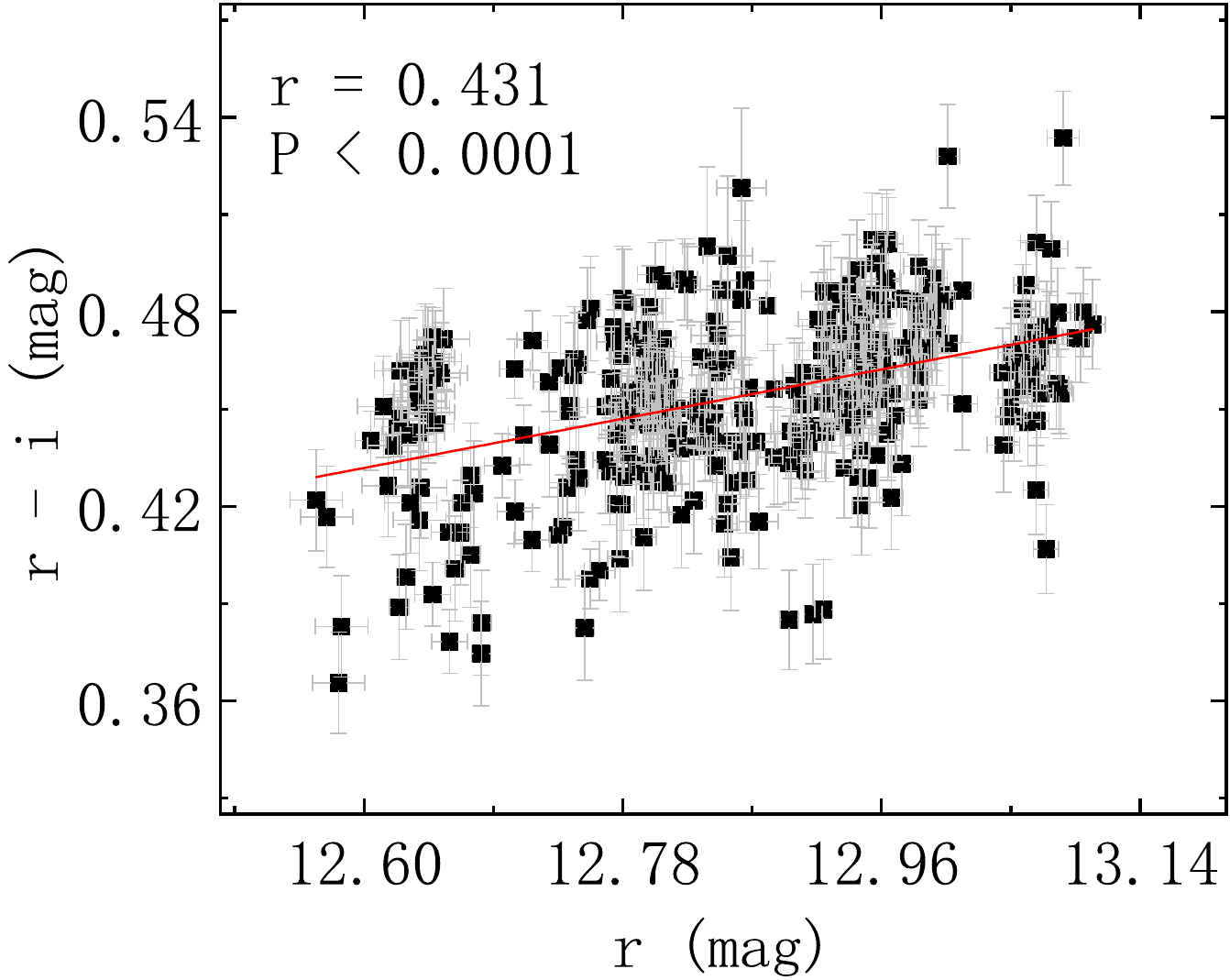}
\caption{$50$ cm correlations between the $g - r$ index and $g$ magnitude, $g - i$ index and $g$ magnitude, and $r - i$ index and $r$ magnitude for long-timescale data sets.
\label{11}}
\end{figure*}

We use the correlation coefficient from error-weighted linear regression analysis to indicate the strength of the BWB trend. The results of Spearman correlation coefficient analysis and the corresponding $p$-values are presented in Tables \ref{table6} and \ref{table7}. The CMDs on each separate night and all observation nights were plotted in Figures \ref{7} $\sim$ \ref{11}, respectively. In Table \ref{table6}, \ref{table7} and Figures \ref{7} $\sim$ \ref{11}, $r$ represents the Spearman correlation coefficient, $p$ represents the chance probability, and $\alpha$ represents the slope. Generally, a positive $r$ indicates a positive correlation, while a negative $r$ indicates a negative correlation. An absolute value of $r$ in the range of $0 \sim 0.1$ indicates no correlation, $0.1 \sim 0.3$ indicates weak correlations, $0.3 \sim 0.5$ indicates moderate correlations, and $0.5 \sim 1.0$ indicates strong correlations \citep{Cohe88}. The $p$-value is a parameter used to determine whether the correlation is significant. We consider the correlation to be reliable only when the derived probability of rejecting the null hypothesis ($p$-value) is less than $0.01$ (i.e., at a $99\%$ significance level), and the absolute value of the correlation coefficient $r$ is greater than $0.2$ \citep{Fang22,Kali23,Li24}. If the $p$-value is less than $0.001$, it indicates a significant correlation; if the $p$-value is greater than $0.05$, it indicates no correlation \citep{Cohe88}.

The relationship between the colour indices and magnitudes in different bands throughout the observation period are shown in Figures \ref{7} $\sim$ \ref{11}, as well as in Table \ref{table6}, \ref{table7}. A least-squares fitting to the $g-r$, $g-i$, and $r-i$ colour indices shows that, with $p < 0.001$ (confidence level $> 99.9\%$), the corresponding correlation coefficients $r$ are all greater than $0.5$, indicating a significant correlation between the colour indices and magnitudes. This demonstrates that BL Lacertae exhibits a significant BWB trend, which means that the optical continuum hardens as the brightness increases, with different regression slopes. Long-term variations in the colour index also follow the BWB trend, with correlation coefficients $r$ of $0.47$, $0.62$, and $0.43$, respectively.

\begin{table*}
   \centering
   \caption{Results of error-weighted linear regression analysis for $50$ cm $g - r$, $g - i$, and $r - i$ bands. Column 1 is the date of the observation; Columns 2 $\sim$ 4, 5 $\sim$ 7, and 8 $\sim$ 10 represent the correlation coefficient, chance probability, and slope for the $g - r$, $g - i$, and $r - i$ bands, respectively.}
   %\label{tab:tab1}
   \renewcommand\arraystretch{1.2}
  \setlength{\tabcolsep}{3.2mm}
   \begin{tabular}{cccccccccccc} % four columns, alignment for each
      \hline
Date    &   $r$  &   $p$ &   $\alpha$    &   $r$  &   $p$ &   $\alpha$   &    $r$  &   $p$ &   $\alpha$   \\
\cmidrule(lr){1-1} \cmidrule(lr){2-4} \cmidrule(lr){5-7} \cmidrule(lr){8-10}
% ---------------------------
50 cm Data	&	$g-r$	&	$g-r$	&	$g-r$	&	$g-i$	&	$g-i$	&	$g-i$	&	$r-i$	&	$r-i$	&	$r-i$	\\
\cmidrule(lr){1-1} \cmidrule(lr){2-4} \cmidrule(lr){5-7} \cmidrule(lr){8-10}
% ---------------------------
2024 Oct 18	&	0.363	&	0.114	&	0.059	&	0.837	&	$<0.0001$	&	0.430	&	0.790	&	$<0.0001$	&	0.417	\\
2024 Oct 29	&	0.455	&	0.007	&	0.238	&	0.534	&	0.001	&	0.775	&	0.561	&	$<0.0001$	&	0.946	\\
2024 Oct 31	&	0.064	&	0.738	&	0.026	&	0.676	&	$<0.0001$	&	0.590	&	0.692	&	$<0.0001$	&	0.485	\\
2024 Nov 02	&	0.690	&	0.018	&	0.927	&	0.670	&	0.033	&	1.121	&	0.735	&	0.009	&	1.463	\\
2024 Nov 03	&	0.099	&	0.612	&	0.030	&	0.752	&	$<0.0001$	&	0.833	&	0.745	&	$<0.0001$	&	0.790	\\
2024 Nov 11	&	0.314	&	0.318	&	0.367	&	0.168	&	0.600	&	0.286	&	0.155	&	0.629	&	0.184	\\
2024 Nov 12	&	0.211	&	0.245	&	0.090	&	0.617	&	$<0.0001$	&	0.269	&	0.623	&	0.623	&	0.265	\\
2024 Nov 13	&	0.764	&	$<0.0001$	&	0.6001	&	0.310	&	0.139	&	0.231	&	-0.081	&	0.704	&	-0.097	\\
2024 Nov 14	&	0.465	&	0.009	&	0.466	&	0.246	&	0.190	&	0.282	&	0.216	&	0.250	&	0.225	\\
2024 Nov 15	&	0.483	&	0.005	&	0.180	&	0.326	&	0.073	&	0.163	&	0.046	&	0.803	&	0.022	\\
2024 Nov 16	&	0.474	&	0.025	&	0.586	&	0.556	&	0.007	&	0.892	&	0.742	&	$<0.0001$	&	1.244	\\
2024 Nov 17	&	0.818	&	$<0.0001$	&	1.041	&	0.575	&	$<0.0001$	&	0.653	&	0.369	&	0.044	&	0.499	\\
2024 Nov 19	&	0.055	&	0.807	&	0.007	&	-0.011	&	0.961	&	-0.002	&	-0.012	&	0.956	&	-0.003	\\
2024 Nov 21	&	0.855	&	0.006	&	0.629	&	0.951	&	$<0.0001$	&	1.459	&	0.966	&	$<0.0001$	&	2.103	\\\hline
\label{table6}
   \end{tabular}
\end{table*}

\begin{table*}
   \centering
   \caption{Results of error-weighted linear regression analysis for $1.6$ m $g - r$ band. Column 1 is the date of the observation; Column 2 is the correlation coefficient; Column 3 is the chance probability; Column 4 is the slope.}
   %\label{tab:tab1}
   \renewcommand\arraystretch{1.2}
  \setlength{\tabcolsep}{6.2mm}
   \begin{tabular}{cccccccccccc} % four columns, alignment for each
      \hline
Date    &   $r$  &   $p$ &   $\alpha$    \\\hline
1.6 m Data	&	$g-r$	&	$g-r$	&	$g-r$	&	\\
2024 Oct 18	&	0.093	&	0.825	&	0.546	&	\\
2024 Oct 20	&	0.868	&	0.005	&	1.074	&	\\
2024 Oct 21	&	0.455	&	0.256	&	0.889	&	\\
2024 Oct 28	&	0.971	&	$<0.0001$	&	0.635	&	\\
2024 Oct 29	&	0.886	&	0.003	&	1.924	&	\\
2024 Oct 30	&	0.336	&	0.514	&	0.396	&	\\
2024 Oct 31	&	0.846	&	0.007	&	0.933	&	\\
2024 Nov 01	&	0.940	&	$<0.0001$	&	1.165	&	\\
2024 Nov 02	&	0.689	&	0.058	&	0.375	&	\\
2024 Nov 03	&	0.822	&	0.012	&	0.791	&	\\
2024 Nov 09	&	0.866	&	0.005	&	1.458	&	\\
2024 Nov 10	&	0.623	&	0.098	&	1.151	&	\\
2024 Nov 11	&	0.919	&	0.001	&	1.192	&	\\
2024 Nov 12	&	0.436	&	0.279	&	0.669	&	\\
2024 Nov 14	&	0.171	&	0.525	&	0.056	&	\\
2024 Nov 15	&	0.657	&	0.076	&	1.419	&	\\
2024 Nov 16	&	0.600	&	0.115	&	0.800	&	\\
2024 Nov 19	&	0.669	&	0.145	&	1.641	&	\\\hline
\label{table7}
   \end{tabular}
\end{table*}

\begin{figure*}
\centering
\includegraphics[scale=.80]{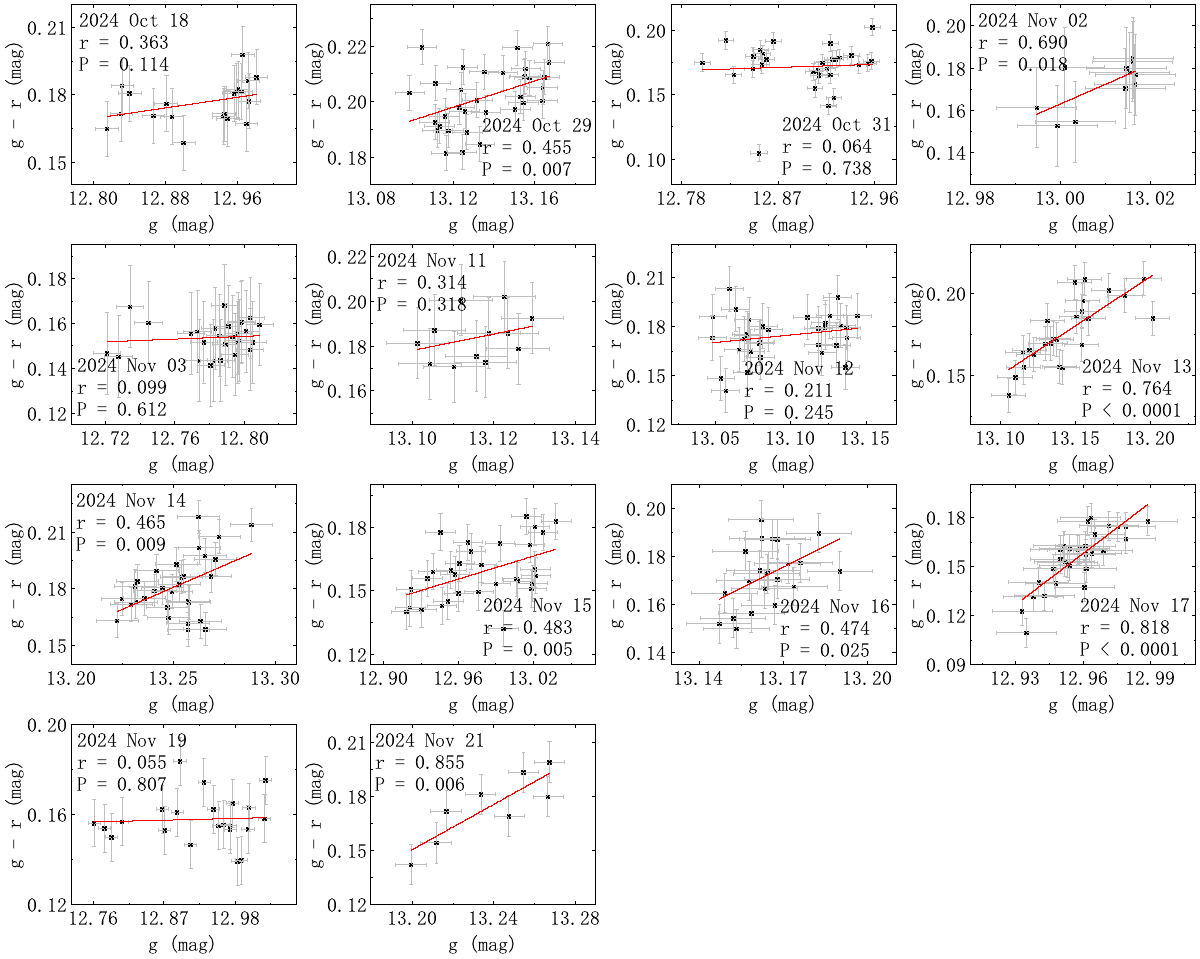}
\caption{$50$ cm correlations between the $g - r$ index and $g$ magnitude for intraday variability.
\label{8}}
\end{figure*}

\begin{figure*}
\centering
\includegraphics[scale=.80]{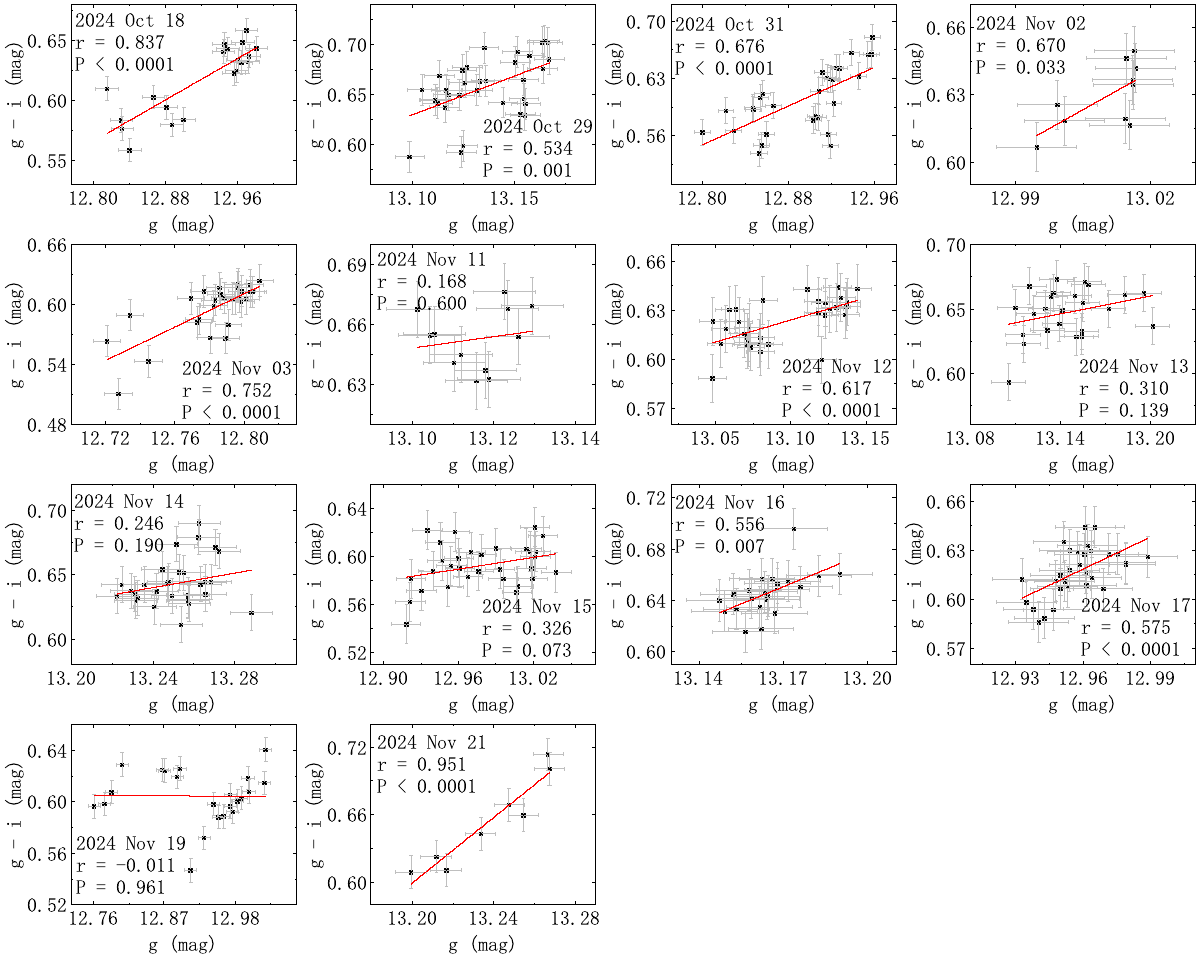}
\caption{$50$ cm correlations between the $g - i$ index and $g$ magnitude for intraday variability.
\label{9}}
\end{figure*}

\begin{figure*}
\centering
\includegraphics[scale=.80]{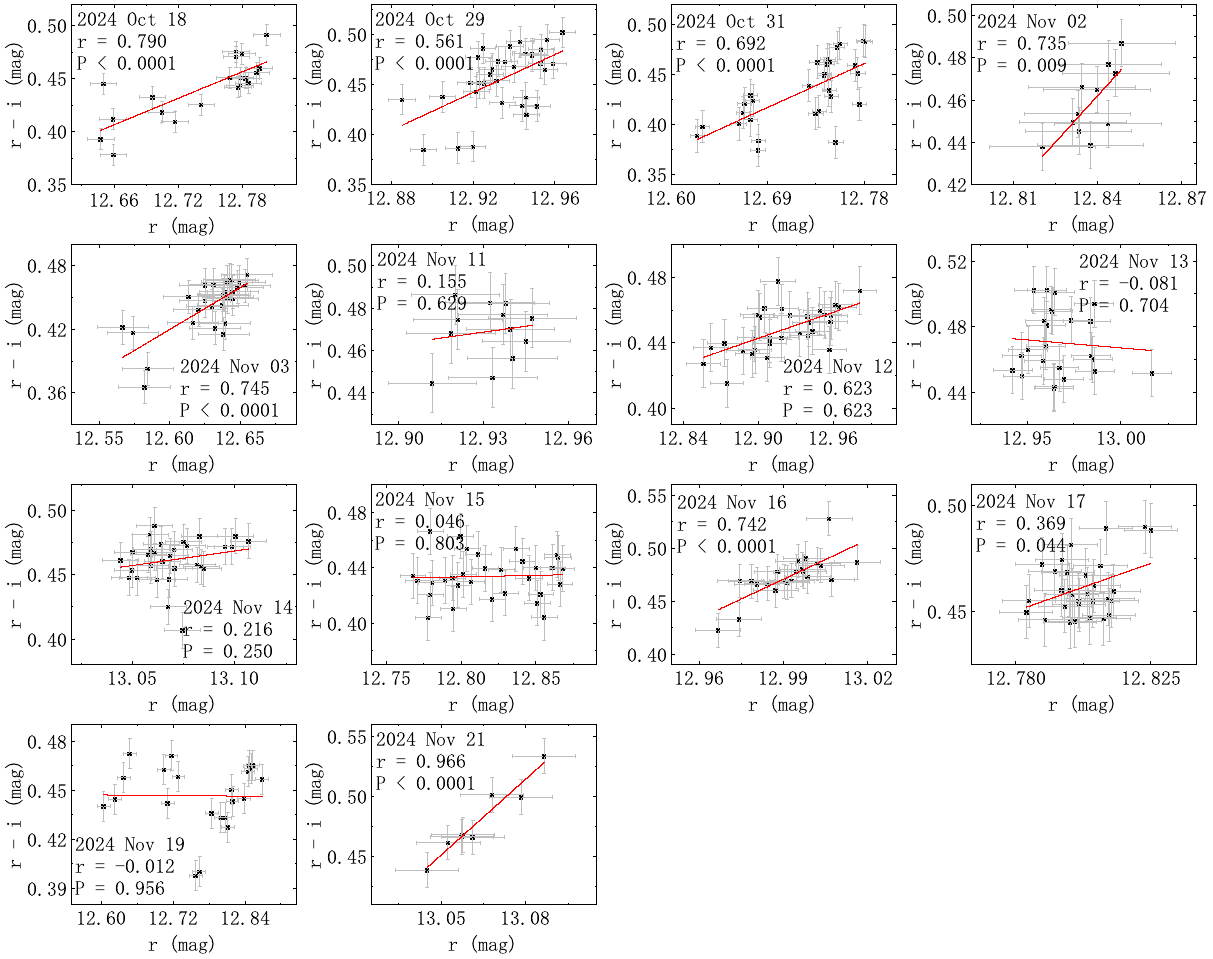}
\caption{$50$ cm correlations between the $r - i$ index and $r$ magnitude for intraday variability.
\label{10}}
\end{figure*}

\subsection{Periodicity Analysis}
\label{sec_lsp}
The Weighted Wavelet Z-transform (WWZ) \citep{Fost96,Bhat17} method is employed to detect and quantify QPOs in both the frequency and time domains, and it can be effectively applied to unevenly sampled and sparse data \citep{Witt05}. In real astronomical systems, the frequencies and amplitudes of QPOs may vary over time. In such cases, the WWZ method is particularly effective in identifying QPOs that evolve and dissipate over time \citep{Bhat17}. Following the studies by \citet{Fost96} and \citet{Bhat17}, we constructed the WWZ spectra using the Morlet mother function for each simulated light curve. The WWZ projects the Morlet wavelet transforms onto three trial functions: $\phi_1(t)=1(t)$, $\phi_2(t)=\cos\left[\omega\left(t-\tau\right)\right]$, $\phi_3(t)=\sin\left[\omega\left(t-\tau\right)\right]$. Additionally, statistical weights $\omega_\alpha=exp\left[-c\omega^2{(t_\alpha-\tau)}^2\right],\;(\alpha=1,2,3)$ are incorporated into the projection, where $c$ is an adjustable parameter \citep{Li21}. The WWZ power is used to estimate the confidence level of a detected periodicity with a frequency of $\omega$ and a time shift of $\tau$, as follows:
\begin{equation}
WWZ=\frac{(N_{eff}-3)V_y}{2(V_x-V_y)},
\label{eq:LebsequeIp6}
\end{equation}
where $N_{eff}$ represents the effective number of data points that contribute to the signal, while $V_x$ and $V_y$ indicate the weighted variations of the uneven data $x$ and the model function $y$, respectively \citep{Fost96}. The factors are defined as follows:
\begin{equation}
\begin{aligned}
& N_{eff}=\frac{{(\sum\omega_\alpha)}^2}{\sum\omega_\alpha^2}, \\
& V_x=\frac{\sum_\alpha\omega_\alpha x^2(t_\alpha)}{\sum_\lambda\omega_\lambda}-\left[\frac{\sum_\alpha\omega_\alpha x(t_\alpha)}{\sum_\lambda\omega_\lambda}\right]^2, \\
& V_y=\frac{\sum_\alpha\omega_\alpha y^2(t_\alpha)}{\sum_\lambda\omega_\lambda}-\left[\frac{\sum_\alpha\omega_\alpha y(t_\alpha)}{\sum_\lambda\omega_\lambda}\right]^2, \\
\label{eq:LebsequeIp7}
\end{aligned}
\end{equation}
where $\lambda$ represents the count of test frequencies. Subsequently, the WWZ periodogram is derived by dissecting the data into observation epochs and the time/frequency domain. The peaks in the WWZ power can be utilized to identify periodic components and track the evolution of both periodicities and amplitudes \citep{Li21}.

The transient periodic fluctuations in blazars usually exhibit a higher level of red noise at low frequencies, which can lead to spurious periodicities in the periodograms \citep{Pres78}. Therefore, during the analysis of blazar periodicity, it is essential to consider the effect of frequency-dependent red-noise \citep{Fan14,Sand16,Bhat17}. This effect can be addressed by the power response method, which characterizes the power spectral density (PSD) \citep{Uttl02}. The random fluctuations of blazars are typically approximated by a power-law PSD: $P(f) \propto f^{{-}\alpha}$, where $P(f)$ represents the power at a temporal frequency $f$, and $\alpha$ represents the spectral slope. The power spectral slope $\alpha$ can be estimated by fitting a linear function to the log-periodogram \citep{Vaug05}. Table \ref{table8} presents the best-fitting PSDs obtained through this process. Subsequently, these values are used to model the red-noise in the optical variability of BL Lacertae. Next, we assessed the confidence level of the QPO. We generated $10,000$ simulated light curves based on the best-fitting PSD model. For candidates with significance levels exceeding $99.7\%$, the number of simulations was increased to $100,000$ to ensure statistical robustness. Subsequently, following the description by \citet{Timm95}, we employed the Monte Carlo method to establish the red-noise background. These artificial light curves have identical sampling intervals, standard deviations, and mean values. For each simulated light curve, the WWZ power spectrum can be obtained \citep{Li23}. Finally, the confidence levels were estimated using the power spectral distribution of the simulated light curves.

\begin{figure*}
\centering
\includegraphics[scale=.32]{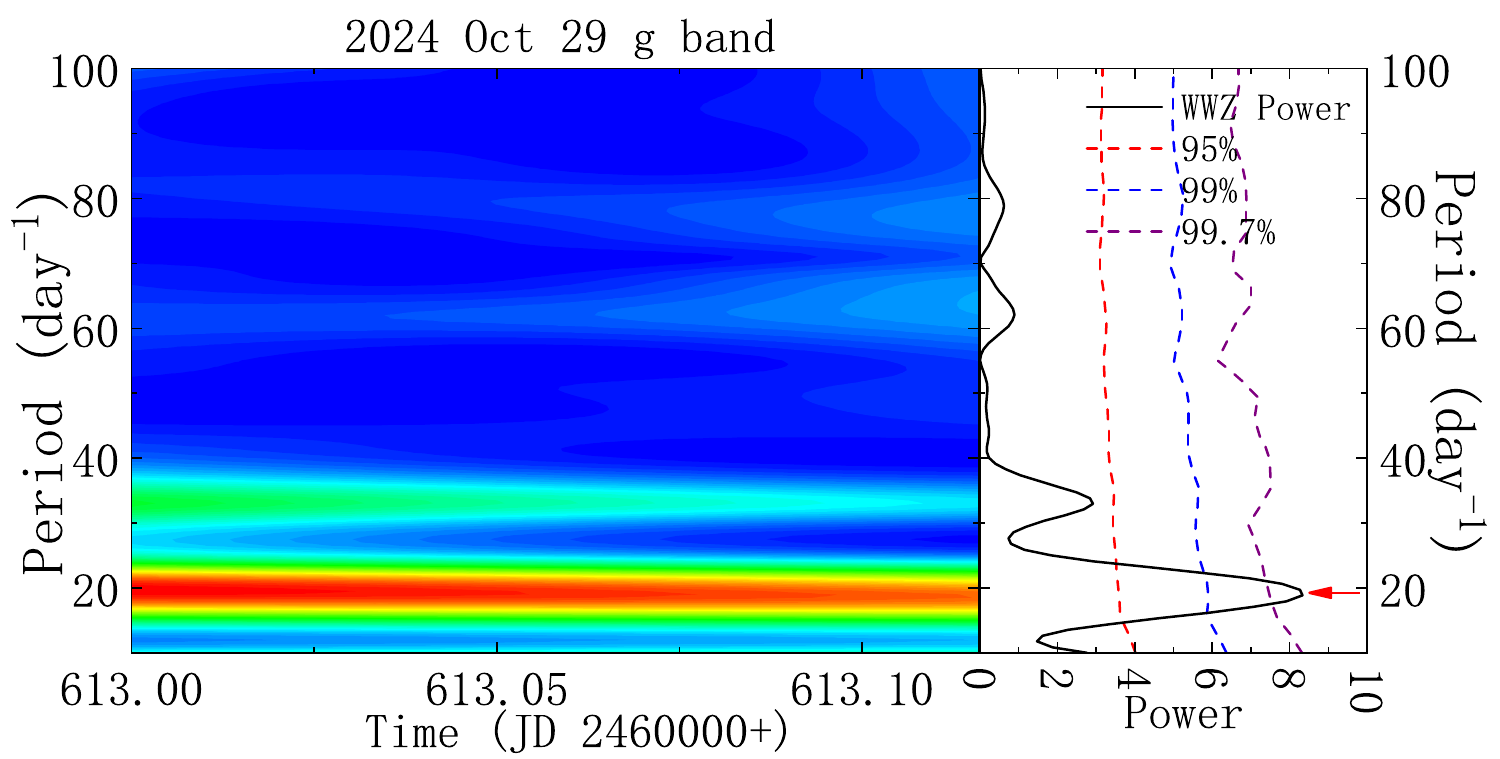}
\includegraphics[scale=.32]{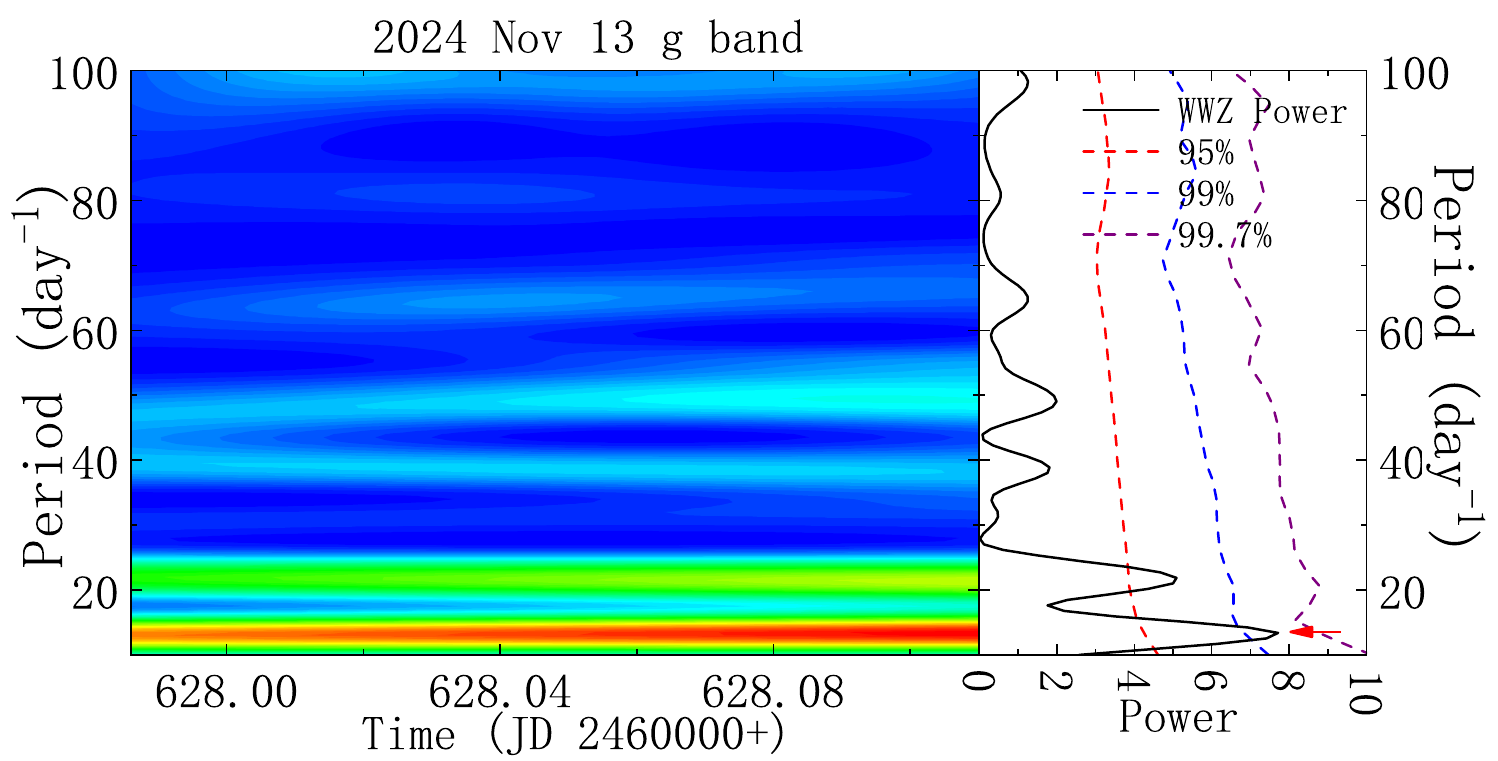}
\includegraphics[scale=.32]{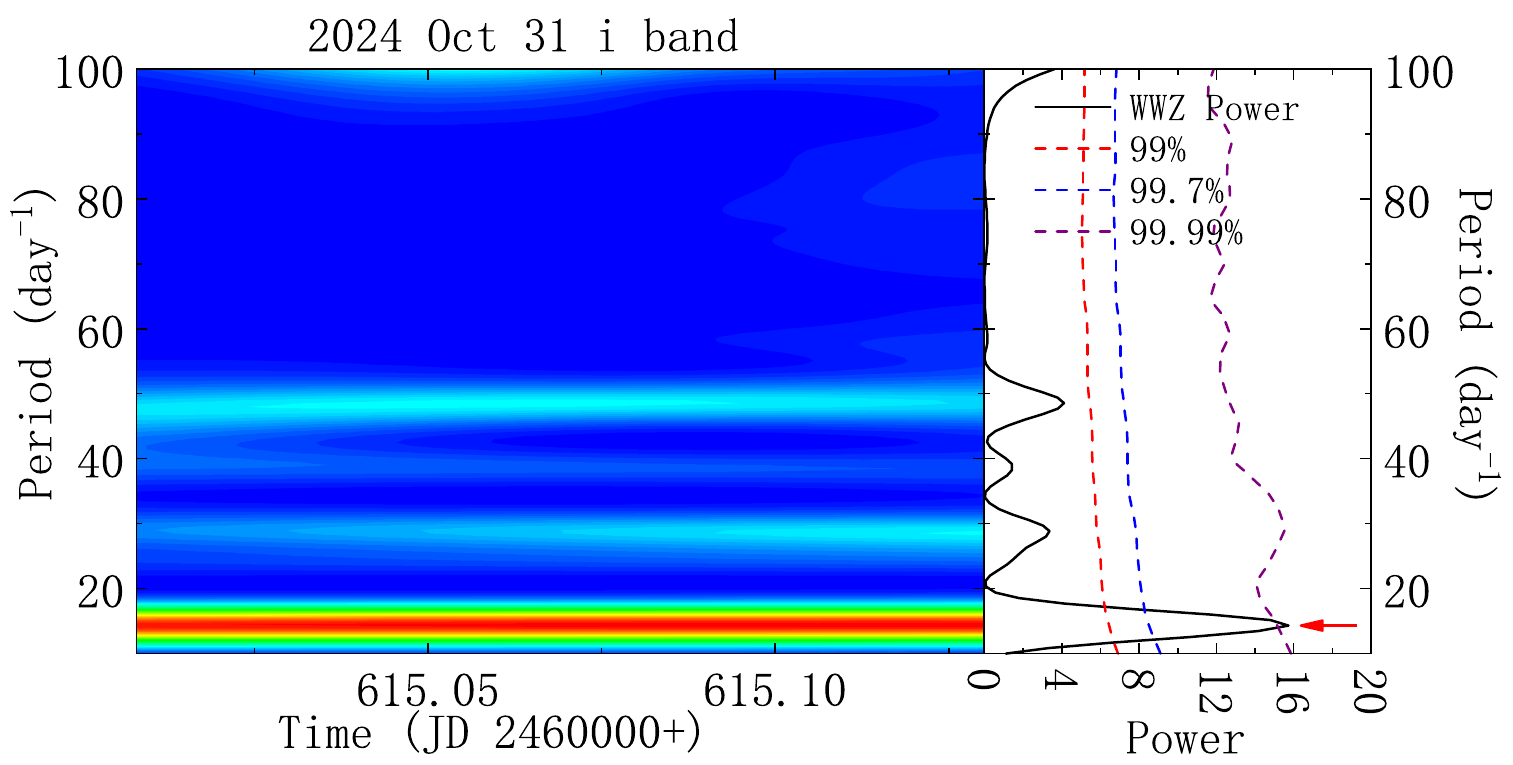}
\includegraphics[scale=.32]{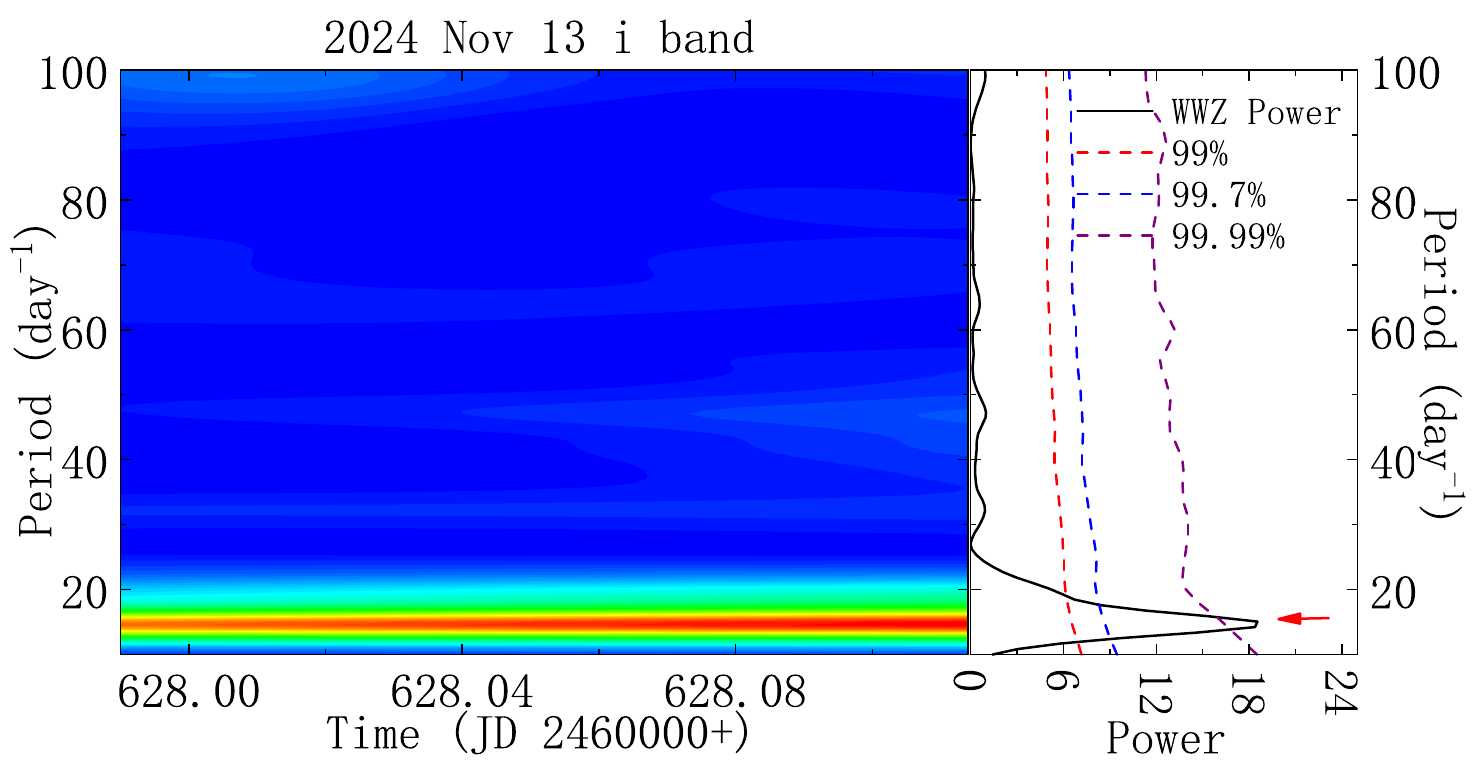}
\includegraphics[scale=.32]{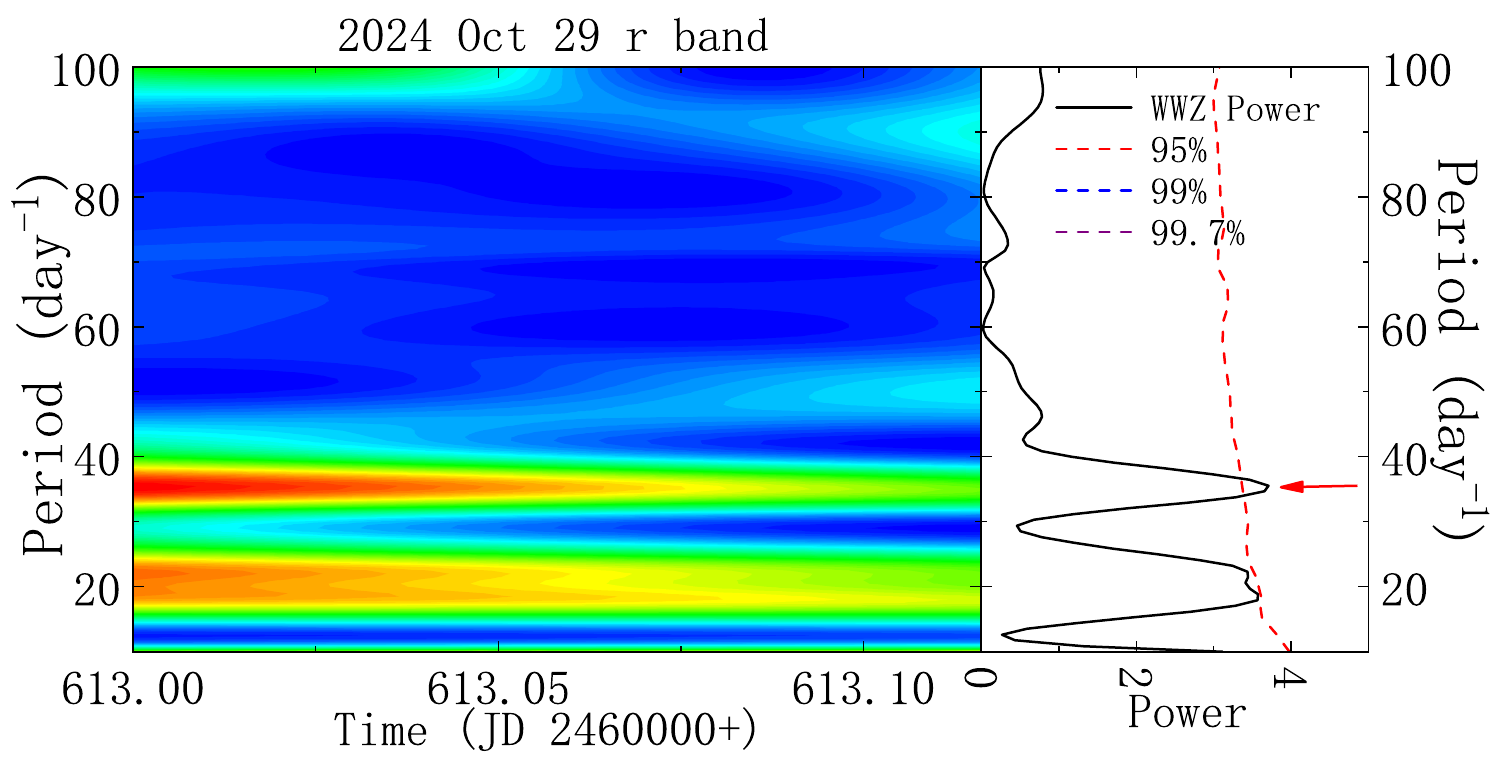}
\includegraphics[scale=.32]{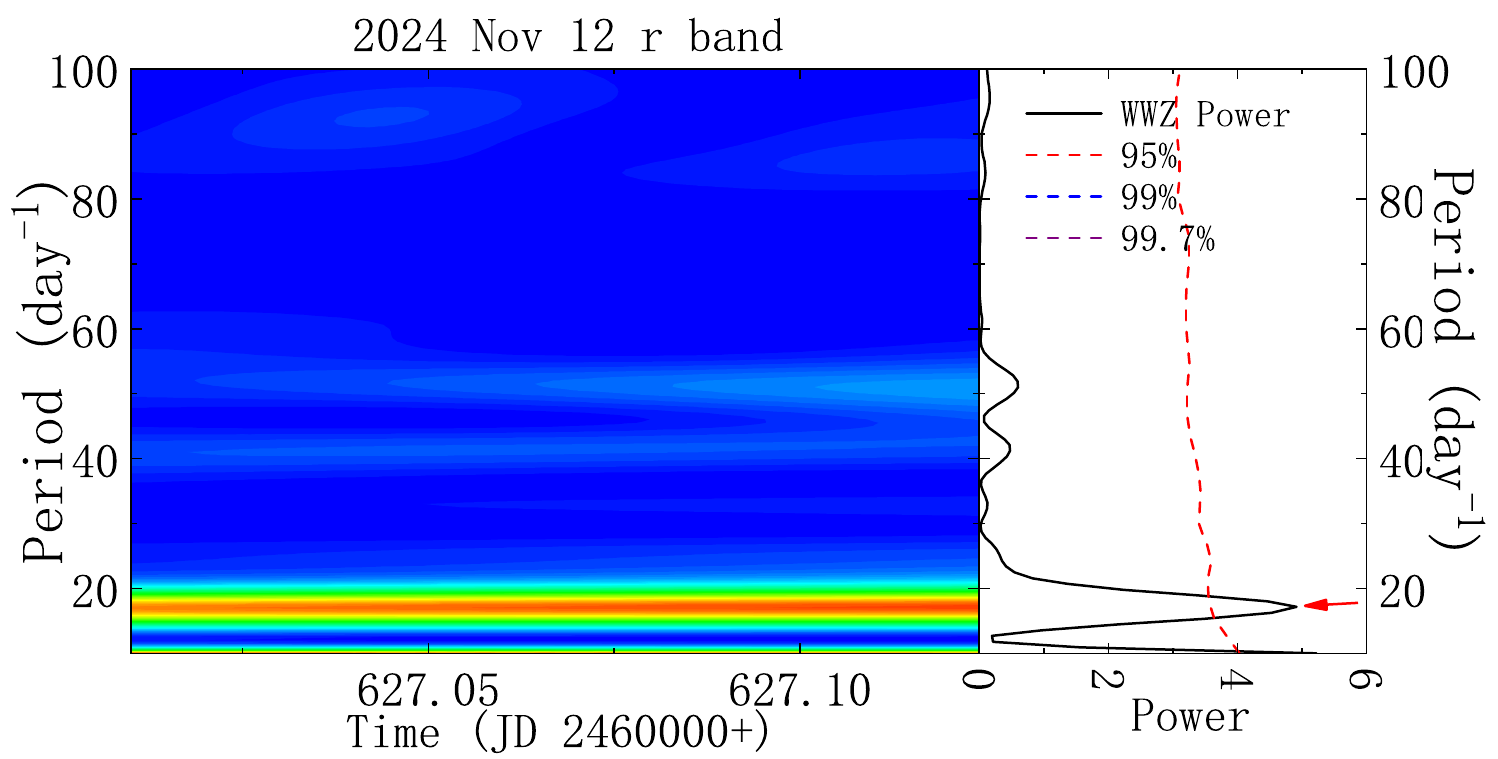}
\includegraphics[scale=.32]{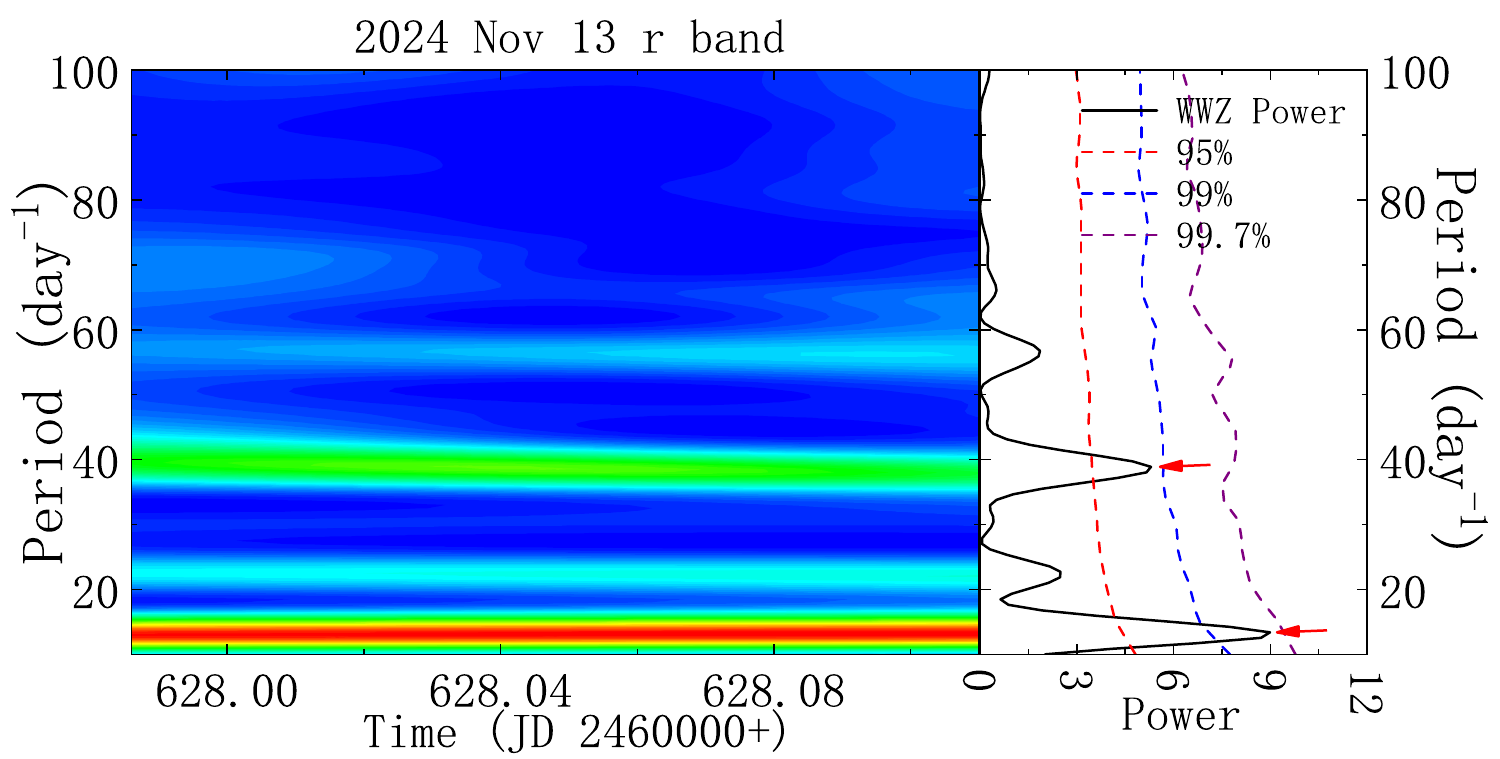}
\caption{2D plane contour of the WWZ power. The red arrow marks the period of the detected QPO. The black solid curve in the side panel represents the time-averaged WWZ power. The red, blue, and purple dashed lines represent the confidence levels.
\label{12}}
\end{figure*}

The analysis results using the WWZ are shown in Figure \ref{12}. The red, blue, and purple curves in Figure \ref{12} correspond to different confidence levels, respectively. The half-width at half-maximum (HWHM) of the power spectrum peak is used to estimate the uncertainty of the periodic modulation signal. As shown in Figure \ref{12}, during the observation of BL Lacertae, the highest power peaks in the WWZ periodogram (marked by red arrows) correspond to different periods of approximately $40$ to $107$ minutes, suggesting the possible presence of QPO signals in the source. Among these, the $i$-band observations on October $31$ and November $13$ show periods of $100.77 \pm 14.56$ minutes and $95.43 \pm 14.51$ minutes, respectively, both with confidence levels exceeding $99.99\%$. The detailed results are listed in Table \ref{table8}.

\begin{table*}
   \centering
   \caption{Results of Quasi-Periodic Oscillation (QPO) analysis. Column 1 is the date of the observation; Column 2 is the band; Column 3 is the power spectral slope $\alpha$; Column 4 is the period and its error; Column 5 is the confidence level $\sigma$.}
   %\label{tab:tab1}
   \renewcommand\arraystretch{1.2}
  \setlength{\tabcolsep}{6.2mm}
   \begin{tabular}{cccccccccccc} % four columns, alignment for each
      \hline
Date  &   Band  &   $\alpha$  &   $p$ (min) &   $\sigma$    \\\hline
2024 Oct 29	&	g	&	0.50	&	76.51 $\pm$ 12.30	   &	99.7\%\\
2024 Nov 13	&	g	&	0.70	&	107.46 $\pm$ 13.85   &	99\%\\
2024 Oct 29	&	r	&	0.50	&	40.46 $\pm$ 3.63     &	95\%\\
2024 Nov 12	&	r	&	0.51	&	84.06 $\pm$ 10.56	   &	95\%\\
2024 Nov 13	&	r	&	0.70	&	65.78 $\pm$ 7.76	   &	95\%\\
2024 Nov 13	&	r	&	0.74	&	107.46 $\pm$ 13.85   &	95\%\\
2024 Oct 31	&	i	&	0.60	&	100.77 $\pm$ 14.56   &	99.99\%\\
2024 Nov 13	&	i	&	0.64	&	95.43 $\pm$ 14.51	   &	99.99\%\\\hline
\label{table8}
   \end{tabular}
\end{table*}

\section{Discussion}
\label{sec_discuss}
During the period of October to November $2024$, we carried out simultaneous photometric monitoring of BL Lacertae over $20$ nights using the Mephisto $1.6$ m telescope, covering the $u$, $v$, $g$, $r$, $i$, and $z$ bands. Additionally, we observed BL Lacertae in the $g$, $r$, $i$, and $z$ bands for $16$ nights using the twin $50$ cm telescopes. The resulting light curves are presented in Figures \ref{1} through \ref{5}.

\subsection{Intraday Variability}
\label{sec_dis_smbhb}
In some studies, the variability amplitude of BL Lacertae objects tends to be greater at higher frequencies \citep{Ghis97,Mass98,Bonn12}. Nevertheless, there have been instances where the variability amplitude does not consistently increase with frequency \citep{Ghos00}. In our results, the variability amplitude in lower-energy bands occasionally exceeds that in higher-energy bands (see Table \ref{table2} $\sim$ \ref{table5}), with no clear correlation between the amplitude and the frequency of the band, which is consistent with the findings of \citet{Gaur15}. Through the F-test and ANOVA-test, we detected IDVs in the light curves across different bands, with $8$ days in the $g$-band, $7$ days in the $r$-band, $6$ days in the $i$-band, and $3$ days in the $z$-band. During this period, the duty cycle (DC) of BL Lacertae was $46.83\%$ in the $g$-band, $41.33\%$ in the $r$-band, $36.13\%$ in the $i$-band, and $19.27\%$ in the $z$-band. The IDV light curves exhibit gradual increasing or decreasing trends, as well as some flares. We found that as the frequency grows, the duty cycle increases. This may be attributed to the fact that high-frequency synchrotron radiation is predominantly produced in more compact regions within the jet, driven by higher-energy electrons. These high-energy electrons release their energy over shorter time-scales, resulting in rapid fluctuations in the emitted radiation \citep{Ghis09,Mars14}.

Between 1999 and 2021, \citet{Webb21} conducted 53 micro-variability observations of BL Lacertae and defined micro-variability as oscillations occurring around a linear increase or decrease in flux level. This definition effectively distinguishes between persistent linear trends throughout the night and low-amplitude oscillations. Such linear trends may be attributed to a background twist in the magnetic field of the jet. They found that several light curves transitioned from ``micro-variability'' to ``no micro-variability'' after subtracting the linear trends. They reported the duty cycles for BL Lacertae, calculated via the Romano method \citep{Rome99}, as 56.67\% and 39.61\% with and without linear trends, respectively. The number of days exhibiting micro-variability decreased from 20 to 13. Following \citet{Webb21}, we performed a linear least-squares fit on our light curves exhibiting significant variations and subtracted the linear trend. Subsequently, the F-test and ANOVA were re-applied to the linearized data. Consistent with their findings, the duty cycles in our results decreased across all bands: 30.43\% in the $g$-band, 19.04\% in the $r$-band, and 12.78\% in the $i$-band, with no micro-variability detected in the $z$-band. The number of days exhibiting micro-variability correspondingly decreased to $5$, $3$, $2$, and $0$ days, respectively. The duty cycle consistently exhibits an increasing trend with frequency. Additionally, we compared our $r$-band (630 nm) DC values with the $R$-band (660 nm) results reported by \citet{Webb21} as their effective wavelengths are similar. Specifically, our calculated $r$-band DC was 41.33\% when the linear trend was included (with linear) and decreased to 19.04\% after the trend was subtracted (without linear). \citet{Webb21} propose that micro-variability is the result of a shock encountering a turbulent region of the jet flow (see Section \ref{sec_krm} for details). If the duty cycle is large, then the jet is mostly well-developed turbulence, and the Reynolds number is much larger. If the duty cycle is moderate, then the turbulence is not well developed with no large-scale eddies. Our relatively lower DC values suggest that the jet conditions during our observations align more closely with the latter case of less-developed turbulence.

We observed that BL Lacertae exhibits various variability time-scales ranging from a few tens of minutes to several tens of minutes on different nights. The shortest time-scale can be associated with the size of the emitting feature in the jet \citep[e.g.,][]{Rome99} and can be utilized to estimate the mass of its central black hole \citep[e.g.,][]{Mill89,Liu15}. According to the model proposed by \citet{Liu15}, the rapid variability of blazars can be used to constrain the black hole mass $M_{\bullet}$, as follows:
\begin{equation}
M_{\bullet} \lesssim \ 5.09 \times 10^4 \frac{\delta\Delta t^{\text{ob}}_{\text{min}}}{1+z} M_{\odot}  \qquad  (j \sim 1),
\label{eq:LebsequeIp8}
\end{equation}
\begin{equation}
M_{\bullet} \lesssim \ 1.70 \times 10 ^4 \frac{\delta\Delta t^{\text{ob}}_{\text{min}}}{1+z} M_{\odot}  \qquad (j = 0),
\label{eq:LebsequeIp9}
\end{equation}
where $\Delta t^{\text{ob}}_{\text{min}}$ is the minimum variability time-scale in seconds, $z$ is the redshift, and $j = J/J_{\text{max}}$ is the dimensionless spin parameter of a black hole, with $J_{\text{max}} = GM^2_{\bullet}/c$ representing the maximum possible angular momentum, where $G$ is the gravitational constant. Equation \ref{eq:LebsequeIp8} applies to Kerr black holes, while Equation \ref{eq:LebsequeIp9} is valid for Schwarzschild black holes.
Using Doppler factor $\delta = 12.17$ \citep{Liod18} and minimum time-scale $\tau = 0.0119$ days (from ACF results), we estimated the black hole mass of BL Lacertae. According to Equations \ref{eq:LebsequeIp8} and \ref{eq:LebsequeIp9}, we derive the upper limits for the black hole mass of BL Lacertae as $M_{\bullet} \lesssim 10^{8.29} M_{\odot}$ for a Kerr black hole and $M_{\bullet} \lesssim 10^{8.77} M_{\odot}$ for a Schwarzschild black hole. \citet{Woo02} estimated the black hole mass of BL Lacertae to be $10^{8.23} M_{\odot}$ using stellar velocity dispersion, which is consistent with the result for the Kerr black hole.

The size of the emission region, assuming it to be a spherical blob, is estimated using the following relation \citep{Rome99}:
\begin{equation}
R\le c \Delta t^{\text{ob}}_{\text{min}} \frac{\delta }{1+z},
\label{eq:LebsequeIp10}
\end{equation}
where $c$ is the speed of light, and $z$ is the redshift. We derive the upper limit for the size of the emission region as $R \le 3.51 \times 10^{14}$ cm. Assuming the jet has a standard geometry, the distance from the emission region to the central supermassive black hole can be estimated using the following formula:
\begin{equation}
R_H\le 2c \Delta t^{\text{ob}}_{\text{min}} \frac{\Gamma^2 }{1+z},
\label{eq:LebsequeIp11}
\end{equation}
where $\Gamma$ is the bulk Lorentz factor. The distance from the central black hole is also constrained to an upper limit of $R_H \le 2.83 \times 10^{15}$ cm ($\Gamma = 7$, from \citet{Rait13}). It is well known that the size of the broad-line region (BLR) is approximately $0.1$ pc ($3.09 \times 10^{17}$ cm). Comparing the distance of the blob with the size of the BLR suggests that the emission region is located deep inside the BLR.

The highest variability time-scale of IDV ($0.0503$ days) corresponds to a radiation region size $R$ of approximately $1.48 \times 10^{15}$ cm, with the distance between this radiation region and the central supermassive black hole ($R_H$) estimated to be around $1.19 \times 10^{16}$ cm. This is consistent with the shock-in-jet model, which predicts that the size or thickness of the emitting zone increases with the distance traveled by the shock along the jet \citep{Blan76}. In the shock-in-jet model, the evolution of the electron energy density distribution of relativistic particles leads to variable synchrotron radiation emission. In this model, shocks accelerate turbulent particles in the jet plasma, which subsequently cool through synchrotron radiation \citep{Mars14,Cala15}. A detailed description of this model will be provided in the following sections.

\subsection{The KRM Model}
\label{sec_krm}
According to the shock-in-jet model, the observed IDV/micro-variability phenomena can be produced through convolution (superposition) of individual synchrotron radiation pulses generated in turbulent jets \citep{Bhat13,Webb21,Webb23}. In this model, the interaction between shock waves and turbulent plasma cells drives electron acceleration within the cells. These accelerated electrons subsequently cool through synchrotron radiation processes, ultimately producing flare events \citep{Bhat13,Webb23}. The synchrotron emission intensity depends on both the electron density and the magnetic field orientation/strength \citep{Webb16,Webb21}. The equations proposed by \citet{Kirk98} (KRM equations) enable calculation of particle distributions at shock fronts under various magnetic field orientations and particle density conditions \citep{Kirk98,Webb10}. The flare profiles derived from the KRM model show remarkable consistency with the observed pulse profiles of IDV/micro-variability in blazar light curves \citep{Bhat11}. Utilizing the model originally developed by \citet{Bhat13} and later refined by \citet{Xu23}, we recalculated the KRM-derived flare profiles and compared them with our IDV/micro-variability light curves. The particle distribution is given by the diffusion equation:
\begin{equation}
\frac{\partial N}{\partial t} + \frac{\partial}{\partial \gamma } \left [ \left ( \frac{\gamma }{t_{acc}} - \beta_s \lambda ^2 \right ) N \right ] + \frac{N}{t_{esc}} = Q\delta (\gamma - \gamma_0),
\label{eq:LebsequeIp12}
\end{equation}
where
\begin{equation}
\beta_{\text{s}} = \frac{4}{3} \frac{\sigma_{\text{T}}}{m_{\text{e}} c} \left (\frac{B^2}{2\mu_0} \right),
\label{eq:LebsequeIp13}
\end{equation}
where $N$ represents the electron number density as a function of energy $\gamma$ and time $t$, where $t_{\text{acc}}$ and $t_{\text{esc}}$ denote the particle acceleration and escape time-scales, respectively. The ratio of these time-scales governs the pulse profile. The synchrotron cooling rate $\beta_{\text{s}}$ incorporates the magnetic field strength $B$, Thomson scattering cross-section $\sigma_{\text{T}}$, vacuum permeability $\mu_0$, electron mass $m_{\text{e}}$, and speed of light $c$. The pulse amplitude is determined by the parameter $Q$, which depends on the magnetic field strength $B$, orientation angle $\theta$, and enhanced electron density \citep{Bhat13,Webb21}. The KRM equations are solved under the condition of a constant injection rate $Q_0$, activated at $t = 0$. The KRM solution defines both the amplitude and shape of individual pulses, corresponding to radiation from discrete turbulent plasma cells \citep{Webb21}. By adjusting the standard pulse width and amplitude to match prominent pulses in the IDV light curves, the emission region size can be constrained \citep{Webb21,Webb23}.

\begin{figure*}
\centering
\includegraphics[scale=0.06]{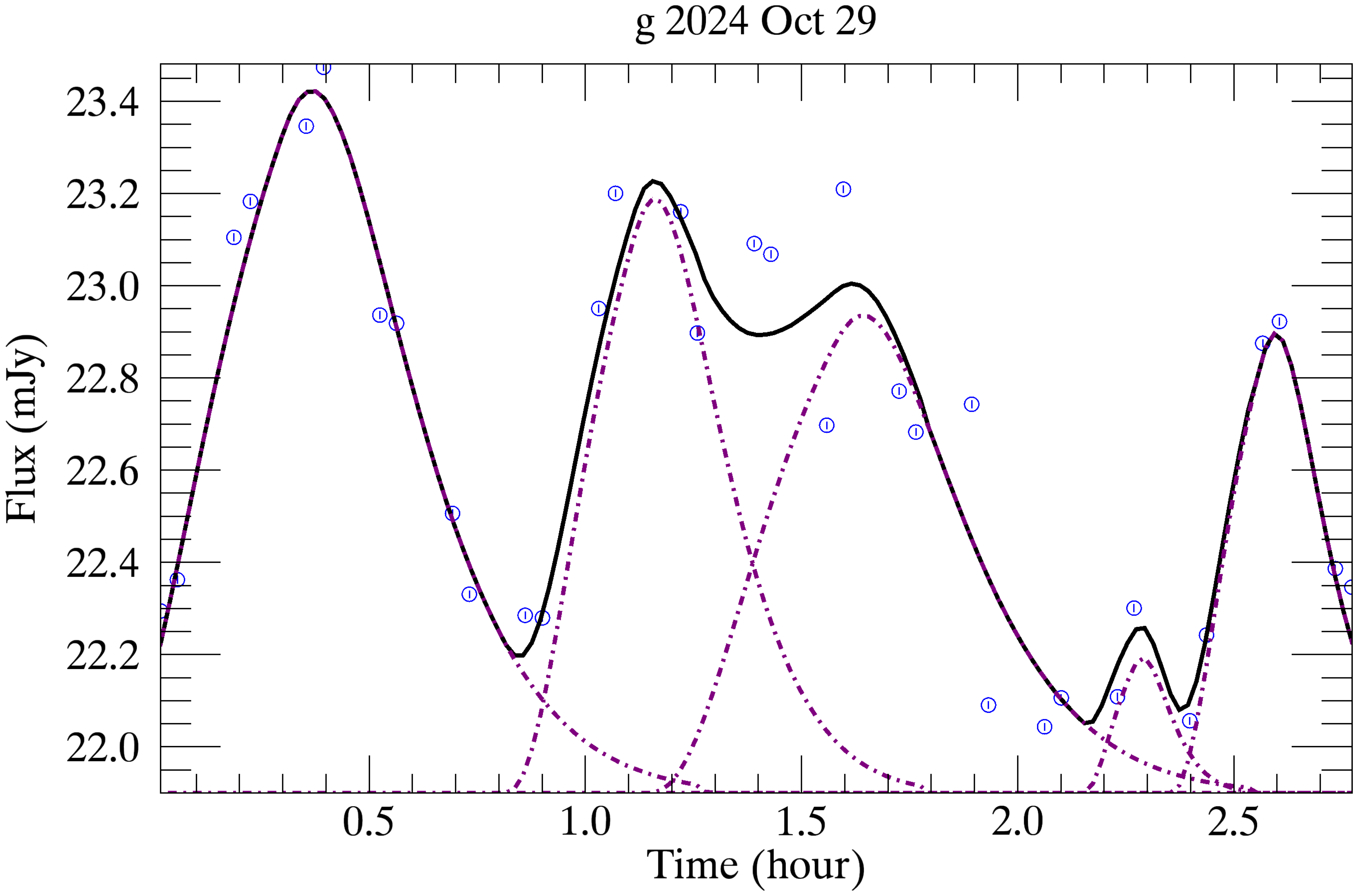}
\includegraphics[scale=0.06]{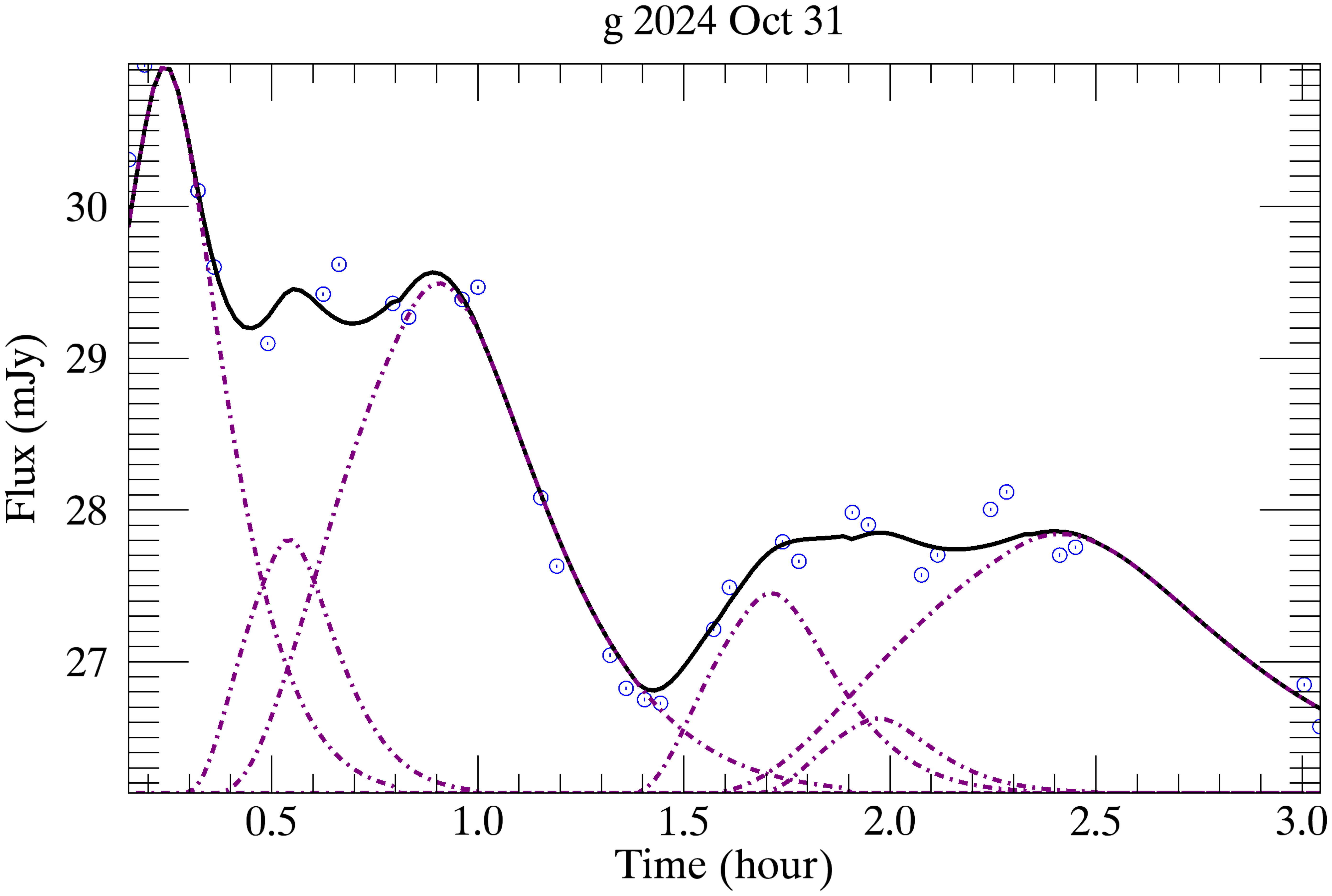}
\includegraphics[scale=0.06]{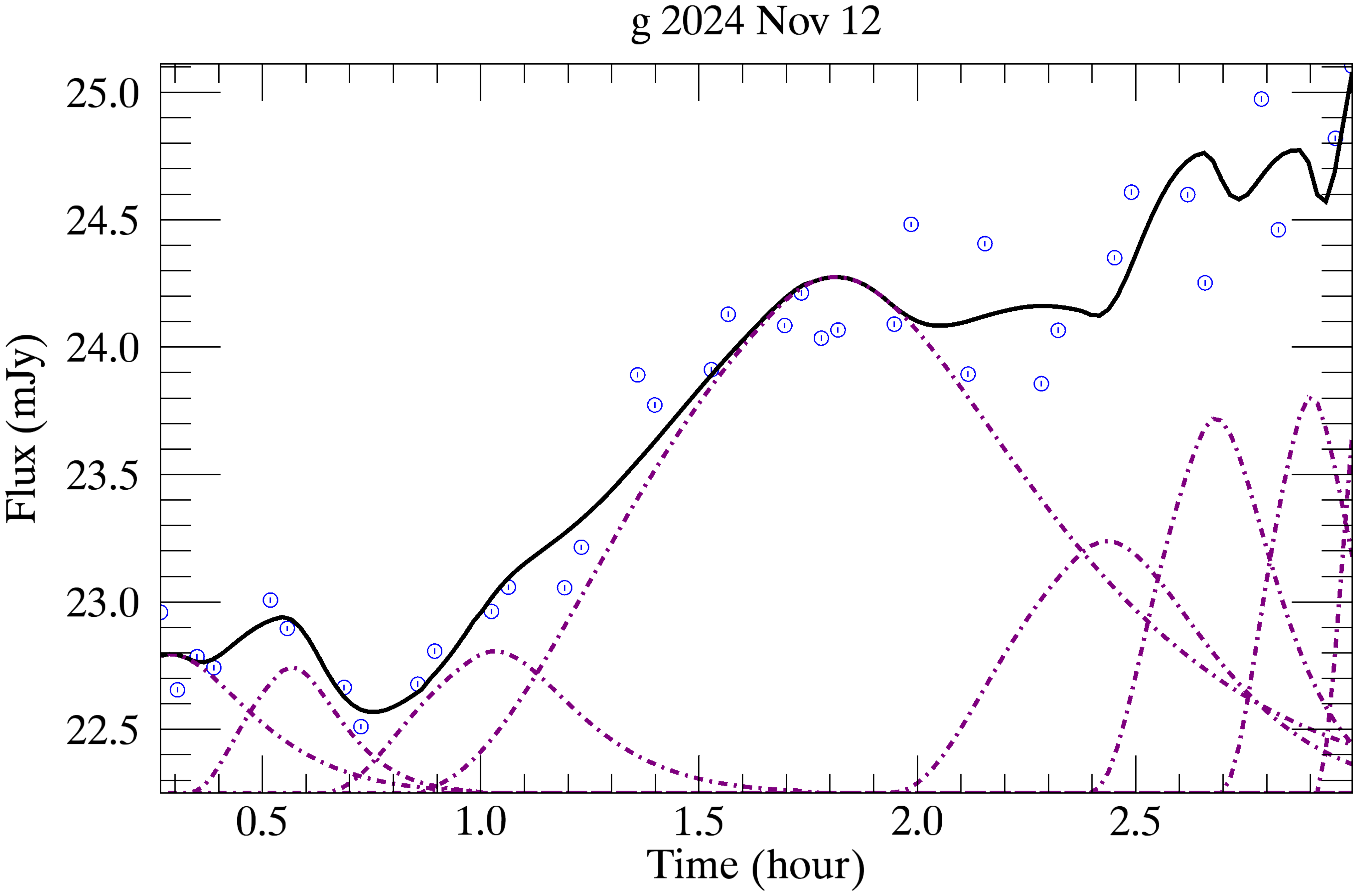}
\includegraphics[scale=0.06]{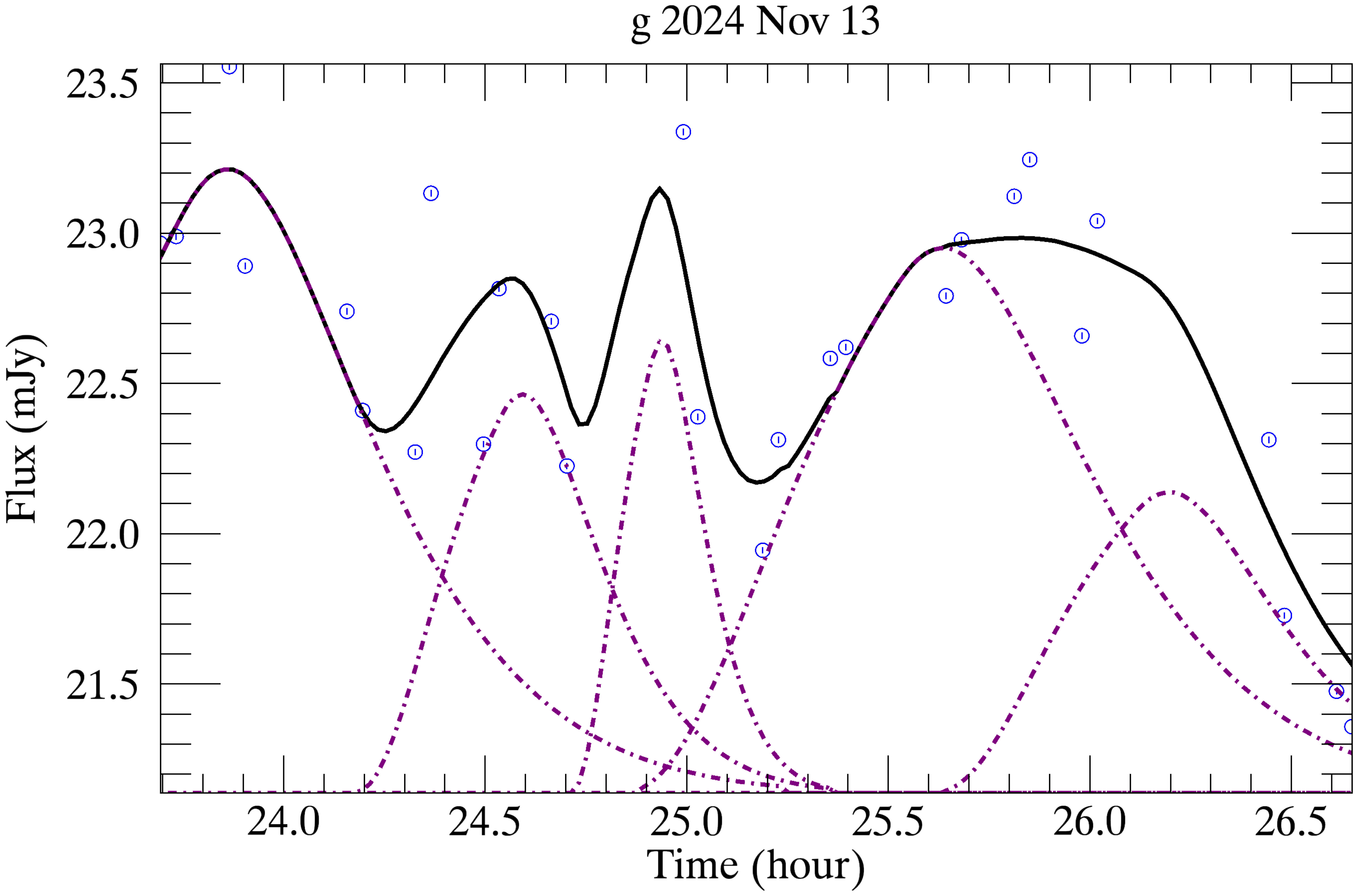}
\includegraphics[scale=0.06]{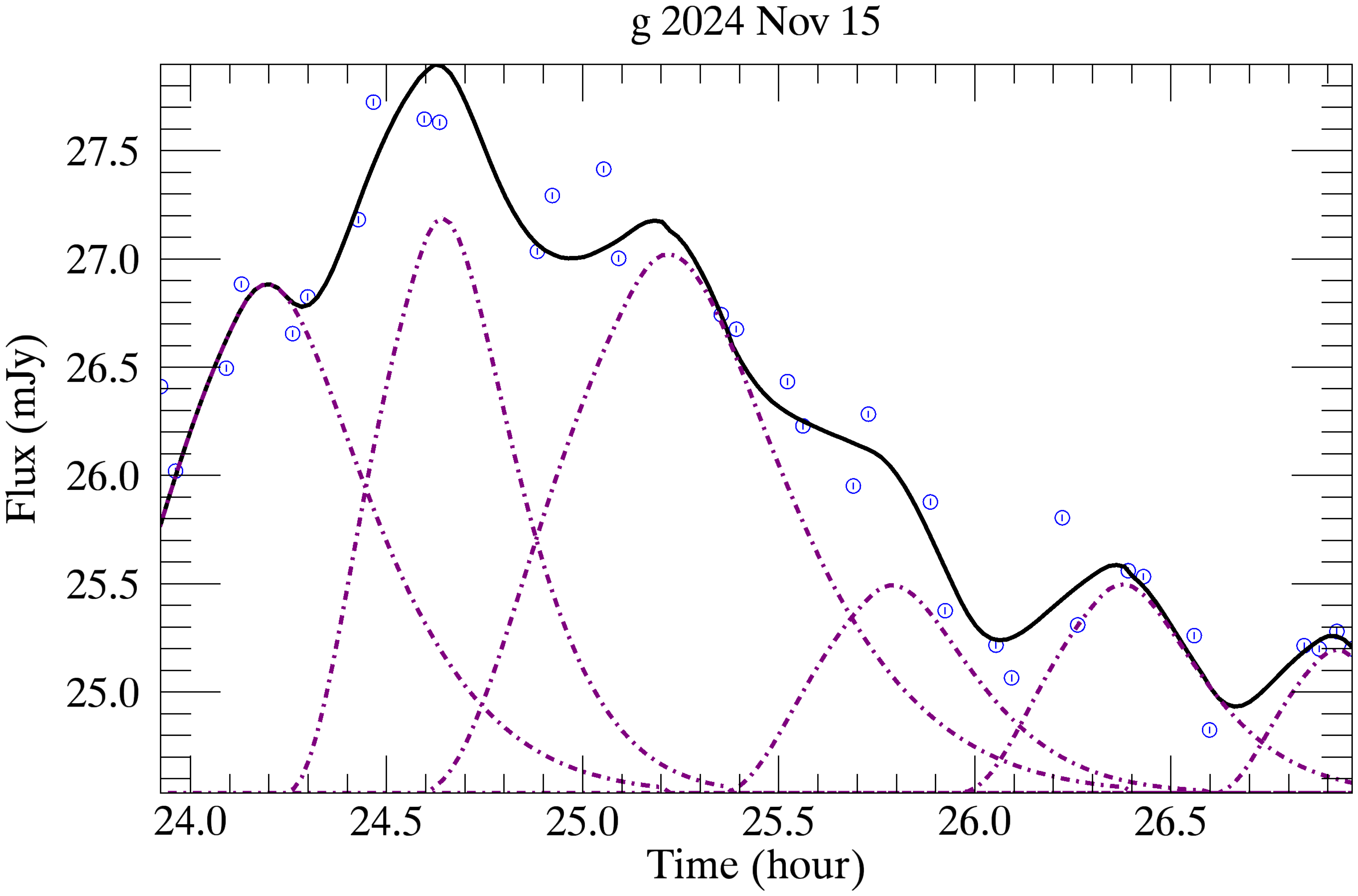}
\includegraphics[scale=0.06]{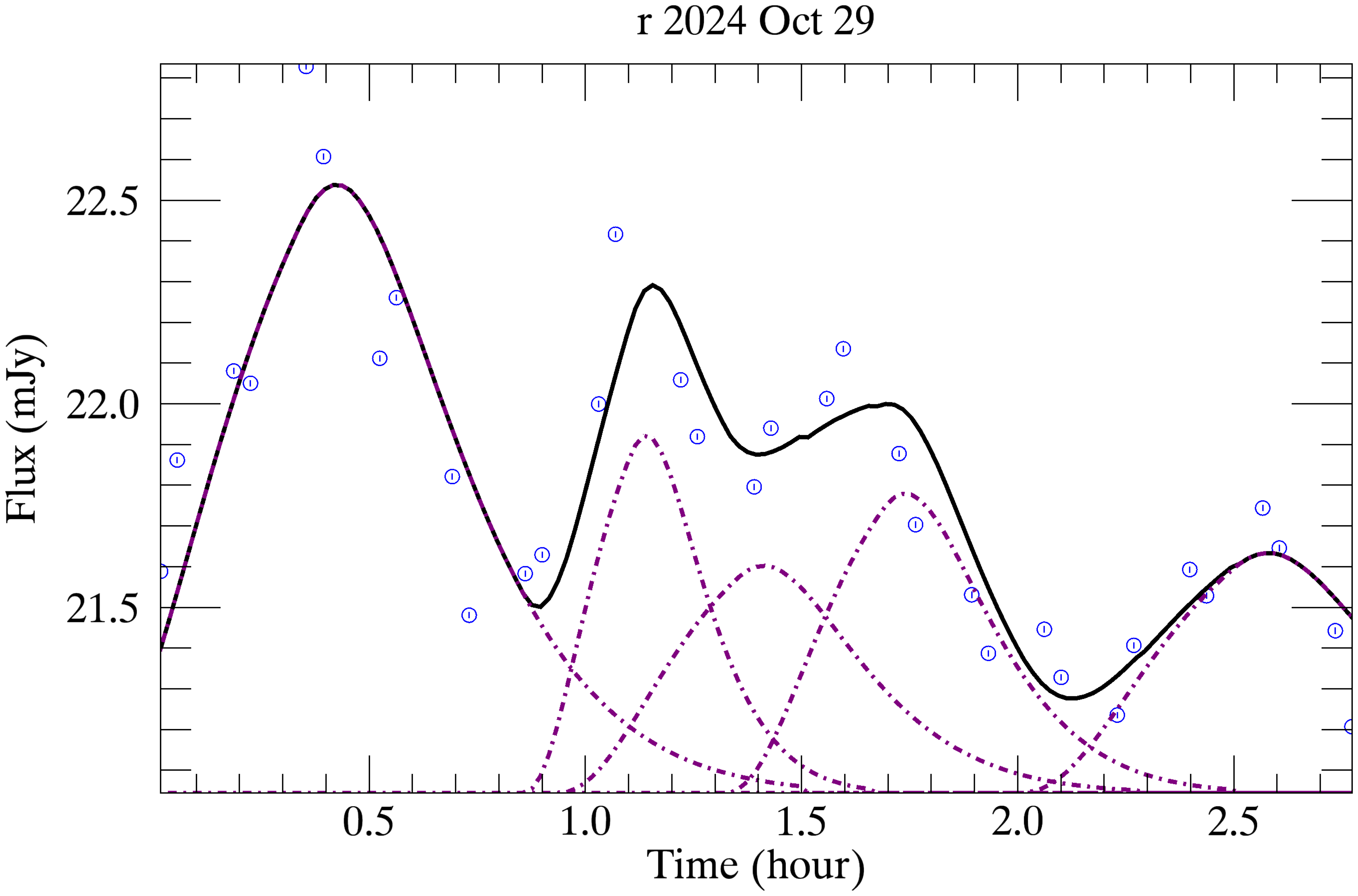}
\includegraphics[scale=0.06]{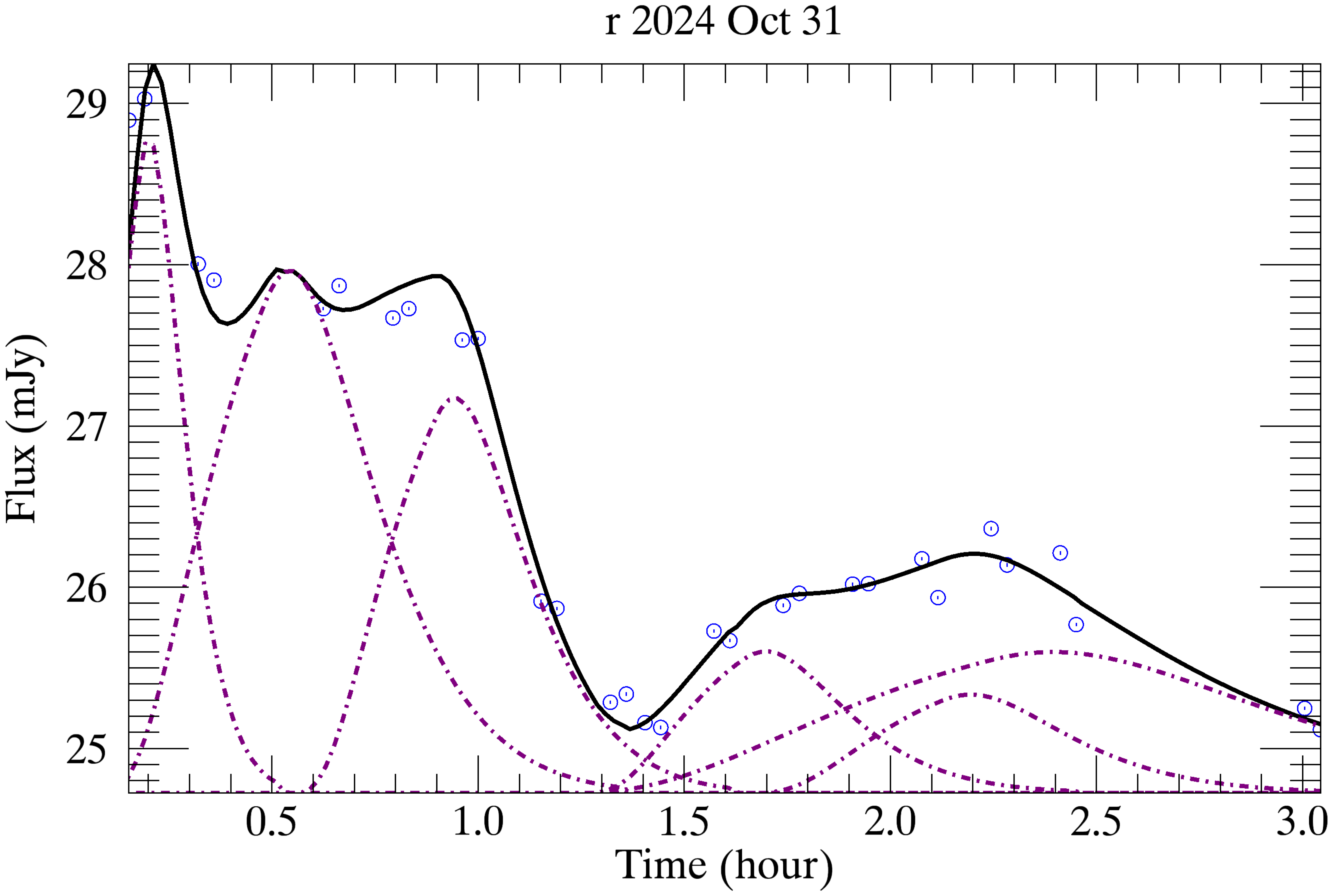}
\includegraphics[scale=0.06]{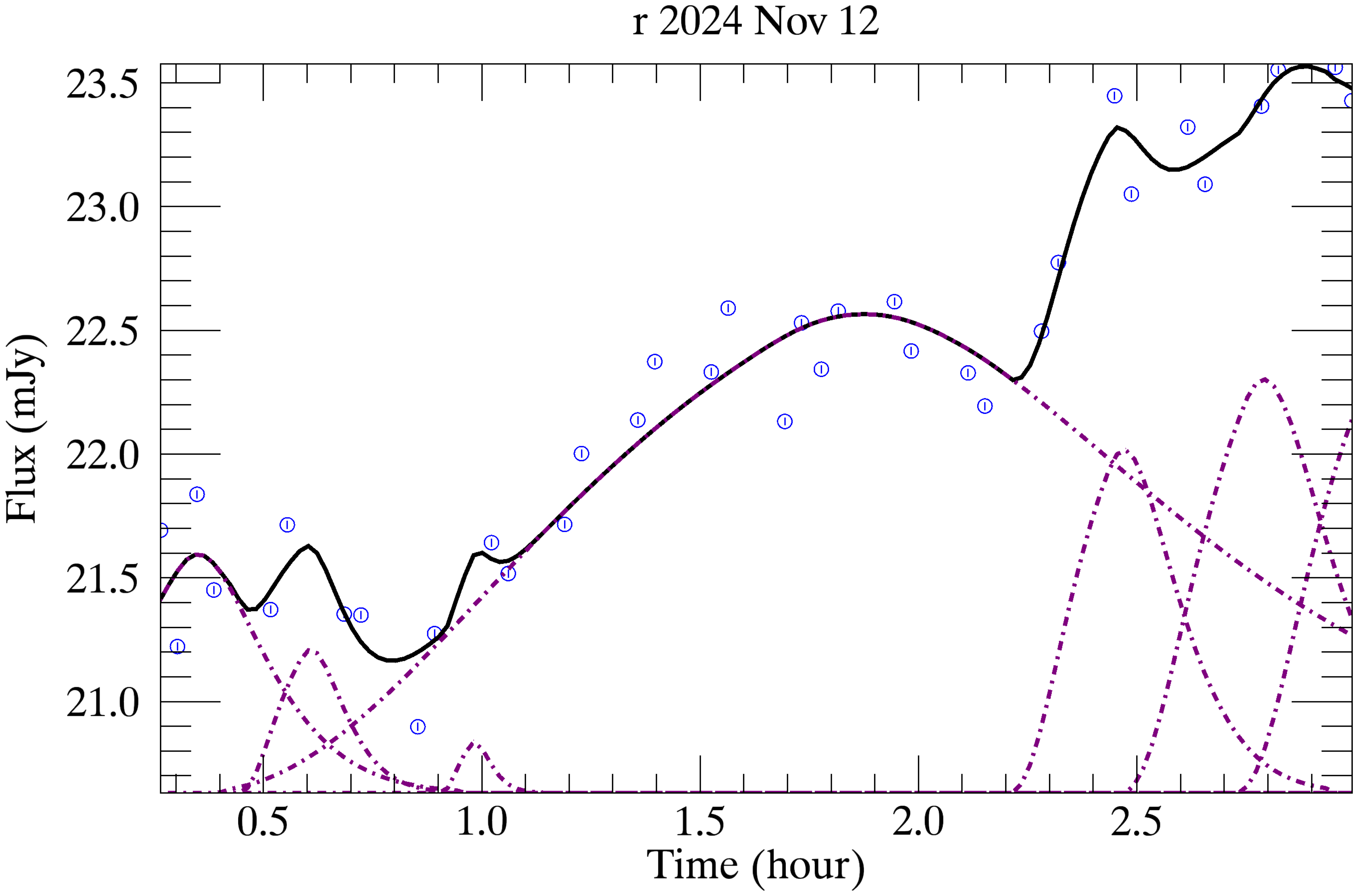}
\includegraphics[scale=0.06]{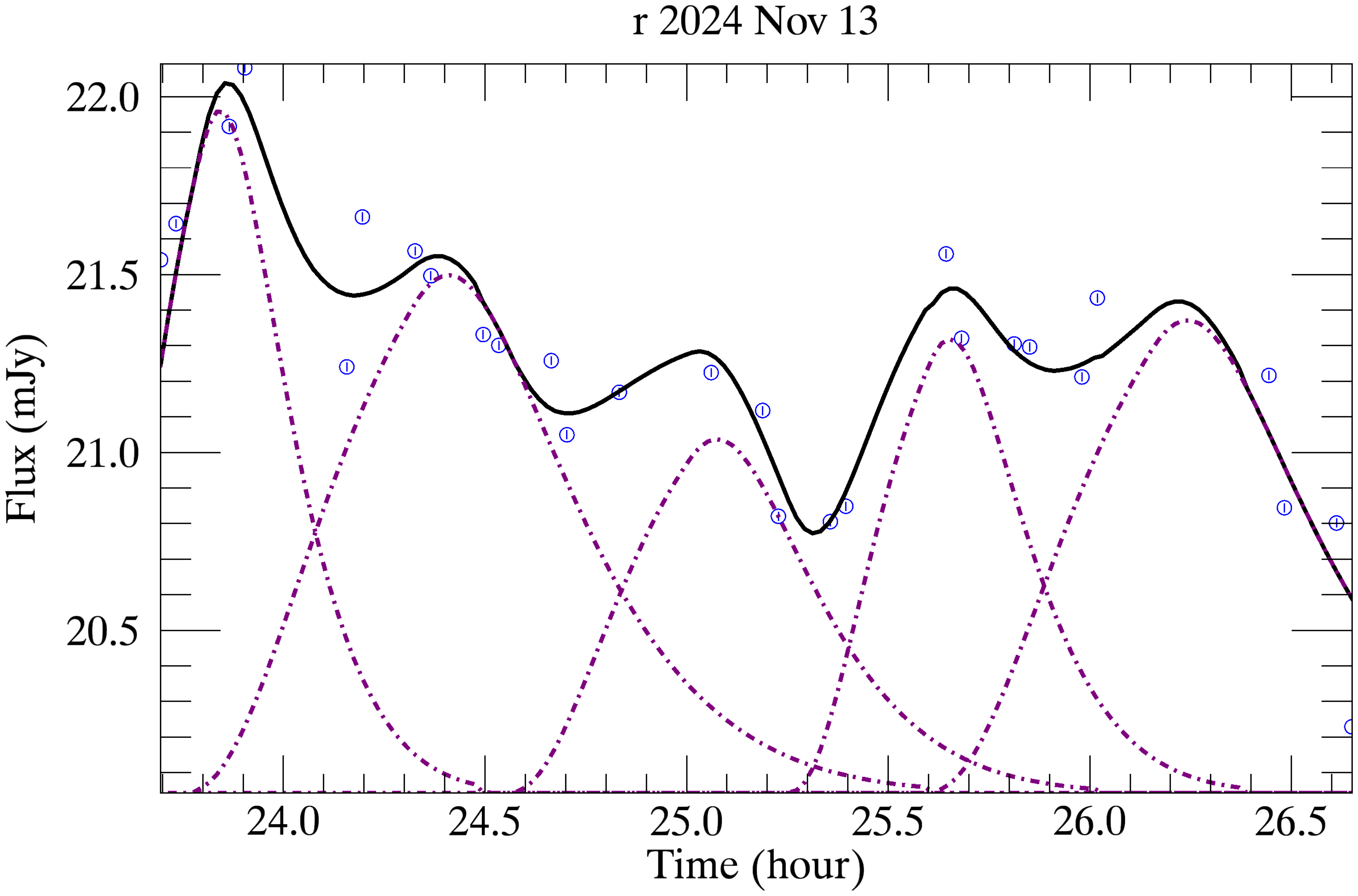}
\includegraphics[scale=0.06]{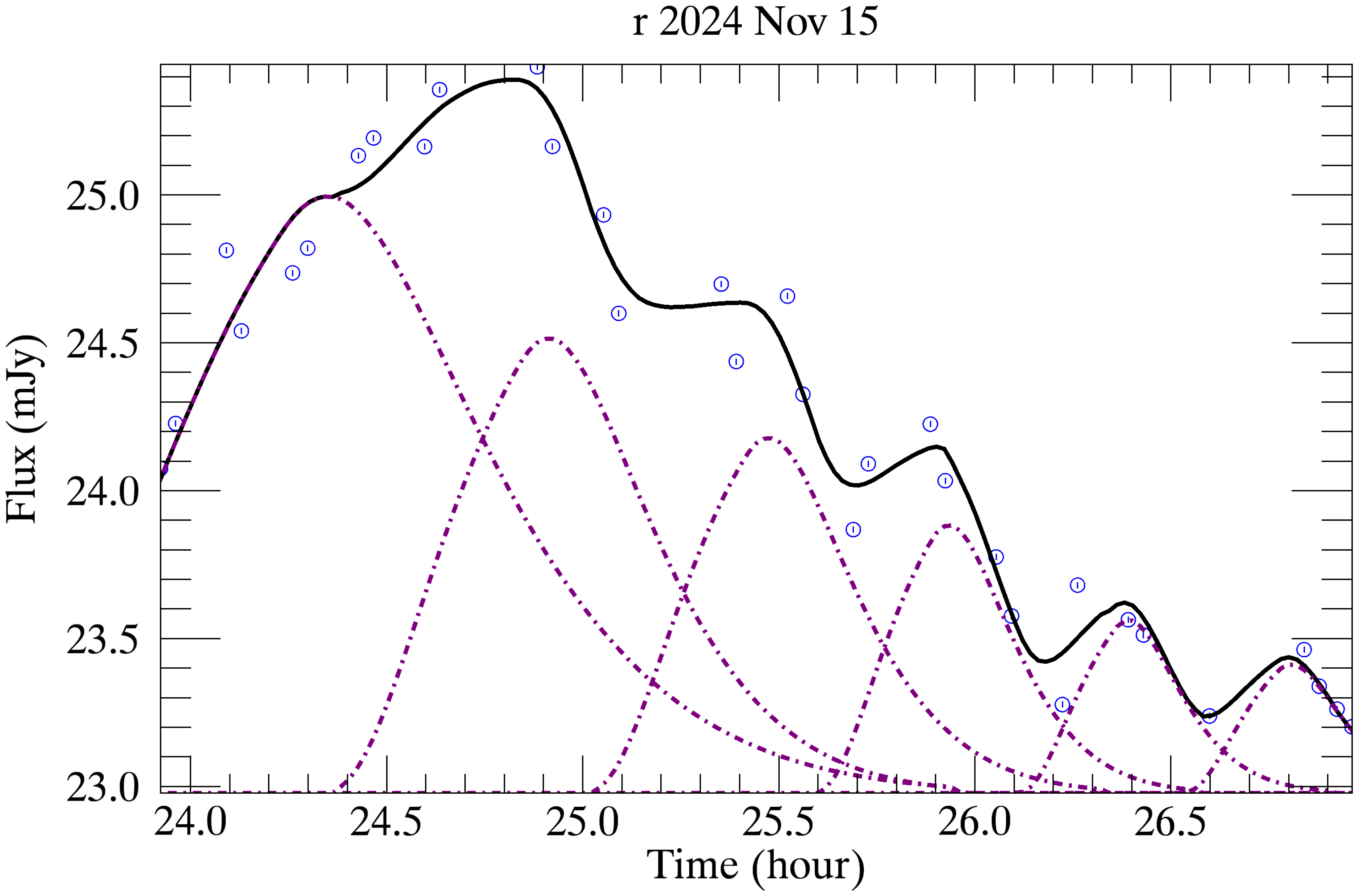}
\includegraphics[scale=0.06]{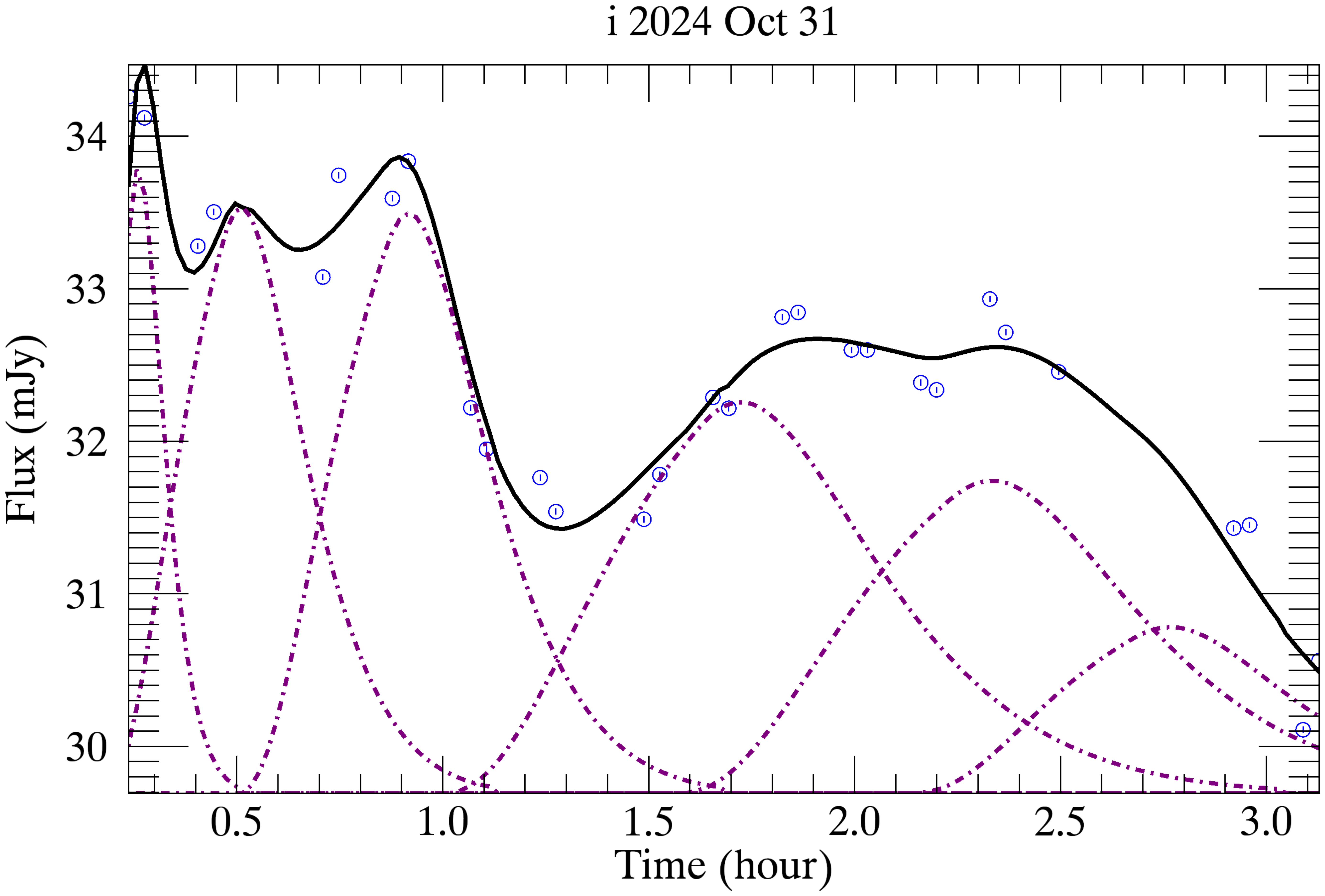}
\includegraphics[scale=0.06]{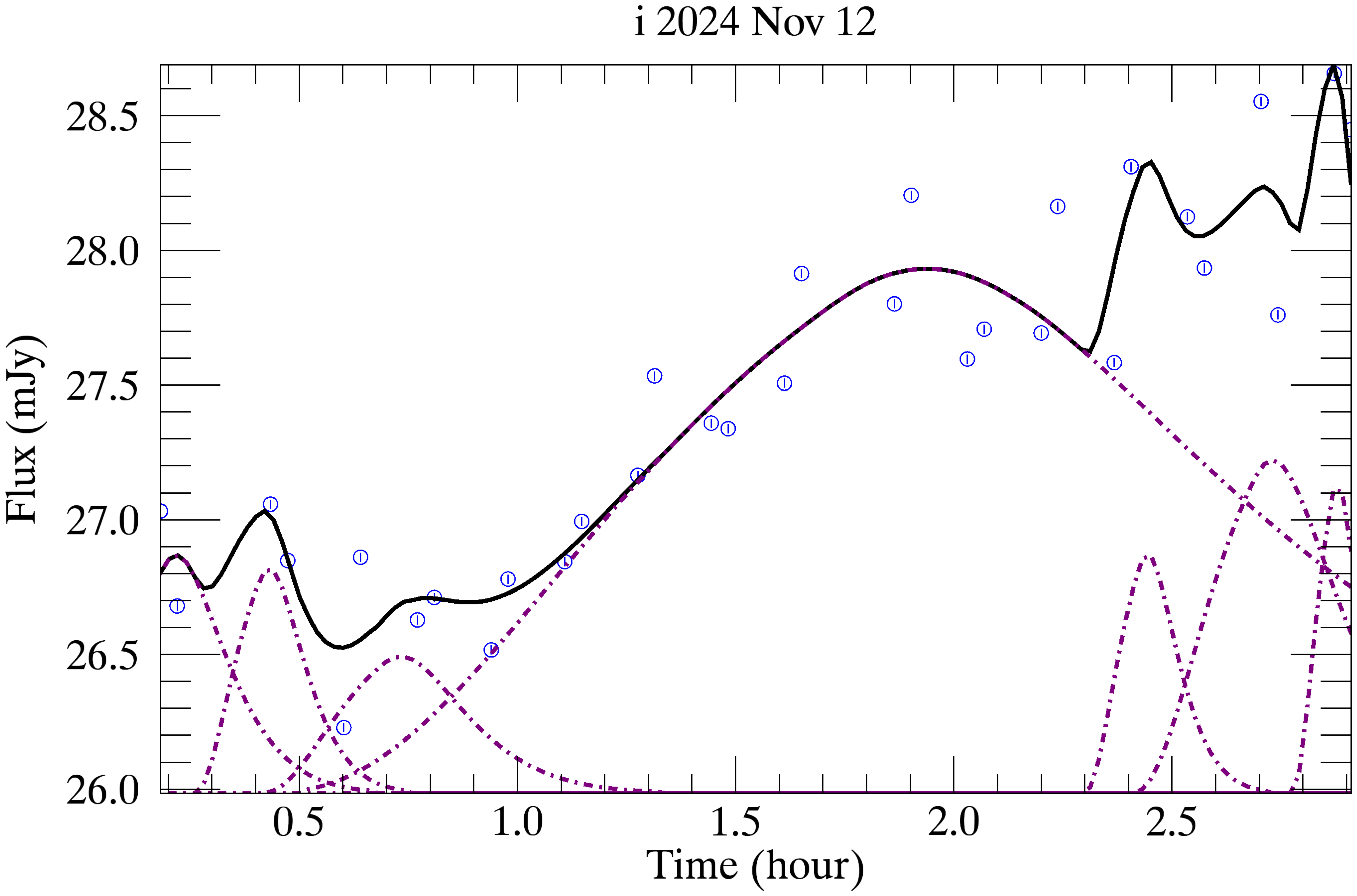}
\includegraphics[scale=0.06]{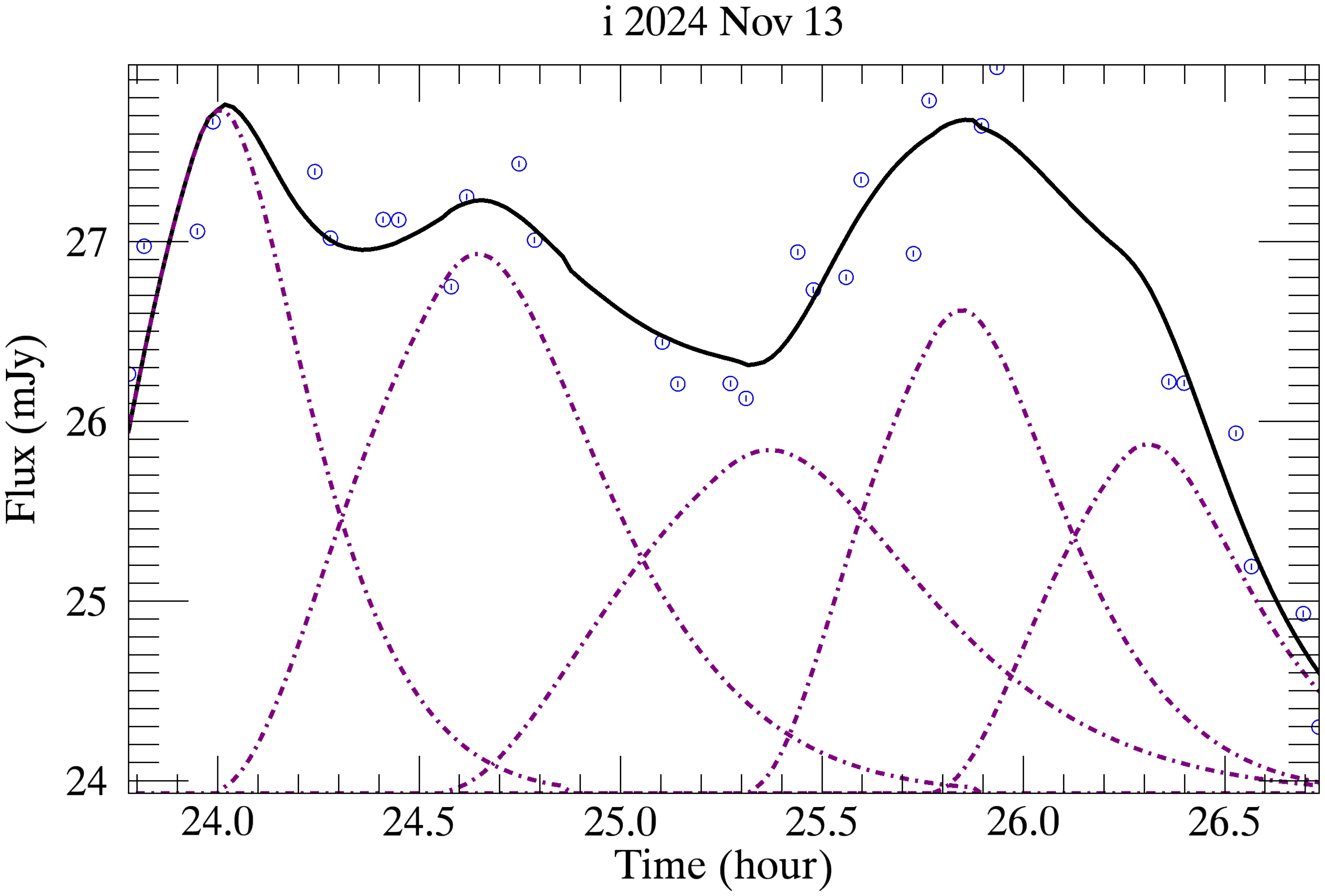}
\includegraphics[scale=0.06]{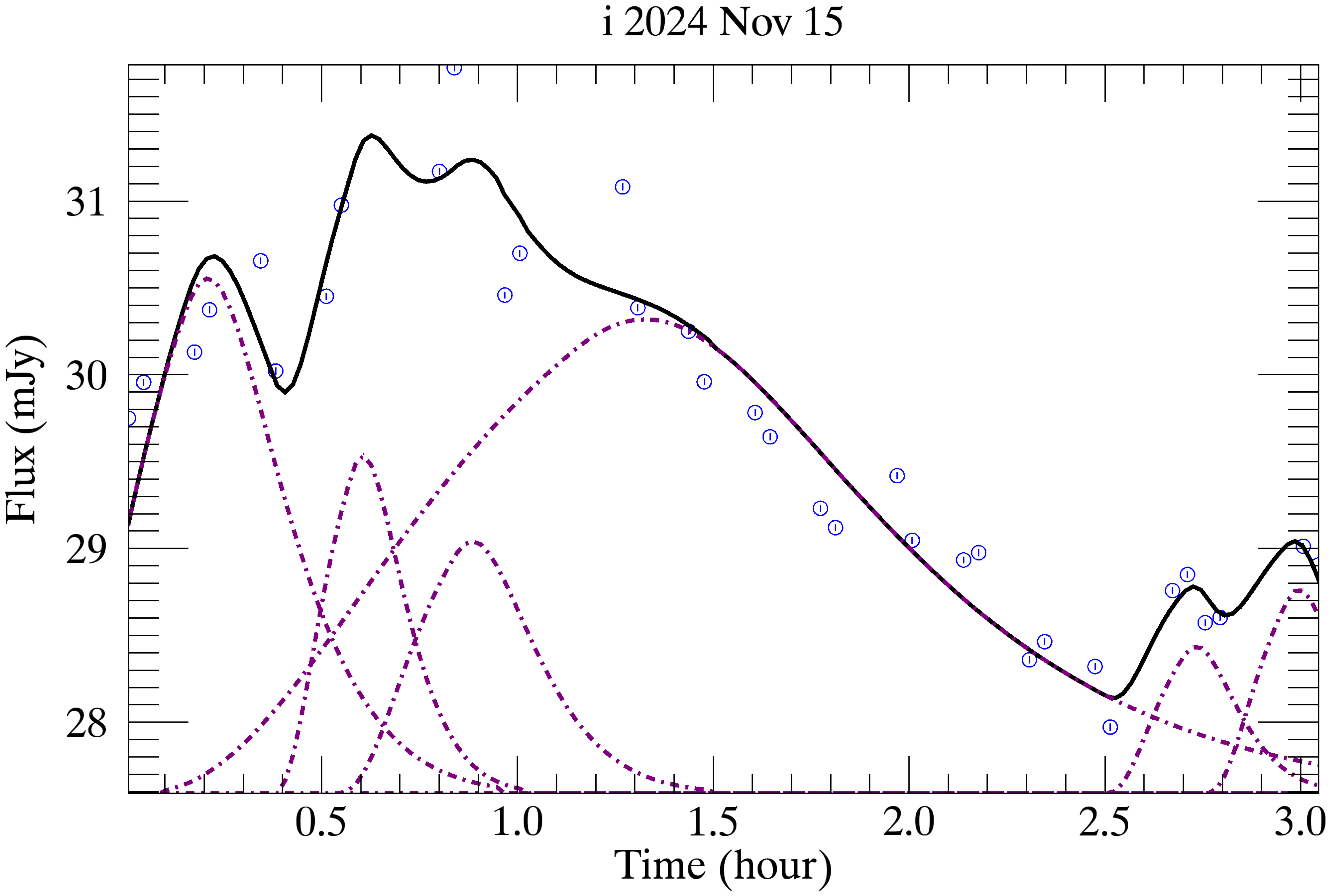}
\caption{IDV fitting results. In each panel, the blue circle points are the original data, the violet dashed lines show the fitted individual flares, and the black solid line is the fitting result.
\label{13}}
\end{figure*}

Figure \ref{13} shows the fitting results of the model. In each subplot, the blue circles represent the observational data, the purple dashed lines correspond to the different flare components obtained through fitting, and the black solid line represents the total fitted curve, which is the sum of all flare components. The pulse parameters used for modeling the light curve are detailed in \citet{Xu23}. The first column of Table \ref{table9} lists the dates of each IDV observation. The second and third columns provide the number of pulses and their average amplitudes, respectively. The fourth column presents the average width of the pulses ($\tau_{\text{flare}}$). The fifth column gives the range of sizes of the emitting regions (or ``cells'') in AU, inferred from the assumed shock speed of $0.1 c$ and the pulse duration. The sixth column lists the number of data points on each date. The last column presents the Pearson correlation coefficients for each light curve and fit. In the framework of the Kolmogorov theory of turbulence \citep{Kolm41}, the Kolmogorov scale represents the smallest spatial scale (``cell'' sizes) in a turbulent flow at which the kinetic energy of turbulence is completely dissipated and converted into thermal energy. Consequently, the minimum spatial scales are determined based on the shortest time-scales of flux variations that can be directly employed to probe the Kolmogorov scale \citep{Agar23}. The results indicate that turbulence is present throughout most regions of the jet. The sizes of all turbulent regions (or ``cells'') range from $1.76$ to $35.24$ AU, with a continuous distribution of cell sizes. The Kolmogorov scale (i.e., the diameter of the smallest emitting region) that we have derived is $1.76$ AU (see Table \ref{table9}). \citet{Meng17} fitted synchrotron pulses to the flare of BL Lacertae. Assuming a shock speed of $0.1 c$, they estimated the size of the smallest turbulent cell to be approximately $1.5 \sim 2.3$ AU, which is consistent with our results. In contrast, the largest cell sizes correspond to the physical scale of the plasma jet or the correlation length within the plasma. When the size of a turbulent cell exceeds this range, it may become unstable \citep{Bhat13,Meng17,Webb21,Webb23}. Alternatively, the large cells may result from unresolved groups of numerous smaller cells. These groups reflect synchrotron pulses generated by compact, fragmented regions within the jet when impacted by a shock \citep{Agar23}.

\begin{table*}
   \centering
   \caption{KRM Fitting Results. Column 1 shows the date of each IDV observation and the band; Columns 2 and 3 give the number of pulses and the average amplitude; Column 4 shows the average width of the pulses ($\tau_{\text{flare}}$); Column 5 indicates the range of cell sizes in AU based on the assumed shock speed of $0.1 c$ and the duration of the pulse; Column 6 gives the number of data points; Column 7 presents the Pearson correlation coefficients for each light curve and fit.}
   %\label{tab:tab1}
   \renewcommand\arraystretch{1.2}
  \setlength{\tabcolsep}{3.8mm}
   \begin{tabular}{cccccccccccc} % four columns, alignment for each
      \hline
Date    &   Pulse Number  &   Amplitude (mJy) &   $\tau_{\text{flare}}$ (h)   &   Cell Size (AU)   &   Data Number   &   Correlation Coefficient  \\\hline
2024 Oct 29 (g)	&	5	&	1.02	&	0.95	&	3.16 $\sim$ 11.38	&	34	&	0.96	\\
2024 Oct 31 (g)	&	6	&	2.22	&	1.19	&	5.69 $\sim$ 19.74	&	31	&	0.99	\\
2024 Nov 12 (g)	&	8	&	1.23	&	1.10	&	3.36 $\sim$ 22.68	&	36	&	0.97	\\
2024 Nov 13 (g)	&	5	&	1.54	&	1.55	&	5.22 $\sim$ 18.25	&	28	&	0.89	\\
2024 Nov 15 (g)	&	6	&	1.67	&	1.28	&	7.18 $\sim$ 14.64	&	34	&	0.98	\\
2024 Oct 29 (r)	&	5	&	0.84	&	1.30	&	6.51 $\sim$ 13.56	&	34	&	0.89	\\
2024 Oct 31 (r)	&	6	&	2.01	&	1.42	&	4.13 $\sim$ 26.30	&	30	&	0.99	\\
2024 Nov 12 (r)	&	8	&	1.36	&	1.11	&	1.76 $\sim$ 35.24	&	36	&	0.98	\\
2024 Nov 13 (r)	&	5	&	1.39	&	1.43	&	8.15 $\sim$ 14.99	&	29	&	0.93	\\
2024 Nov 15 (r)	&	6	&	1.11	&	1.31	&	6.37 $\sim$ 19.95	&	34	&	0.99	\\
2024 Oct 31 (i)	&	6	&	2.90	&	1.38	&	3.02 $\sim$ 17.40	&	29	&	0.97	\\
2024 Nov 12 (i)	&	8	&	1.18	&	1.02	&	2.39 $\sim$ 34.33	&	32	&	0.95	\\
2024 Nov 13 (i)	&	5	&	2.66	&	1.71	&	10.72 $\sim$ 19.30	&	30	&	0.94	\\
2024 Nov 15 (i)	&	6	&	1.85	&	1.26	&	5.15 $\sim$ 29.53	&	34	&	0.96	\\\hline
\label{table9}
   \end{tabular}
\end{table*}

\subsection{Simultaneous Colour Variations}
\label{sec_color}
The BWB trend has been observed in BL Lacertae on both short and long time-scales \citep[e.g.,][]{Stal06,Gaur19}. In our observations, we found the BWB trend. For results with a $p$-value less than $0.001$ (at a $99.9\%$ significance level), all exhibited a strong BWB trend (with correlation coefficient $r > 0.5$), and the slopes varied between $0.27$ and $2.10$ (see Table \ref{table6}, Table \ref{table7} and Figure \ref{7} $\sim$ \ref{11}). For results with a $p$-value less than $0.01$ (at a $99\%$ significance level), all displayed moderate, weak, or strong BWB trends (with correlation coefficient $r$ ranging from $0.1$ to $1.0$), and the slopes varied between $0.18$ and $1.46$. In comparison, the colour analysis results for the entire sample (spanning approximately one month) indicate that BL Lacertae exhibits an almost achromatic behavior, with a slope ranging from $0.03$ to $0.14$. \citet{Vill04} found that the long-term variability of BL Lacertae is nearly achromatic, while the short-term variability demonstrates chromatic behavior. Furthermore, their analysis revealed a strong BWB trend with a slope of approximately $0.4$ during an intraday flare. Our results are consistent with those of \citet{Vill04}.

Due to the radiatively inefficient accretion discs in BL Lacs \citep{Ghis97}, their emission is predominantly jet-dominated. According to the jet emission model, the colour behavior characteristics are significantly influenced by the electron energy distribution. The BWB behavior of IDV implies that, within a specific emitting region (a distinct ``cell'' or sub-region of emission), relativistic particles are rapidly evolving. This results in different cells sequentially dominating the overall observed emission at various times. Within these cells, relativistic electrons are rapidly injected or accelerated, followed by a radiative cooling process \citep{Bach23}. Specifically, if the injected electrons have higher or lower energies compared to the cooled electrons, it leads to BWB or RWB trends in colour variation, respectively \citep{Kirk98,Mast02}. When shocks accelerate electrons, the population of high-energy electrons increases. These high-energy electrons radiate more intensely at higher frequencies, causing the spectral peak to shift toward the blue end. This process naturally gives rise to the observed BWB trend. The reasons for the observed complex colour-magnitude behaviors (e.g., varying or near-zero slopes) are as follows: as high-energy electrons gradually cool, low-frequency radiation may dominate, potentially exhibiting a weakening or even opposite trend. Alternatively, if the shock-accelerated plasma blobs are weaker or more diffuse, the changes in high-frequency and low-frequency radiation become more comparable, resulting in relatively flatter slopes \citep{Kirk98,Vill02,Bonn12}. On the long time-scales, high-energy electrons can also decelerate due to synchrotron radiation or inverse Compton scattering. This dynamic equilibrium leads to a relatively stable ratio between high-energy and low-energy electrons, resulting in a steady spectral energy distribution and consequently producing flat colour variations \citep{Kirk98,Bonn12}. The achromatic colour behavior of BL Lacertae on long time-scales can also be explained by the geometric model. In this model, the rotation of an inhomogeneous helical jet induces variations in the viewing angle, subsequently resulting in variable Doppler boosting effects \citep{Vill02}.

\begin{figure*}
\centering
\includegraphics[scale=.18]{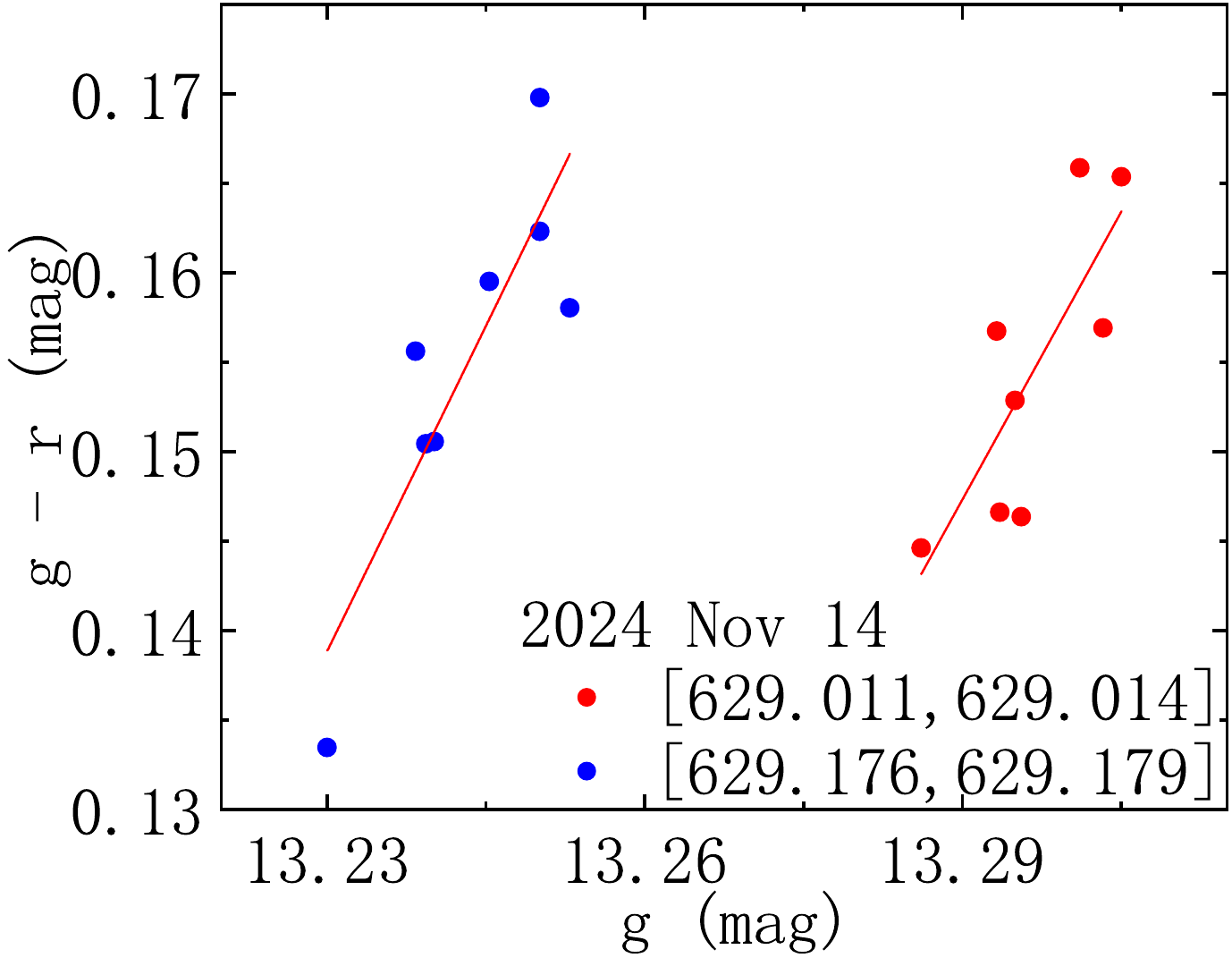}
\includegraphics[scale=.18]{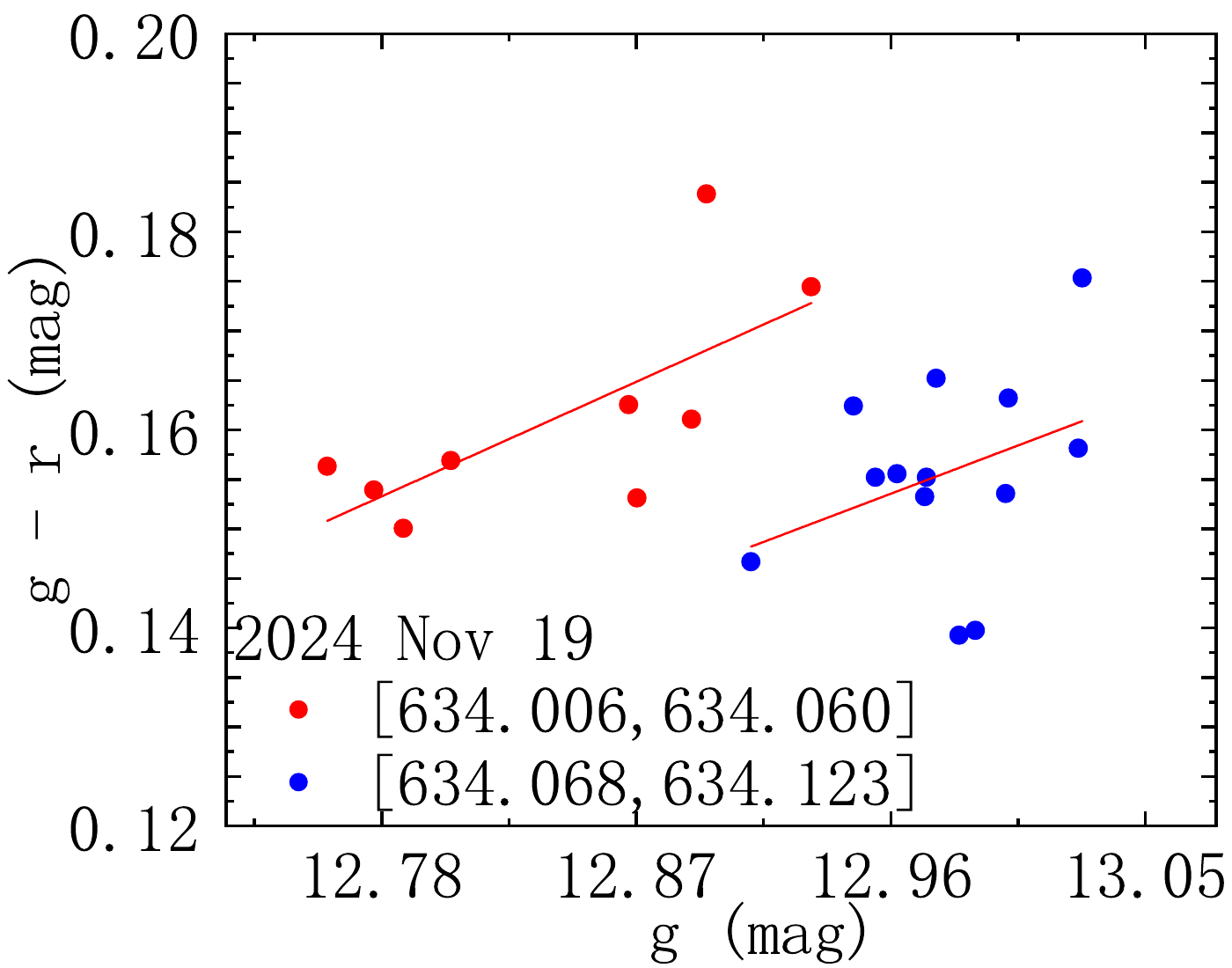}
\includegraphics[scale=.18]{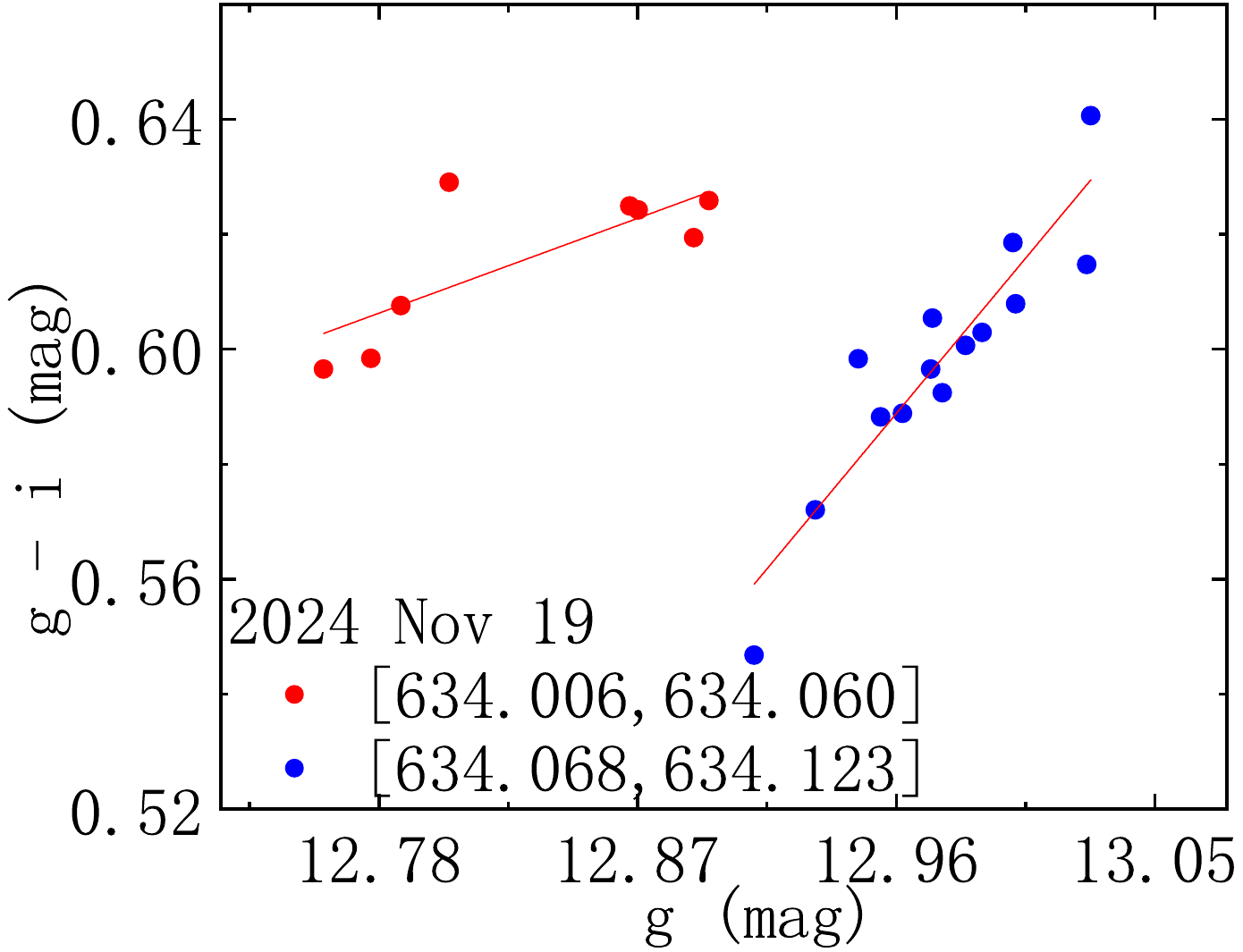}
\includegraphics[scale=.18]{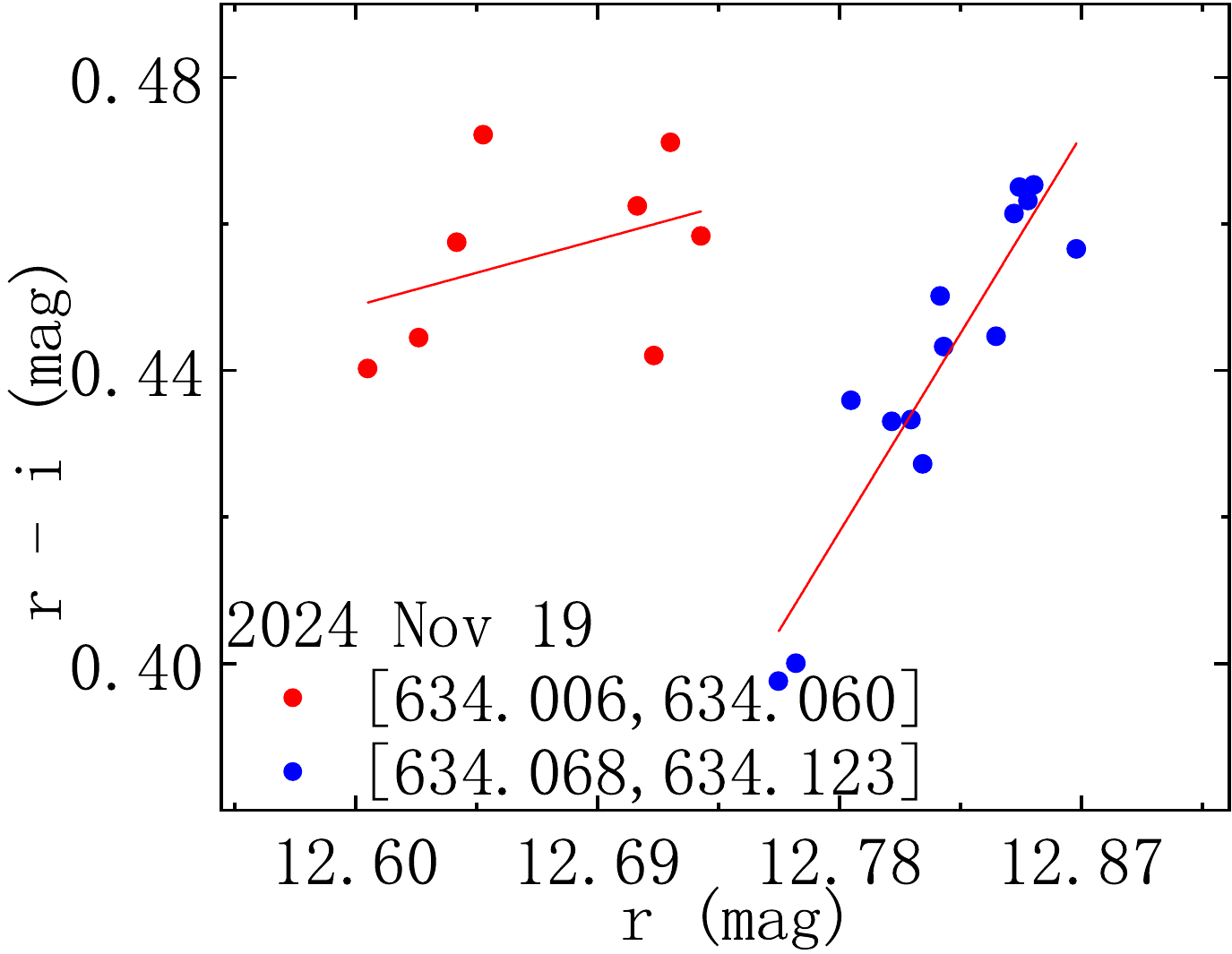}
\caption{The correlation between magnitude and colour index, where the coloured dots represent distributions of different stages, the red lines indicate the linear fitting, and the rectangular boxes mark the time range of observations (JD +2460000). The first plot on the left shows the correlation between $g$ magnitude and the colour index $g - r$ observed on November 14, 2024 ($1.6$ m telescope). The three plots on the right correspond to observations on November 19, 2024, showing the correlations of $g$ magnitude versus $g - r$, $g - i$, and $r$ magnitude versus $r - i$, respectively ($50$ cm telescope).
\label{14}}
\end{figure*}

During our observations on November $14$ and $19$, we detected a time evolution of colour, evidenced by the clustering of the $g-r$, $g-i$, and $r-i$ colour indices into two distinct branches. In Figure \ref{14}, we plotted the CMDs that exhibit distinct branching structures. To enhance visibility, we omitted the error bars. In this figure, the two branches (represented by red and blue points respectively), which belong to two different observation periods ($\mathrm{JD} + 2460000$), are clearly visible. During two distinct temporal phases on November $19$ ($50$ cm telescope), the fitted slopes showed variations of $0.13$ and $0.11$ for $g-r$, $0.18$ and $0.60$ for $g-i$, and $0.10$ and $0.60$ for $r-i$. On November 14 ($1.6$ m telescope), two evolutionary phases exhibited fitted slopes of $1.21$ and $1.07$ for $g-r$. This colour evolution over time has also been reported by \citet{Kali23} and \citet{Yuan23}. They suggest that the primary cause of this phenomenon could be particle acceleration induced by shock waves within the jet. \citet{Kali23} propose that the electron energy distribution of the newly observed emission components (or ``cells'') is either softer (i.e., a reduced proportion of high-energy electrons) or harder (i.e., an elevated proportion of high-energy electrons) compared to previous observations. This characteristic results in two distinct tracks in the CMD, both of which adhere to the BWB trend.

\subsection{Magnetic reconnection model of the period}
\label{sec_dis_reconn}
Magnetic reconnection occurring within magnetically dominated jets can account for the transient QPOs observed in blazars \citep{Mizu11,Dong20}. When the shocks within the jet interact with an inhomogeneous medium, hydrodynamic instabilities give rise to turbulence behind the shock front \citep{Mizu11}. This turbulence locally amplifies the magnetic field in the form of filamentary structures. The charged particles experience acceleration in a turbulent plasma environment containing rapidly migrating magnetic filaments. A turbulent plasma with fast-moving magnetic filaments is likely a site for the acceleration of charged particles. As the propagating shock continuously interacts with the inhomogeneous medium, the turbulent magnetic field downstream of the shock front intensifies, creating conditions conducive to magnetic reconnection events. Strong magnetic reconnection could generate mini-jets localized, transient, and highly collimated plasma outflows―whose structural configuration resembles the nested-jet model proposed by \citet{Gian09}. When oppositely directed magnetic field lines interact, a significant amount of magnetic energy is released, which accelerates particles and produces the observed rapid optical variability. The distorted magnetic field continues to evolve, giving rise to a quasi-periodic kink structure within the jet. This structure then transforms into a region of compressed, moving plasma with enhanced emission, which may exhibit QPO signatures upon observation. \citet{Jors22} reported that a quasi-periodic feature with a period of approximately $0.57$ days was detected in BL Lacertae based on optical and $\gamma$ - ray observational data acquired between March and December $2020$. During this period, they detected a kink structure within the jet, indicating the presence of a tightly helical magnetic field, and proposed that the kink instability was responsible for the generation of this QPO.

According to the magnetohydrodynamic simulations of relativistic plasma jets (RMHD) simulations conducted by \citet{Dong20} and \citet{Acha21}, kink instabilities lead to a quasi-periodic release of energy, resulting in QPOs. The time-scales of these QPOs are correlated with the development of kink instabilities. Quantifying the transverse motion of the kinked region enables calculation of its growth rate \citep{Mizu09}. \citet{Dong20} estimated that the kink growth time ($\tau_{\text{KI}}$) is equal to the ratio of the transverse displacement of the jet from its centre ($R_{\text{KI}}$) to the average propagation velocity. The transverse displacement of the kink is approximately equal to the size of the emission blob \citep{Jors22}. They found that the growth time of the kink instability is consistent with the period of the QPOs they estimated in their simulations. In the observer's frame, the period ($T_{\text{obs}}$) associated with the kink instability is given by the following formula \citet{Dong20}:
\begin{equation}
P_{\text{obs}} = R_{\text{KI}}/ v_{\text{tr}} \delta ,
\label{eq:LebsequeIp14}
\end{equation}
where $\delta$ is the Doppler factor, $R_{\text{KI}}$ represents the size of the emission region in the co-moving frame (transverse displacement of the strongest kinked region), and $v_{\text{tr}}$ is the average transverse velocity of the kink \citep{Dong20}. The kink instability is related to the properties of the jet, which include its inclination angle, velocity (which in turn affects the Doppler factor), and the size of the emission region. Based on the typical transverse velocity $v_{\text{tr}} \approx 0.16 c$ \citep{Dong20}, the Doppler factor of the jet $\delta = 12.17$ \citep{Liod18}, and the upper - limit estimate of the emission region size of $3.51 \times 10^{14}$ cm derived from our IDV analysis, the corresponding period in the observer's frame should be $100.14$ minutes, which aligns with the time-scales of the QPOs we discovered in this study. Therefore, it is reasonable to suggest that the quasi-period of approximately $100$ minutes we observed can be attributed to the growth of quasi-periodic kinks originating in the inner regions of the jets. It is important to note that kink instabilities are not persistent physical processes in relativistic jets \citep{Dong20}.

To rigorously validate the authenticity of a photometric periodicity, the data sample must span at least six times the proposed period duration \citep{Kidg92}. The duration of observations each night falls short of the required sixfold criterion, thereby limiting the reliability of the inferred approximately $100$ minutes oscillatory signal. However, the currently detected results can only serve as preliminary evidence for the presence of periodic oscillatory behavior with a period of approximately $100$ minutes. To confirm this periodic phenomenon, additional observations that exceed the minimum required observation duration are necessary.

\section{Conclusion}
\label{sec_conclusion}
\vspace{0.4in}
We present simultaneous multi-colour photometric data of BL Lacertae (2018 data points) during October to November $2024$ following its brightest flaring event. After analyzing the flux variations, correlations between magnitude and colour, and quasi-periodic oscillations, our main results are summarized as follows.

(1) The F-test and ANOVA methods indicate that IDV was detected in different bands, specifically, on eight days in the $g$-band, seven days in the $r$-band, six days in the $i$-band, and three days in the $z$-band. We found that as the frequency grows, the duty cycle increases. After subtracting the linear trends, the duty cycle decreased, but it continued to exhibit the same increasing trend with frequency.

(2) Based on the shortest variability time-scale derived from ACF analysis, we obtained the upper limits of the black hole mass as $M_{\bullet} \lesssim 10^{8.29} M_{\odot}$ for a Kerr black hole and $M_{\bullet} \lesssim 10^{8.77} M_{\odot}$ for a Schwarzschild black hole. Our mass estimate for the Kerr case is consistent with the value reported by \citet{Woo02}. Furthermore, the size of the emission region is constrained to $R \le 3.51 \times 10^{14}$ cm, and its distance from the central supermassive black hole is estimated to be $R_H \le 2.83 \times 10^{15}$ cm. These results suggest that the emission region is located deep within the broad-line region.

(3) We use the turbulent jet model to investigate the mechanism of IDV. The fitted light curves match the actual IDV curves well. The Kolmogorov scale that we have derived is $1.76$ AU.

(4) According to the results from simultaneous colour variations, a general BWB trend was detected on intraday time-scales. For the entire sample, BL Lacertae exhibits an almost achromatic behavior. These colour variations can be explained by the jet emission model.

(5) We identified a highly significant (confidence level $>99.99\%$) QPO with a period of approximately $100.77$ minutes, which is consistent with the time-scale of $100.14$ minutes predicted by the magnetic reconnection model.

%Acknowledgements %
%%%%%%
\section*{Acknowledgements}
This work is supported by the National Key Research and Development Program of China (No. 2024YFA1611603) and the Yunnan Key Laboratory of Survey Science (No. 202449CE340002). Mephisto is developed at and operated by the South-Western Institute for Astronomy Research of Yunnan University (SWIFAR-YNU), funded by the ``Yunnan University Development Plan for World-Class University'' and ``Yunnan University Development Plan for World-Class Astronomy Discipline''. This work is also supported by the National Natural Science Foundation of China (grants 11863007, 12063005, 12063007, 11703078), the Yunnan Province Foundation (2019FB004), the Program for Innovative Research Team (in Science and Technology) in University of Yunnan Province (IRTSTYN), and Yunnan Local Colleges Applied Basic Research Projects (2019FH001-12). We acknowledge the science research grants from the China Manned Space Project with No. CMS-CSST-2021-A06. The authors acknowledge support from the ``Science \& Technology Champion Project'' (202005AB160002) and from two ``Team Projects'' - the ``Innovation Team'' (202105AE160021) and the ``Top Team'' (202305AT350002), all funded by the ``Yunnan Revitalization Talent Support Program''. This work is partially supported by a program of the Polish Ministry of Science under the title ``Regional Excellence Initiative'', project No. RID/SP/0050/2024/1.

\section*{Data availability}
The data underlying this article will be shared on reasonable request to the corresponding author.

%%%%%%%%%%%%%%%%%%%%%%%%%%%

%\attachfile{Optical1426data.txt}

\end{document}